\documentclass[11pt,reqno]{amsart}

\usepackage{amsmath}
\usepackage{amssymb}
\usepackage{amsfonts}
\usepackage{amsthm}
\usepackage{a4wide}
\usepackage{bbm}
\usepackage{graphics}
\usepackage{color}
\usepackage[sans]{dsfont}

\usepackage{float}
\usepackage{hyperref}
\usepackage{protosem}
\usepackage{linearA}
\usepackage{pgfkeys}
\usepackage{longtable}
\usepackage{pdflscape}
\usepackage{cancel}
\usepackage{dsfont}
\usepackage{setspace}
\usepackage{array}
\usepackage{hyperref}
\usepackage[toc,page]{appendix}
\usepackage{amssymb}
\usepackage{epstopdf}
 \usepackage{multirow}
\usepackage{epsfig}
\usepackage{lscape}
\usepackage{yhmath}
\usepackage[english]{babel}
\usepackage[latin1]{inputenc}
\usepackage[T1]{fontenc}
\usepackage{wasysym}
\usepackage{tikz,pgflibraryshapes}
\usepackage{pstricks,pst-plot,pstricks-add}
\usepackage{enumerate}
\usetikzlibrary{arrows}
\usetikzlibrary{snakes}
\usepackage{textcomp}

\newtheorem{theorem}{Theorem}[section]
\newtheorem{lemma}[theorem]{Lemma}

\newtheorem{definition}[theorem]{Definition}

\newtheorem{remark}[theorem]{Remark}

\numberwithin{equation}{section}

\def\XXint#1#2#3{{\setbox0=\hbox{$#1{#2#3}{\int}$}
     \vcenter{\hbox{$#2#3$}}\kern-.5\wd0}}

\renewcommand{\eqref}[1]{Eq.~(\ref{#1})}

\renewcommand{\d}{\mathrm{d}}

\newcommand{\captionfonts}{\footnotesize}

\makeatletter  
\long\def\@makecaption#1#2{%
  \vskip\abovecaptionskip
  \sbox\@tempboxa{{\captionfonts #1: #2}}%
  \ifdim \wd\@tempboxa >\hsize
    {\captionfonts #1: #2\par}
  \else
    \hbox to\hsize{\hfil\box\@tempboxa\hfil}%
  \fi
  \vskip\belowcaptionskip}
\makeatother   

\title[Quantum Probability]{Quantum Probability Theory and the Foundations of Quantum Mechanics}

\begin{document}

\author[J. Fr\"ohlich]{J\"urg Fr\"ohlich}
\address[J. Fr{\"o}hlich]{Institut f{\"u}r Theoretische Physik, HIT K42.3, ETH Z{\"u}rich, CH-8093 Z{\"u}rich, Switzerland}
\email{juerg@phys.ethz.ch}
\author[B. Schubnel]{Baptiste Schubnel}
\address[B. Schubnel]{Departement Mathematik, ETH Z{\"u}rich, CH-8092 Z{\"u}rich, Switzerland}
\email{baptiste.schubnel@math.ethz.ch}

\maketitle
\small
\tableofcontents

\normalsize
\section{A glimpse of Quantum Probability Theory and of a Quantum Theory of Experiments}
\label{sec1}

By and large, people are better at coining expressions than at filling them with interesting, concrete contents. Thus, it may not be very surprising that there are many professional probabilists who  may have heard the expression but do not appear to be aware of the need to develop \textit{"quantum probability theory"} into a thriving, rich, useful field featured at meetings and conferences on probability theory. Although our aim, in this essay, is not to contribute new results on quantum probability theory, we hope to be able to let the reader feel the enormous potential and richness of this field. What we intend to do, in the following, is to
contribute some novel points of view to the \textit{"foundations of quantum mechanics"},
using mathematical tools from "quantum probability theory" (such as the theory of operator algebras).

The "foundations of quantum mechanics" represent a notoriously thorny and enigmatic subject. Asking twenty-five grown up physicists to present their views on the foundations of quantum mechanics, one can expect to get the following spectrum of reactions \footnote{This story is purely fictional, but quite plausible.}: Three will refuse to talk -- alluding to the slogan "shut up and calculate" -- three will say that the problems encountered in this subject are so difficult that it might take another 100 years before they will be solved; five will claim that the "Copenhagen Interpretation",  \cite{Cop}, has settled all problems, but they are unable to say, in clear terms, what they mean; three will refer us to Bell's book \cite{Bell}  (but admit they have not understood it completely); three confess to be "Bohmians" \cite{Bohm} (but do not claim to have had an encounter with Bohmian trajectories); two claim that all problems disappear in the Dirac-Feynman path-integral formalism \cite{Di,Fey};  another two believe in "many worlds" \cite{Everett} but make their income in our's, and two advocate "consistent histories" \cite{Griff}; two swear on QBism \cite{QB}, (but have never seen "les demoiselles d'Avignon"); two are convinced that the collapse of the wave function \cite{GRW} - spontaneous or not - is fundamental; and one thinks that one must appeal to quantum gravity to arrive at a coherent picture, \cite{Pen}.

 \textit{ Almost all of them are convinced that theirs is the only sane point of view} \footnote{and that Heisenberg's 1925 paper \cite{Hei} cannot be understood.}. Many workers in the field have lost the ability to do technically demanding work or never had it.   Many of them are knowingly or unknowingly envisaging an extension of quantum mechanics -- but do not know how it will  look like. But some claim that "quantum mechanics cannot be extended" \cite{Col-Re}, (perhaps unaware of the notorious danger of "no-go theorems"). 
 
 At least fifteen of the views those twenty-five physicists present logically contradict one another. Most colleagues are convinced that somewhat advanced mathematical methods are superfluous in addressing the problems related to the foundations of quantum mechanics, and they turn off when they hear an expression such as "$C^{*}$-algebra" or "type-III factor".
Well, it might just turn out that they are wrong! What appears certain is that the situation is somewhat desperate, and this may explain why people tend to become quite emotional when they discuss the foundations of quantum mechanics; (see, e.g., \cite{Werner_blog}).

When the senior author had to start teaching quantum mechanics to students, many years ago, he followed the slogan "shut up and calculate" -- until he decided that the situation described above, namely the fact that we do not really understand, in a coherent and conceptual way, what that most successful theory of physics called "quantum mechanics" tells us about Nature, represents an intellectual scandal.

Our essay will, of course, not remove this scandal. But we hope that, with some of our writings, (see also \cite{Fr-Schub},\cite{Fr-Schub2}), we may be able to contribute some kind of intellectual "screw driver" useful in helping to unscrew \footnote{"d\'evisser les probl\`emes" (in reference to A. Grothendieck)} the enigmas at the root of the scandal, before very long. We won't attempt to extend or "complete" quantum mechanics (although we bear people no grudge who try to do so, and we wish them well). We are convinced that starting from simple, intuitive, general principles ("information loss" and "entanglement generation") and then elucidating the $mathematical$  $structure$ inherent in quantum mechanics will lead to a better understanding of its deep message. (Of course, we realize that our hope is lost on people who are convinced that the mysteries surrounding the interpretation of quantum mechanics can be unravelled without any use of somewhat advanced mathematical concepts.)

Just to be clear about one point: We are not claiming to present any "revolutionary" new ideas; and we do not claim or expect to get much credit for our attempts.

But, by all means, let's get started! Quantum mechanics is "quantum", and it is intrinsically "probabilistic" \cite{Einstein,Born}. We should therefore expect that it is intimately connected to quantum probability theory,   hence to "non-commutative measure theory", etc. However, in the end, \textit{"quantum mechanics is quantum mechanics and everything else is everything else!"} \footnote{\textit{"The one thing to say about art is that it is one thing. Art is art-as-art and everything else is everything else."} Ad Reinhardt, \cite{AdRei}}

\subsection{Might quantum probability theory be a subfield of (classical) probability theory?}

  And -- if not -- what's different about it? These questions are  related to one concerning the existence of \textit{hidden variables}. The first convincing results on  hidden variables were brought forward by Kochen and Specker \cite{K-Sp} and (independently) by Bell \cite{Bell}. These matters are so well known, by now, that we do not repeat them here. The upshot is that, loosely speaking, quantum probability theory cannot be imbedded in classical probability theory (except in the case of a two-level system). 
  
  The deeper problems of quantum mechanics can probably only be understood if we admit a notion of $time$, introduce  time-evolution, proceed to consider \textit{ repeated  measurements}, i.e., time-ordered sequences of observations or measurements resulting in a time-ordered sequence of  events, and understand \textit{in which way information gets lost for ever}, in the course of time evolution. (We believe that this will lead to an acceptable "ontology" of quantum mechanics \cite{Goldstein, Bohm}) not involving any fundamental role of the "observer".)
  
   In both worlds, the classical and the quantum world, physical quantities or (potential) properties are represented by self-adjoint operators, $a=a^{*}$, and possible events by spectral projections, $\Pi$, or certain products thereof (POVM's; see Appendix A to Section 4, and Subsection 5.4). A successful measurement or observation of a physical quantity or property represented by an operator $a=a^{*}$ results in \textit{one} of several possible events, $\Pi_1,...,\Pi_k$ (spectral projections of $a$), with the properties that
 \begin{equation}
 \label{1.0}
 \begin{split}
  &(i) \text{ } \text{ } \text{ } \Pi_{\alpha}^2=\Pi_{\alpha}=\Pi_{\alpha}^{*} ,  \text{ }\alpha=1,...,k,\\
  &(ii) \text{ } \text{ } \Pi_{\alpha} \Pi_{\beta}=\delta_{\alpha \beta} \Pi_{\alpha},\\
  &(iii) \text{ }  \sum_{\alpha=1}^{k} \Pi_{\alpha}= \mathds{1}.
  \end{split}
  \end{equation}
   
   Suppose we carry out a sequence of \textit{mutually "independent"} measurements or observations of physical quantities, $a_1,...,a_n$, ordered in time, i.e., $a_1$ before $a_2$ before $a_3$ ... before $a_n$ ($a_1 \prec a_2 \prec ... \prec a_n$). A physical theory should enable us to predict the probabilities for all possible "histories",
   \begin{equation*}
   h_1^n(\underline{\alpha})= \lbrace \Pi_{\alpha_1}^{(1)},..., \Pi_{\alpha_n}^{(n)} \rbrace,
   \end{equation*}
of events, where $ \Pi_{\alpha_1}^{(i)},..., \Pi_{\alpha_{k_i}}^{(i)} $ are the possible events resulting from a successful measurement of $a_i$, $i=1,...,n$. -- On the basis of what prior knowledge? Well, we must know the time evolution of physical quantities and the "state", $\omega$, of the system, $S$, we observe. That means that, given a state $\omega$, there should exist a functional, $\text{Prob}_{\omega}$, that associates with each history $ \lbrace \Pi_{\alpha_1}^{(1)},..., \Pi_{\alpha_n}^{(n)} \rbrace$ -- but for what $family$ of histories, i.e., for which properties $a_1,...,a_n$? -- a probability
\begin{equation}
0 \leq \mu_{\omega}(\alpha_1,...,\alpha_n) \equiv \text{Prob}_{\omega}  \lbrace \Pi_{\alpha_1}^{(1)},..., \Pi_{\alpha_n}^{(n)} \rbrace \leq 1.
\end{equation}
By property (iii) in \eqref{1.0}, 
\begin{equation}
\underset{\alpha_1,...,\alpha_n}{\sum} \mu_{\omega}(\alpha_1,...,\alpha_n)=1,
\end{equation}
because $\text{Prob}_{\omega}$ is normalized such that $\text{Prob}_{\omega}\lbrace \mathds{1}, \mathds{1},... \rbrace=1$. In a classical theory, the projections $ \lbrace \Pi_{\alpha_i}^{(i)} \rbrace_{\alpha_i=1}^{k_i}$, $i=1,...,n$, are characteristic functions on a measure space, $M_S$, and a state, $\omega$, is a probability measure on $M_S$. It then follows from property  (iii) that
\begin{equation}
\label{1.3}
\sum_{\alpha=1}^{k_i} \text{Prob}_{\omega} \lbrace \Pi_{\alpha_1}^{(1)},...,\Pi_{\alpha}^{(i)},...,\Pi_{\alpha_n}^{(n)} \rbrace=   \text{Prob}_{\omega} \lbrace \Pi_{\alpha_1}^{(1)},...,\Pi_{\alpha_{i-1}}^{(i-1)},\Pi_{\alpha_{i+1}}^{(i+1)},...,\Pi_{\alpha_n}^{(n)} \rbrace.
\end{equation}
 for arbitrary $i=1,...,n$.
 
 If we consider a quantum mechanical system with finitely many degrees of freedom then the projections $\lbrace \Pi_{\alpha_i}^{(i)} \rbrace$ are orthogonal projections on a separable Hilbert space, $\mathcal{H}$, and, by Gleason's theorem \cite{Gleason}, $\omega$ is given by a density matrix, $\rho_{\omega}$, on $\mathcal{H}$. Moreover, according to \cite{Lue,Schw,Wig, Kochen},
 \begin{equation}
 \label{1.4}
 \text{Prob}_{\omega}\lbrace \Pi_{\alpha_1}^{(1)},..., \Pi_{\alpha_n}^{(n)}  \rbrace=\text{Tr}_{\mathcal{H}} \left( \Pi_{\alpha_n}^{(n)} ... \Pi_{\alpha_1}^{(1)} \rho_{\omega}  \Pi_{\alpha_1}^{(1)}... \Pi_{\alpha_n}^{(n)}     \right).
 \end{equation}
 
 The problem with  \eqref{1.4}  is that, most often, it represents physical and probability-theoretical nonsense. For example, it is usually left totally unclear what physical quantities or properties of $S$ will be measurable (i.e., which family of histories will become observable), given a time evolution $\tau_{t,s}$ and a state $\omega$. But such problems do not stop people from studying \eqref{1.4} again and again -- and we are no exception. To address one of  the key problems with \eqref{1.4}, we study an example.
 
 We consider a monochromatic beam of light, which, according to Einstein \cite{Einst.1905}, consists of individual photons of fixed frequency. We then bring three filters into the beam that produce linearly polarized light. The direction of polarization is given by an angle $\theta$ that can be varied by rotating the filter around the axis defined by the beam; see Figure \ref{Fig1}.

\begin{center}
\begin{tikzpicture}
\label{Fig1}
        
    \def\a{4}  
      \def\b{-1.75}    
        
     \draw[-](0,2)--(2,1.5);   
       \draw[-](0,2)--(-0.5,-2);   
         \draw[-](-0.5,-2)--(1.6,-2.4);   
           \draw[-](2,1.5)--(1.6,-2.4);   
           
            \draw[-,rotate=20](\a,2+\b)--(2+\a,1.5+\b);   
       \draw[-,rotate=20](\a,2+\b)--(-0.5+\a,-2+\b);   
         \draw[-,rotate=20](-0.5+\a,-2+\b)--(1.6+ \a,-2.4+\b);   
           \draw[-,rotate=20](2+\a,1.5+\b)--(1.6+\a,-2.4+\b);

            \def\c{8}  
      \def\d{5}

         \draw[-,rotate=-30](\c,2+\d)--(2+\c,1.5+\d);   
       \draw[-,rotate=-30](\c,2+\d)--(-0.5+\c,-2+\d);   
         \draw[-,rotate=-30](-0.5+\c,-2+\d)--(1.6+ \c,-2.4+\d);   
           \draw[-,rotate=-30](2+\c,1.5+\d)--(1.6+\c,-2.4+\d);  
           
   \draw[-](-1.3,-0.6)--(-0.35,-0.6); 
     \draw[-,dotted](0.8,-0.6)--(-0.35,-0.6); 
     \draw[-](0.8,-0.6)--(0.8,2.3);
  \draw[-](0.8,-0.6)--(4.2,-0.6); 
   \draw[-,dotted](5.4,-0.6)--(4.2,-0.6); 
      \draw[-](5.4,-0.6)--(5.4,2.3);
 \draw[-](5.4,-0.6)--(8.4,-0.6); 
  \draw[-,dotted](9.9,-0.6)--(8.4,-0.6); 
     \draw[-](9.9,-0.6)--(9.9,2.3);
 \draw[->](9.9,-0.6)--(13.5,-0.6);

    \draw[->](0.8,-0.6)--(1.4,1);
      \draw[->](5.4,-0.6)--(4.8,1);
      \draw[->](9.9,-0.6)--(11,1);
    
\draw[->] (.8,0.8) arc (90:30:.45cm);
\draw[->] (9.9,0.8) arc (90:40:0.9cm);
\draw[->] (5.4,0.2) arc (90:-240:0.7cm);

   \draw (0.8,1) node[right] { $\theta_1$};    
    \draw (5.4,-1.4) node[right] { $\theta_2$};    
     \draw (9.9,1) node[right] { $\theta_3$};

\draw[decorate, decoration={snake, segment length=3mm, amplitude=2mm}] (-1.3,-0.6)--(-0.35,-0.6);     
  \draw[decorate, decoration={snake, segment length=3mm, amplitude=2mm}] (0.8,-0.6)--(4.2,-0.6);        
   \draw[decorate, decoration={snake, segment length=3mm, amplitude=2mm}] (5.4,-0.6)--(8.4,-0.6); 
    \draw[decorate, decoration={snake, segment length=3mm, amplitude=2mm}] (9.9,-0.6)--(13,-0.6);

       \draw (5.,-3) node[right] { Figure 1.1};   
       
        \end{tikzpicture}

\end{center}

With the filter $i$, we associate two possible events
\begin{align*}
\Pi_{+}^{(i)} &\leftrightarrow \text{ a  photon passes through filter } i \\
\Pi_{-}^{(i)}=\mathds{1}-\Pi_{+}^{(i)} & \leftrightarrow \text{ a  photon does \textit{not} pass through filter } i.
 \end{align*} 
 Experimentally, one finds that, for any initially unpolarized beam of light, (meaning that the photons are all prepared in a state 
 $\omega_0 \propto \frac{1}{2} \text{Tr}_{\mathbb{C}^2} ( \cdot )$),
 \begin{equation}
 \label{1.5}
 \text{Prob}_{\omega_0} \lbrace \Pi_{+}^{(i)},\Pi_{+}^{(j)} \rbrace= \frac{1}{2} \cos^2(\theta_i-\theta_j), \text{ } i<j,
 \end{equation}
 if only filters $i$ and $j$ are present, with $1 \leq i<j \leq 3$. It follows from \eqref{1.5} that
 \begin{equation}
 \label{1.6}
 \text{Prob}_{\omega_0} \lbrace \Pi_{+}^{(i)},\Pi_{-}^{(j)} \rbrace= \frac{1}{2} \sin^2(\theta_i-\theta_j), \text{ } i<j,
 \end{equation}
 the probability that a photon passes the first filter, $i$, being $1/2$, because the initial beam is \textit{unpolarized} (or circularly polarized). Formulae (\ref{1.5}) and (\ref{1.6}) can be tested experimentally by intensity measurements before and after each filter. If the projections $\Pi^{(i)}_{\pm}$ were characteristic functions on a measure space,$M_{\text{photon}}$, then we would have that
 \begin{equation}
 \label{1.7}
 \text{Prob}_{\omega_0}  \lbrace \Pi_{+}^{(1)},\Pi_{-}^{(3)} \rbrace \leq   \text{Prob}_{\omega_0}  \lbrace \Pi_{+}^{(1)},\Pi_{-}^{(2)} \rbrace +  \text{Prob}_{\omega_0}  \lbrace \Pi_{+}^{(2)},\Pi_{-}^{(3)} \rbrace. 
 \end{equation}
 For, 
  \begin{equation}
  \begin{split}
 \label{1.8}
 \text{Prob}_{\omega_0}  \lbrace \Pi_{+}^{(1)},\Pi_{-}^{(3)} \rbrace &= \text{Prob}_{\omega_0}  \lbrace \Pi_{+}^{(1)},\Pi_{-}^{(2)},\Pi_{-}^{(3)} \rbrace +  \text{Prob}_{\omega_0}  \lbrace \Pi_{+}^{(1)},\Pi_{+}^{(2)},\Pi_{-}^{(3)} \rbrace \\
 &   \leq \text{ right side of \eqref{1.7}},
 \end{split} 
 \end{equation}
 where \eqref{1.8} follows from the sum rule (\ref{1.3}), and the upper bound (\ref{1.7}) from the trivial inequality $0 \leq \Pi_{\pm}^{(i)} \leq 1 $. Plugging expression (\ref{1.6}) into (\ref{1.7}). we conclude that 
 \begin{equation}
 \label{1.9}
 \frac{1}{2}\sin^2(\theta_1-\theta_3) \leq \frac{1}{2}\sin^2(\theta_1-\theta_2)+\frac{1}{2}\sin^2(\theta_2-\theta_3).
 \end{equation}
 
 Setting $\theta_1=0$, $\theta_2=\pi/6$ and $\theta_3=\pi/3$, \eqref{1.9} would imply that $3/8 \leq 1/8+1/8$, which is obviously wrong! What is going on? It turns out that the sum rule (\ref{1.8}) is violated. The reason is that the projections $\Pi_{\pm}^{(2)}$ and  $\Pi_{\pm}^{(3)}$ \textit{ do not commute}. This fact is closely related to \textit{ non-vanishing interference} between  $\Pi_{+}^{(2)}$ and  $\Pi_{-}^{(2)}$ analogous to the interference encountered in the \textit{double-slit experiment}. Interference between 
 $\Pi_{+}^{(2)}$ and  $\Pi_{-}^{(2)}$ is measured by 
 \begin{equation}
 \mathcal{I}( \Pi_{+}^{(2)},\Pi_{-}^{(2)} \mid  \Pi_{\alpha}^{(1)},\Pi_{\beta}^{(3)}):=\text{Tr}_{\mathcal{H}} (\Pi_{\beta}^{(3)}  \Pi_{+}^{(2)}   \Pi_{\alpha}^{(1)} \rho_{\omega_0}    \Pi_{\alpha}^{(1)}   \Pi_{-}^{(2)} \Pi_{\beta}^{(3)} ).
 \end{equation}
 Choosing $\alpha=+$ and $\beta=-$ (for example), we find a non-vanishing interference term, which explains why the sum rule (\ref{1.8}) is violated. What is the message? The first filter, 1, may be interpreted as \textit{"preparing"} the photons in the beam hitting the filter 2 to be linearly polarized as prescribed by the angle $\theta_1$. In our experimental set-up there is \textit{no} instrument measuring whether a photon has passed filter 2, or not. The \textit{only} measurement is made after filter 3, where either a photon triggers a Geiger counter to click, or there is no photon triggering the Geiger counter. Let us denote the probability for the first event (Geiger counter clicks)  by $p_+$, the second by $p_-$. The histories contributing to $p_-$ are 
 
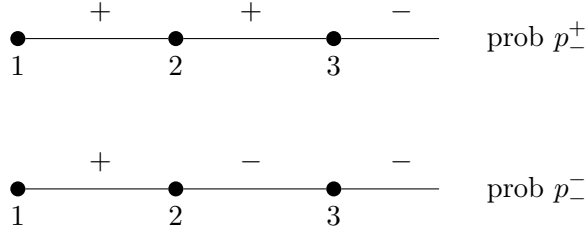
\begin{center}
\begin{tikzpicture}
        
    \def\a{-2}  

     \draw[-](0,0)--(1.9,0);   
       \draw[-](2.1,0)--(4,0);   
         \draw[-](4.2,0)--(5.5,0);

         \draw[-](0,\a)--(1.9,\a);   
       \draw[-](2.1,\a)--(4,\a);   
         \draw[-](4.2,\a)--(5.5,\a);

          \fill[black] (2,0) circle (1mm);
           \fill[black] (4.1,0) circle (1mm);
            \fill[black] (-0.1,0) circle (1mm); 
            
                \fill[black] (2,\a) circle (1mm);
           \fill[black] (4.1,\a) circle (1mm);
            \fill[black] (-0.1,\a) circle (1mm);

          \draw (-0.1,-0.1) node[below] {$1 $};  
            \draw (4.1,-0.1) node[below] {$3 $};  
              \draw (2,-0.1) node[below] {$2 $}; 
              
                \draw (-0.1,-0.1+\a) node[below] {$1 $};  
            \draw (4.1,-0.1+\a) node[below] {$3 $};  
              \draw (2,-0.1+\a) node[below] {$2 $};

               \draw (1,0.1) node[above] {$+ $};  
            \draw (5,0.1) node[above] {$- $}; 
              \draw (3,0.1) node[above] {$+ $}; 
              
                   \draw (1,0.1+\a) node[above] {$+ $};  
            \draw (5,0.1+\a) node[above] {$- $}; 
              \draw (3,0.1+\a) node[above] {$- $};

             \draw (6,0) node[right] {prob $p_{-}^{+}$}; 
              \draw (6,0+\a) node[right] {prob $p_{-}^{-}$};   
    
        \end{tikzpicture}
\end{center}
 with $p_-=p_{-}^{+}+p_{-}^{-}$. These two histories show interference. Given that a photon has passed filter 1, expressions (\ref{1.5}) and (\ref{1.6}) appear to imply that
 \begin{equation}
 \begin{split}
 p_{-}^{+}&= \cos^2(\theta_1-\theta_2) \sin^2(\theta_2-\theta_3)\\
p_{-}^{-}&= \sin^2(\theta_1-\theta_2).
 \end{split}
 \end{equation}
 The unique history contributing to $p_+$appears to be 
 
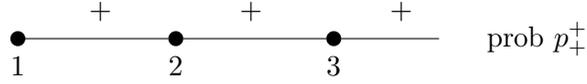
\begin{center}
\begin{tikzpicture}
        
    \def\a{-2}  

     \draw[-](0,0)--(1.9,0);   
       \draw[-](2.1,0)--(4,0);   
         \draw[-](4.2,0)--(5.5,0);

          \fill[black] (2,0) circle (1mm);
           \fill[black] (4.1,0) circle (1mm);
            \fill[black] (-0.1,0) circle (1mm);

          \draw (-0.1,-0.1) node[below] {$1 $};  
            \draw (4.1,-0.1) node[below] {$3 $};  
              \draw (2,-0.1) node[below] {$2 $};

               \draw (1,0.1) node[above] {$+ $};  
            \draw (5,0.1) node[above] {$+$}; 
              \draw (3,0.1) node[above] {$+ $};

             \draw (6,0) node[right] {prob $p_{+}^{+}$}; 
             
                 \end{tikzpicture}
\end{center}
 with
 $$p_{+}^{+}=\cos^2(\theta_1-\theta_2) \cos^2(\theta_2-\theta_3),$$
 and, indeed,
 $$p_{+}^{+}+p_{-}^{+}+p_{-}^{-}=1.$$
 These findings can be accounted for by associating with the event $"+"$ the operator 
 $$X_+=\Pi_{+}^{(3)} \Pi_{+}^{(2)}$$
 and with the event $"-"$ the operators
 $$X_{-}^{+}= \Pi_{-}^{(3)} \Pi_{+}^{(2)} \qquad \text{and} \qquad X_{-}^{-}= \Pi_{-}^{(2)}.$$
 Then, 
 $$X_{+}^{*} X_{+} + (X_{-}^{+})^*X_{-}^{+}+  (X_{-}^{-})^*X_{-}^{-}=\mathds{1}.$$
 It should however be noted that 
  $$X_{+}X_{+}^{*}  +X_{-}^{+} (X_{-}^{+})^*+ X_{-}^{-} (X_{-}^{-})^* \neq \mathds{1}.$$
For this reason, some people may prefer to replace $X_+$ by the pair $X_1:=\Pi_{+}^{(3)} \Pi_{+}^{(2)}$, $X_2:=\Pi_{+}^{(3)} \Pi_{-}^{(2)}$, and to set $X_3:=X_{-}^{+}$, $X_4:= \Pi_{-}^{(3)} \Pi_{-}^{(2)}$. Then,
\begin{equation}
\sum_{\alpha=1}^{4}X^{*}_{\alpha} X_{\alpha}=  \sum_{\alpha=1}^{4}X_{\alpha} X^{*}_{\alpha}=\mathds{1}.
 \end{equation}
 The family $(X_1,X_2,X_3,X_4)$ is called (the "square root" of) a \textit{positive operator-valued measure} (POVM); (see \cite{Perez}, and Subsects. \ref{sec4.3} and 5.4). Note that 
 \begin{equation*}
 \text{Tr}_{\mathcal{H}}(X_2 \rho_{\omega_0} X_2^{*})=\text{Prob}_{\omega_0} \lbrace \Pi_{-}^{(2)}, \Pi_{+}^{(3)} \rbrace 
 \end{equation*}
 corresponds to the "virtual history"
 
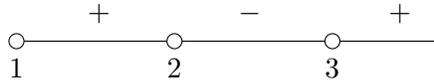
\begin{center}
\begin{tikzpicture}
        
    \def\a{-2}  

     \draw[-](0,0)--(1.9,0);   
       \draw[-](2.1,0)--(4,0);   
         \draw[-](4.2,0)--(5.5,0);

          \draw[black] (2,0) circle (1mm);
            \draw[black] (4.1,0) circle (1mm);
            \draw[black]  (-0.1,0) circle (1mm);

          \draw (-0.1,-0.1) node[below] {$1 $};  
            \draw (4.1,-0.1) node[below] {$3 $};  
              \draw (2,-0.1) node[below] {$2 $};

               \draw (1,0.1) node[above] {$+ $};  
            \draw (5,0.1) node[above] {$+$}; 
              \draw (3,0.1) node[above] {$-$};

                 \end{tikzpicture}
\end{center}
 which cannot be interpreted classically. This should not bother us, because no measurement is carried out between filters 2 and 3.

There is a more drastic way to present these findings: Consider N filters in series, the $ j^{th}$ filter being rotated through an angle $j\pi/2N$. The probability for an initially vertically polarized photon $(\theta_0=0)$ to be transmitted through all the filters is then given by 
\begin{equation}
p_+=\text{Prob}_{\omega_{\theta_0=0}} \lbrace \Pi^{(1)}_{+},..., \Pi^{(N)}_{+} \rbrace = \left(\cos \left( \frac{\pi}{2N} \right) \right)^{2N}\underset{N \rightarrow \infty}{\longrightarrow} 1.
\end{equation}
If, however, all filters, except for the $N^{th}$ one, are removed, then
\begin{equation}
p_+':=\text{Prob}_{\omega_{\theta_0=0}} \lbrace \Pi_{+ }^{(N)} \rbrace =\cos^2 \left( \frac{\pi}{2}\right) =0.
\end{equation}
If $\Pi_{+}^{(1)},...,\Pi_{+}^{(N)}$ were "classical events", i.e., non-negative random variables, then one would have that $p_+ \leq p_+'.$ (See \cite{Bell,Maassen} for closely related arguments.)

Actually, the discussion presented above, although often repeated, is somewhat misleading. The only measurement takes place \textit{after} the last filter and is supposed to determine whether a photon has passed all the filters, or not.  The corresponding physical quantity corresponds to the operators  $\Pi_{\pm}^{(N)}$, where $N$ is the label of the last filter, and the measurement consists in verifying whether a Geiger counter placed after the last filter has clicked, or not. The filters have \textit{nothing} to do with measurements, but determine (or, at least, affect) the form of the \textit{ time evolution} of the photons. The use of POVM's in discussing experiments like the ones above is not justified at a fundamental, conceptual level. It merely substitutes for a more precise understanding of time-evolution that involves including the filters in a quantum-mechanical description. It appears that, often, POVM's are used to cover up a lack of understanding of the time-evolution of large quantum systems. The role they play in a quantum theory of experiments is briefly described in Subsect. 5.4.

A more compelling way of convincing oneself that quantum probability cannot be imbedded in classical probability theory than the one sketched above consists in studying \textit{correlation} \textit{matrices} of families of (non-commuting) possible events in two \textit{independent} systems. One then finds that the numerical range of possible values of the matrix elements of such correlation matrices is strictly larger in quantum probability theory than in classical probability theory, as discovered by Bell \cite{Bell,Tsirelson}. See \cite{K-Sp} for an alternative approach.

\subsection{The quantum theory of experiments}
\label{1.2}
We return to considering a system, $S$, and suppose that $n$ consecutive measurements have been carried out \textit{successfully}, with the $i^{th}$ measurement described by spectral projections $\Pi_{\alpha}^{(i)}=(\Pi_{\alpha}^{(i)})^{*}$, $\alpha=1,...,k_i$, of a physical quantity $a_i=a^{*}_i$, with
\begin{equation}
\Pi^{(i)}_{\alpha} \Pi^{(i)}_{\beta}= \delta_{\alpha \beta} \Pi^{(i)}_{\alpha}, \qquad \sum_{\alpha=1}^{k_i} \Pi_{\alpha}^{(i)} = \mathds{1},
\end{equation}
for all $i$. (We could also use POVM's, instead of projections, but let's not!)  The probability of a history $\lbrace \Pi^{(1)}_{\alpha_1},...,\Pi^{(n)}_{\alpha_n} \rbrace$ in a state $\omega$ of $S$ given by a density matrix $\rho_{\omega}$ is then given by formula (\ref{1.4}), above.   The measurements can be considered to be successful only if the \textit{sum rules} (\ref{1.3}) \textit{ are very nearly satisfied, for all i}. Whether this is true, or not, can be determined by studying the interference between different histories. Given a state $\omega$, we define $N \times N$ matrices, $P^{\omega}= (P^{\omega}_{\underline{\alpha}, \underline{\alpha}'})$, $N=k_{1}$ $^{ ...}$\text{ }$k_{n}$, by 
\begin{equation}
\label{1.14}
P^{\omega}_{\underline{\alpha}, \underline{\alpha}'}:= \omega \left( \Pi_{\alpha_1}^{(1)} ... \Pi_{\alpha_n}^{(n)} \Pi_{\alpha'_n}^{(n)}...\Pi_{\alpha'_1}^{(1)} \right)= \text{Tr} \left( \Pi_{\alpha'_n}^{(n)}...\Pi_{\alpha'_1}^{(1)} \rho_{\omega} \Pi_{\alpha_1}^{(1)} ... \Pi_{\alpha_n}^{(n)}  \right),  
\end{equation}
where $\omega(a)$ is the expectation of the operator $a$ in the state $\omega$. Measurements of the quantities $a_1,...,a_n$ can be considered to be successful \textit{only} if $P^{\omega}$ is approximately \textit{diagonal}, i.e.,
 \begin{equation}
 \label{1.15}
 \vert P^{\omega}_{\underline{\alpha}, \underline{\alpha}'} \vert \ll \frac{1}{2} \left( P^{\omega}_{\underline{\alpha}, \underline{\alpha}}+ P^{\omega}_{\underline{\alpha}', \underline{\alpha}'} \right), 
 \end{equation}
 which is customarily called "\textit{decoherence}"; see, e.g., \cite{Hepp, GM-Hartle, blanchard2003decoherence, HMaass}. All this is discussed in much detail in Sections 4.3 and 5. In particular, we will show that decoherence is a consequence of \textit{"entanglement} \textit{generation"} between the system $S$ and its environment $E$ and of \textit{"information} \textit{loss"}, meaning that the original state of $S \vee E$ \textit{cannot} be fully reconstructed  from the results of arbitrary measurements carried out after some time $T$, long after the interactions between $S$ and $E$ have set in; see Sect. 5, and \cite{J.F2,J.F., Buchh-Roberts}. In local relativistic quantum theory with massless particles (photons), the kind of information loss alluded to here is a general consequence of Huyghens' principle \cite{Buchholz} and of "Einstein causality". It appears already in classical field theory. In local relativistic quantum theory it becomes manifest in the circumstance that the algebra of operators representing physical quantities measurable by a localized observer \textit{after} some time $T$  does \textit{not} admit any \textit{pure} states. See \cite{Buchh-Roberts}.
 
\begin{center}
\begin{tikzpicture}
        
    \def\a{0.3}  
    \def\b{0.165}

     \draw[->](0,0)--(2,0);   
     \draw[->](0,0)--(0,2);   
      \draw[->](0,0)--(-1,-1);

   \draw[-](5,-0.5)--(3,2); 
     \draw[-](5+\a,-0.5)--(3+\a,2); 
     \draw[-](5,-0.5)--(7,2); 
     \draw[-](5+\a,-0.5)--(7+\a,2); 
   \draw[-](5,-0.5)--(5+\a,-0.5);
 
    \draw[->,decorate, decoration={snake, segment length=3mm, amplitude=0.65mm}] (5+\b,-0.5)--(3+\b,2); 
      \draw[->,decorate, decoration={snake, segment length=3mm, amplitude=0.65mm}]  (5+\b,-0.5)--(7+\b,2); 
      
       \fill[black,opacity=0.25]   (5.35,1.8+\b) to[bend left=20]  (5.39,1.5+\b) to[bend left=-20]  (5.35,1.3+\b)  to[bend left=20]  (5.39,1.1+\b)   to[bend left=-20]  (5.35,0.9+\b)  to[bend left=20]  (5.39,0.7+\b)    to[bend left=0]    (5.35,0.5+\b)to[bend left=60] (5.05,0.5+\b) to[bend left=0] (5.05,0.5+\b)   to[bend left=0] (5.05,0.7+\b)  to[bend left=20] (5.05,0.9+\b)   to[bend left=-20]  (5.05,1.1+\b)   to[bend left=20] (5.05,1.3+\b)  to[bend left=-20] (5.05,1.5+\b)   to[bend left=20] (5.05,1.8+\b)       ;
        \draw[-](5.2,0.4)--(6.5,2); 
        \draw[-](5.2,0.4)--(3.9,2); 
        \draw[-,dashed] (6.5,2) to[bend left=30]    (3.9,2);
         \draw[-,dashed] (6.5,2) to[bend left=-15]    (3.9,2);

         \draw[-,color=gray] (5.05,1.8+\b)   to[bend left=30]     (5.35,1.8+\b);
          \draw[-,color=gray] (5.05,1.8+\b)to[bend left=-15]     (5.35,1.8+\b);
          
           \fill[black,opacity=0.35] (5.05,1.8+\b)   to[bend left=30]     (5.35,1.8+\b)to[bend left=-15]     (5.35,1.8+\b);

     \draw[-,dotted](5.2,0.45)--(6.4,2); 
     \draw[-,dotted](5.2,0.45)--(4,2); 
      \draw[-,dotted](5.2,0.42)--(6.45,2); 
     \draw[-,dotted](5.2,0.42)--(3.95,2); 

\draw[-,dotted](5.3,0.5)--(6.3,2); 
\draw[-,dotted](5.3,0.6)--(6.2,2); 
\draw[-,dotted](5.4,0.7)--(6.1,2.1); 
\draw[-,dotted](5.4,0.7)--(6.1,2.1); 
\draw[-,dotted](5.45,0.75)--(6,2.12); 

\draw[-,dotted](5.1,0.5)--(4.1,2); 
\draw[-,dotted](5.1,0.6)--(4.2,2); 
\draw[-,dotted](5.,0.7)--(4.3,2.1); 
\draw[-,dotted](5.,0.7)--(4.3,2.1); 
\draw[-,dotted](4.95,0.75)--(4.4,2.12);

        \draw(2,0) node[below]{$x$};
         \draw(0,2) node[right]{$t$};
           \draw(-1,-1) node[right]{$\text{ }y$};

          \draw(6.3,-0.5) node[right]{$t_0$};
         \draw(6.3,0.4) node[right]{$t^*$};
           \draw[-,color=gray](5.3,-0.5)--(6.3,-0.5);
             \draw[-,color=gray](5.3,0.4)--(6.3,0.4);
        
         \draw(5.2,0.8) node[above]{$\mathcal{O}$};
         
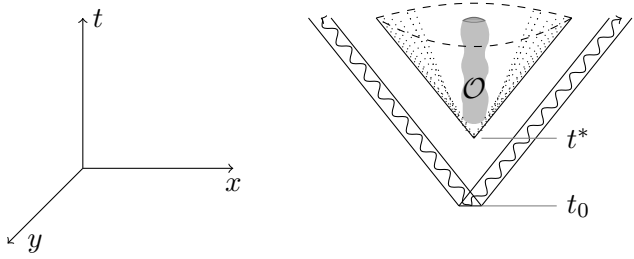
           \draw(5,-2) node[above]{ \footnotesize The event at time $t_0<t^*$ involving photons can never be observed by the observer  $\mathcal{O}$};

        \end{tikzpicture}
\end{center}
 
The \textit{key problem} in a quantum theory of experiments (or measurements/observations) is, however, to find out \textit{which physical quantities will be measured} (i.e., what potential properties of a system will become \textit{"empirical"} properties, or what families of histories of events can be expected to be observed) in the course of time, given the choice of a system, $S$, coupled to an environment, $E$, of a specific time evolution of $S\vee E$, and of a fixed state, $\omega$, of $S\vee E$. This is sometimes referred to as the problem of eliminating the mysterious role of the "observer" from quantum mechanics (making \textit{many worlds} superfluous),  and of determining the \textit{"primitive ontology"} of quantum mechanics, \cite{Goldstein}. This problem will be reckoned with in Subsects. 5.3 and 5.4.

 One customarily distinguishes between \textit{"direct} (or von Neumann) \textit{measurements"} and (indirect, or) \textit{"non-demolition measurements"} carried out on a physical system $S$. It may be assumed that it is clear what is meant by a direct measurement. A non-demolition measurement is carried out by having a sequence of "probes" $(E_k)$ interact with the system $S$, one after another, with the purpose of measuring a physical quantity, $a=a^*$, of $S$ with (for simplicity) finite point spectrum, $\text{spec}(a)=\lbrace \alpha_1,...,\alpha_n \rbrace$. If $S$ is in an eigenstate, $\mid \alpha_i \rangle$, of $a$ corresponding to the eigenvalue $\alpha_i$ right before it starts to interact with the $k^{th}$ probe, $E_k$, the time-evolution of the composed system, $S \vee E_k$, is assumed to leave $\vert \alpha_i \rangle$ invariant but changes the state of $E_k$ in a manner that depends \textit{non-trivially} on $\alpha_i$, for each $i=1,...,n$. This leads to entanglement between $S$ and $E_k$, $k=1,2,3,...$ If, for simplicity, it is assumed that the probes $E_1,E_2,E_3,...$ are all \textit{independent} of one another and that $E_k$  interacts with $S$ strictly \textit{after} $E_{k-1}$ and strictly \textit{before} $E_{k+1}$, then the state of $S$ \textit{decohers} exponentially rapidly with respect to the basis $\vert \alpha_1 \rangle,...,\vert \alpha_n \rangle$, as $k \rightarrow \infty$. More precisely, if $\rho^{(k)}$ denotes the state of $S$ after its interaction with $E_k$ and before its interaction with $E_{k+1}$, with 
 \begin{equation}
 \label{1.16}
 \rho_{\alpha_i,\alpha_j}^{(k)}:= \langle \alpha_i \vert \rho^{(k)} \vert \alpha_j \rangle,
 \end{equation}
 then
 \begin{equation}
 \label{1.17}
  \rho_{\alpha_i,\alpha_j}^{(k)} \longrightarrow \delta_{\alpha_i \alpha_j} \rho_{\alpha_i,\alpha_i},
 \end{equation}
 exponentially rapidly. This is easily verified; (see Subsect. 5.6). A more subtle result on decoherence involving \textit{correlated} probes that lead to \textit{memory effects} has been established in \cite{DeR-Fr}.
 
 One might ask what happens if a \textit{direct measurement} is carried out on \textit{every probe} $E_k$ after it has interacted with $S$, $k=1,2,3,...$. (We assume, for simplicity, that all probes $E_k$ are identical, independent and identically prepared, and that they are all subject to the \textit{same} direct measurement). Then one can show that, under natural non-degeneracy conditions, the state, $\rho^{(k)}$, of $S$, after the passage of $k$ probes $E_1,...,E_k$, \textit{converges to an eigenstate of $a$}, i.e.,  
 \begin{equation}
 \label{1.18}
 \rho^{(k)}  \longrightarrow \vert \alpha_i \rangle \langle \alpha_i \vert,
 \end{equation}
 as $k \rightarrow \infty$, for some $i$, and the probability of approach of $\rho^{(k)}$ to  $\vert \alpha_i \rangle \langle \alpha_i \vert$   is given by $\rho_{\alpha_i,\alpha_i}$. This important result has been derived by M. Bauer and D. Bernard in \cite{Ba-Be}  as a corollary of the \textit{Martingale Convergence Theorem}; (see \cite{Adler} for earlier ideas in this direction.) The convergence claimed in \eqref{1.18} is remarkable, because it says that, asymptotically as $k \rightarrow \infty$, a \textit{pure state} (some eigenstate of $a$) is approached; i.e., a very long sequence of \textit{indirect} (non-demolition) \textit {measurements} carried out on $S$ always results in a \textit{"fact"} (namely, the state of $S$ approaches an eigenvector of the quantity $a$ that one intends to measure).  Somewhat related results ("approach to a groundstate") for more realistic models have been proven in \cite{F-G-S,DeR-K,Fr-Schub3}. \footnote{A result of the form of \eqref{1.18} was conjectured by J.F. in the 90's. But the proof remained elusive.}
 
 In order to control the rate of convergence in \eqref{1.17} and \eqref{1.18}, it is helpful to make use of various notions of \textit{quantum entropy}; (see, e.g., \cite{Roeps,Al-Fannes}).
 
 Some details concerning (indirect) non-demolition measurements and some remarks concerning interesting applications are sketched in Subsect. 5.6; (but see \cite{Guerlin,Ba-Be,Adler,Mott,Fr-Schub2}).
 
 \subsection{Organization of the paper}
 In Section 2, we introduce an abstract algebraic framework for the formulation of mathematical models of physical systems that is general enough to encompass classical and quantum mechanical models. We attempt to clarify what kind of predictions a model of a physical system ought to enable us to come up with. Furthermore, we summarize some important facts about operator algebras needed in subsequent sections.
 
 In Section \ref{section3}, we describe \textit{classical} models of physical systems within our algebraic framework and explain in which sense, and why, they are "realistic" and "deterministic".
 
 In Section 4, we study a general class of quantum-mechanical models of physical systems within our general framework. We explain what some of the key problems in a quantum theory of observations and measurements are. 
 
 The most important section of this essay is Section 5. We attempt to elucidate the roles played by entanglement between a system and its environment and of information loss in understanding "decoherence" and "dephasing", which are key mechanisms in a quantum theory of measurements and experiments; see also \cite{Hepp, Bell, GM-Hartle, HMaass}. In particular, we point out that the state of the composition of a system with its environment can usually not be reconstructed from measurements long after interactions between the system and its environment have set in; ("information loss"). We also discuss the problem of "time in quantum mechanics" and sketch an answer to the question when an experiment can be considered to have been completed successfully; ("when does a detector click?"). Put differently, the "primitive ontology" of quantum mechanics is developed in Subsects. 5.3 and 5.4. Finally, in Subsection 5.6, we briefly develop the theory of indirect non-demolition measurements, following \cite{Ba-Be} .
 
An outline of \textit{relativistic quantum theory} and of the role of space-time in relativistic quantum theory
has been sketched in lectures and will be presented elsewhere; (see also \cite{Bannier}).
 
 The main weakness of this essay (which might be fatal) is that we do not (and cannot) discuss sufficiently many simple, convincing examples illustrating the power of the general ideas presented here. This would simply take too much space. But examples will be discussed in \cite{Fr-Schub2,Fr-Schub3}.
  
  \subsection*{Acknowledgements} A rough first draft of this paper has been written during J.F.'s stay at the School of Mathematics of the Institute for Advanced Study (Princeton), 2012/2013. His stay has been supported by the 'Fund for Math' and the 'Monell Foundation'. He is deeply grateful to Thomas C. Spencer for his most generous hospitality. He acknowledges useful discussions with Ph. Blanchard, P. Deift, S. Kochen and S. Lomonaco. He thanks D.Bernard for drawing his attention to \cite{Ba-Be} and W. Faris for correspondence. He is grateful to D. Buchholz, D. D\"urr, S. Goldstein, J. Yngvason and N. Zanghi for numerous friendly and instructive discussions, encouragement and for the privilege to occasionally disagree in mutual respect and friendship.

\section{Models of Physical Systems}
\label{sec2}
In this section, we sketch a somewhat abstract algebraic framework suitable to formulate \textit{mathematical models of physical systems}. Our framework is general enough to encompass \textit{classical and quantum-mechanical} models. 

Throughout most of this essay, we consider \textit{non-relativistic} models of physical systems, so that, in principle, all "observers" have access to the \textit{same} observational data. For this reason, reference to "observers" is superfluous in the framework to be exposed here. This is radically different in causal relativistic models.

In every model of a physical system, $S$, one specifies $S$ in terms of (all) its \textit{"potential properties"}, i.e., in terms of \textit{"physical quantities"} or \textit{"observables"} characteristic of $S$; see, e.g., \cite{Kochen}. No matter whether we consider classical or quantum-mechanical systems, "physical quantities" are represented, mathematically, by \textit{bounded, self-adjoint, linear operators}. Thus, a system $S$ is specified by a list
\begin{equation}
\mathcal{P}_{S}=\{ a_i\}_{i \in I_S}
\end{equation}
of physical quantities, $a_i=a_{i}^{*}$, characteristic of  $S$ that can be observed or measured in experiments. 

In classical physics, a physical quantity, $a$, is given by a real-valued (measurable or continuous) function on a topological space, $M_S$, which is the \textit{"state space"} of $S$ (the phase space if $S$ is Hamiltonian). Quantum-mechanically, more general linear operators are encountered, and, as is well known, the operators in $\mathcal{P}_S= \{ a_i\}_{i \in I_S}$ need \textit{not} all commute with one another. It is natural to assume that if $a \in \mathcal{P}_S$ is a physical quantity of $S$ then so is any polynomial, $p(a)$, in $a$ with real coefficients.  It is, however, not very plausible that arbitrary real-linear combinations and/or symmetrized products of distinct elements in  $\mathcal{P}_S$ would belong to   $\mathcal{P}_S$. But, in non-relativistic physics, it has turned out to be reasonable to view  $\mathcal{P}_S$ as a \textit{self-adjoint subset of an operator algebra}, $\mathcal{A}_S$, usually taken to be a $C^{*}-$ or a von Neumann algebra, in terms of which a model of $S$ can be formulated. Physicists tend to be scared when they hear expressions like 'C*-' or 'von Neumann algebra'. Well, they shouldn't!

\subsection{Some basic notions from the theory of operator algebras}  
\label{sec2.1}
In order to render this paper comprehensible to the non-expert, we summarize some basic definitions and notions from the theory of operator algebras; for further details see \cite{Opalg}. 

An algebra, $\mathcal{A}$, over the complex numbers is a complex vector space equipped with a multiplication: If $a$ and  $b$ belong to $\mathcal{A}$, then 
\begin{itemize}
\item $\lambda a + \mu b \in \mathcal{A}, \quad \lambda, \mu \in \mathbb{C}$,
\item $a \cdot b \in \mathcal{A}$,
\end{itemize}
where "$\cdot$" denotes multiplication in $\mathcal{A}$. One says that $\mathcal{A}$ is a $^*$\textit{algebra} iff there exists an anti-linear involution, $^*$, on $\mathcal{A}$, i.e., $^*: \mathcal{A} \rightarrow \mathcal{A}$, with $(a^{*})^{*}=a$, for all $a \in \mathcal{A}$, such that 
$$(\lambda a + \mu b)^{*} = \overline{\lambda} a^*+ \overline{\mu} b^*,$$ 
where $\overline{\lambda}$ is the complex conjugate of $\lambda \in \mathbb{C}$, and 
$$(a \cdot b)^*=b^*\cdot a^*.$$
The algebra $\mathcal{A}$ is a \textit{normed algebra} (Banach algebra) if it comes with a norm $\|(\cdot) \|$ satisfying
\begin{itemize}
\item   \begin{equation*} 
\|(\cdot) \| : \mathcal{A} \rightarrow [0,\infty[
\end{equation*}
\item   \begin{equation} \label{2.2} \| a\| =0, \text{ for } a \in \mathcal{A} \implies a=0 \end{equation}
\item   ($\mathcal{A}$ is \textit{complete} in $\|(\cdot) \|$, i.e., every Cauchy sequence in $\mathcal{A}$ converges to an element of $\mathcal{A}$). 
\end{itemize}
A Banach algebra, $\mathcal{A}$, is a $C^*-$\textit{algebra} iff
\begin{equation}
\label{2.3}
\|a^* \cdot a\| = \|a  \cdot a^* \|=\|a\|^2, \text{ } \forall a \in \mathcal{A}.
\end{equation}

We define the \textit{centre}, $\mathcal{Z}_{\mathcal{A}}$, of $\mathcal{A}$ to be the subset of $\mathcal{A}$  given by
\begin{equation}
\label{2.4}
\mathcal{Z}_{\mathcal{A}}:=\{  a \in \mathcal{A} \mid  a \cdot b = b \cdot a , \forall b \in \mathcal{A}\}.
\end{equation}

A \textit{state}, $\omega$, on a $^*$algebra $\mathcal{A}$ with identity $\mathds{1}$ is a linear functional $\omega: \mathcal{A} \rightarrow \mathbb{C}$ with the properties that
\begin{equation}
\label{2.5}
\omega(a^*)=\overline{\omega(a)}, \qquad \omega(a^* a ) \geq 0,
\end{equation}
 for all $a \in \mathcal{A}$, and
 \begin{equation}
 \label{2.6}
 \omega(\mathds{1})=1.
 \end{equation}
 
 A \textit{representation}, $\pi$, of a $C^*$-algebra $\mathcal{A}$ on a complex Hilbert space, $\mathcal{H}$, is a $^*$homomorphism from $\mathcal{A}$ to the algebra, $\mathcal{B}(\mathcal{H})$, of all bounded linear operators on $\mathcal{H}$; i.e., $\pi$ is linear, $\pi (a \cdot b)=\pi(a) \cdot \pi(b)$, $\pi(a^*)=(\pi(a))^*$, and $\|\pi(a) \| \leq \|a \|$, (where $\|A\|$ is the operator norm of a bounded linear operator $A$ on $\mathcal{H}$).
 
 With a $C^*$-algebra $\mathcal{A}$ and a state $\omega$ on $\mathcal{A}$ we can associate a Hilbert space, $\mathcal{H}_{\omega}$, a unit vector $\Omega \in \mathcal{H}_{\omega}$, and a representation, $\pi_{\omega}$, of $\mathcal{A}$ on $\mathcal{H}_{\omega}$ such that $\{ \pi_{\omega }(a) \Omega \mid a \in \mathcal{A}\}$  is dense in   $\mathcal{H}_{\omega}$ (i.e. $\Omega$ is \textit{cyclic} for $\pi_{\omega}(\mathcal{A})$), and
 \begin{equation} 
\label{2.7}
\omega(a)=\langle \Omega, \pi_{\omega}(a)\Omega \rangle,
\end{equation}
where $\langle \cdot ,  \cdot \rangle$ is the scalar product on $\mathcal{H}_{\omega}$. This is the so-called \textit{Gel'fand-Naimark-Segal  (GNS) construction}.

A theorem due to Gel'fand and Naimark says that every $C^*-$algebra, $\mathcal{A}$, can be viewed as a norm-closed subalgebra of $\mathcal{B}(\mathcal{H})$ closed under $^*$, for some Hilbert space $\mathcal{H}$.
 
Thus, consider a $C^*$-algebra $\mathcal{A} \subset \mathcal{B}(\mathcal{H})$, for some Hilbert space $\mathcal{H}$. We define the commuting algebra, or commutant, $\mathcal{A}'$, of $\mathcal{A}$ by
\begin{equation}
\label{2.8}
\mathcal{A}':= \{ a \in \mathcal{B}(\mathcal{H}) \mid a \cdot b =b \cdot a, \forall b \in \mathcal{A} \}.
\end{equation}
The double commutant of $\mathcal{A}$, $\mathcal{A}''$, is defined by
\begin{equation}
\label{2.9}
\mathcal{A}'' \equiv (\mathcal{A}')'= \{ a \in \mathcal{B}(\mathcal{H}) \mid a \cdot b = b \cdot a, \forall b \in \mathcal{A}'\} \supseteq \mathcal{A}. 
\end{equation}
It turns out that $\mathcal{A}'$ and $\mathcal{A}''$ are closed in the so-called weak $^*$ topology of $\mathcal{B}(\mathcal{H})$; i.e., if $\{ a_i \}_{i \in I}$  is a sequence (net) of operators in $\mathcal{A}'$ (or in $\mathcal{A}''$), with
\begin{equation*}
\langle \varphi, a_i \psi \rangle \rightarrow \langle \varphi, a \psi \rangle , \quad  \text{ as } i \rightarrow \infty,
\end{equation*}
for \textit{all} $\varphi, \psi \in \mathcal{H}$, where $a \in \mathcal{B}(\mathcal{H})$, then $a \in \mathcal{A}'$ (or $a \in \mathcal{A}''$, respectively). 
$^*$Subalgebras of $\mathcal{B}(\mathcal{H})$ that are closed in the weak $^*$ topology are called \textit{von Neumann algebras} (or $W^*$-algebras).

Thus, if $\mathcal{A}$ is a $C^*$-algebra contained in $\mathcal{B}(\mathcal{H})$, for some Hilbert space $\mathcal{H}$, then $\mathcal{A}'$ and $\mathcal{A}''$ are von Neumann algebras. A von Neumann algebra $\mathcal{M} \subseteq \mathcal{B}(\mathcal{H})$ is called a $factor$ iff its centre, $\mathcal{Z}_{\mathcal{M}}$, consists of multiples of the identity operator $\mathds{1}$.

A von Neumann factor $\mathcal{M}$ is said to be of type $I$ iff $\mathcal{M}$ is isomorphic to $\mathcal{B}(\mathcal{H}_0)$, for some Hilbert space $\mathcal{H}_0$. A general von Neumann algebra, $\mathcal{N}$, is said to be of type I  iff $\mathcal{N}$ is a direct sum (or integral) over its centre, $\mathcal{Z}_{\mathcal{N}}$, of factors of type I. A $C^*$-algebra $\mathcal{A}$ is called a type-I $C^*$-algebra, iff, for every representation $\pi$, of $\mathcal{A}$ on a Hilbert space $\mathcal{H}$, 
$$\pi(\mathcal{A}):=\{\pi(a) \mid a \in \mathcal{A}\}$$ 
has the property that $\pi(\mathcal{A})''$ is a von Neumann algebra of type I. (For mathematical properties of type-I $C^*$-algebra see \cite{Glimm}, and for examples relevant to quantum physics see \cite{BuGr}).

A $^*$\textit{automorphism}, $\alpha$, of a $C^*$-algebra $\mathcal{A}$ is a linear isomorphism from $\mathcal{A}$ onto $\mathcal{A}$ with the properties
\begin{equation}
\label{2.10}
\begin{split}
\alpha(a \cdot b)&= \alpha(a) \cdot \alpha(b),\\
\alpha(a^*)&=(\alpha(a))^*,
\end{split}
\end{equation}
for all $a,b \in \mathcal{A}$.

It is clear what is meant by $\mathcal{A} \subseteq \mathcal{B}$, where $\mathcal{A}$ and $\mathcal{B}$ are $C^*-$ or von Neumann algebras. We define
\begin{equation}
\label{2.11}
\mathcal{A}' \cap \mathcal{B}:=\{ b \in \mathcal{B} \mid b \cdot a =a \cdot b, \text{ } \forall a \in \mathcal{A}\},
\end{equation}
the "relative commutant" of $\mathcal{A}$ in  $\mathcal{B}$.

Given a set $\mathcal{P}=\{ a_i\}_{i \in I}$ of operators in a $C^*$-algebra $\mathcal{B}$, we define $\langle \mathcal{P} \rangle$  to be the $C^*$-subalgebra of $\mathcal{B}$ generated by $\mathcal{P}$, i.e., the norm-closure of arbitrary finite complex-linear combinations of arbitrary finite products of elements in the set $\{ a_i,a_{i}^{*} \} _{i \in I}$, where $^*$ is the $^*$ operation on $\mathcal{B}$.

A trace $\tau: \mathcal{M}_{+} \rightarrow [0,\infty]$ on a von Neumann Algebra $\mathcal{M}$ is a  function defined on the positive cone, $\mathcal{M}_{+}$, of positive elements of $\mathcal{M}$ (i.e., elements $x \in \mathcal{M}$ of the form $x=y^*y$, $y\in \mathcal{M}$)  that satisfies the properties
\begin{align*}
&(i) \quad \tau(x+y)=\tau(x)+\tau(y), \qquad x,y \in \mathcal{M}_{+}\\
&(ii) \quad \tau(\lambda x)= \lambda \tau(x), \qquad \quad \quad \quad   \lambda \in \mathbb{R}_{+}, \text{ } x \in \mathcal{M}_+\\
&(iii)  \quad \tau(x^*x)=\tau(xx^*), \qquad   \quad \quad x \in \mathcal{M}.
\end{align*}

A trace $\tau$ is   said to be \textit{finite} if $\tau(\mathds{1})<+ \infty$. It can  then be uniquely extended by linearity to a \textit{state} $\tau$ on  $\mathcal{M}$. Conversely, any state $\tau$ on $\mathcal{M}$ enjoying the property 
\begin{equation}
\label{trace}
\tau(a\cdot b) = \tau(b\cdot a), \qquad \forall a, b \in \mathcal{M},
\end{equation}
defines a finite trace on $\mathcal{M}$. We say that   $\tau$ is faithful  if $\tau(x)>0$ for any non-zero element $x \in \mathcal{M}_+$.  A trace $\tau$ is said to be  \textit{normal}  if $\tau(\sup x_i) = \sup \tau(x_i)$ for every bounded net $(x_i )_{i\in I}$ of positive elements in $\mathcal{M}$, and \textit{semifinite},  if, for any $x \in \mathcal{M}_+$, $x \neq 0$, there exists $y \in \mathcal{M}_+$, $0<y \leq x$, such that $\tau(y)<\infty$. 
Traces play an important role in the classification of von Neumann algebras. It can be shown that a von Neumann algebra $\mathcal{M}$ is a direct sum (or direct integral) of factors of type $\text{I}_n$ and type $\text{II}_1$ if and only if it admits a faithful finite normal  trace; see \cite{Opalg}. Similarly,  $\mathcal{M}$ is a direct sum (or direct integral)  of type $\text{I}$, type $\text{II}_1$ and type $\text{II}_{\infty}$ factors iff it admits a faithful semifinite normal trace.   We use these results in Section \ref{sec5} to characterize the centralizer of a state $\omega$.

For the time being, we do not have to know more about operator algebras than what has just been reviewed here. We can test our understanding of the notions introduced above on the example of direct sums of full finite-dimensional matrix algebras (block-diagonal matrices) and by doing some exercises, e.g., reproducing a proof of the GNS construction, or applying this material to group theory.

\subsection{The operator algebras used to describe a physical system}
\label{sec2.2}
We have said that (a model of) a physical system, $S$, is specified by a list
\begin{equation*}
\mathcal{P}_S = \{a_i\}_{i \in I_S}
\end{equation*}
of \textit{physical quantities} or \textit{potential properties}, $a_i=a_{i}^*$ ($i \in I_S$), characteristic of $S$ that can be observed or measured in experiments. (What is meant by this will hopefully become clear later, in Sections \ref{sec4} and \ref{sec5}). We assume that $\mathcal{P}_S$ is a self-adjoint subset of a $C^*-$algebra. As explained in Sect. \ref{sec2.1}, we may then consider 
\begin{equation}
\label{2.12}
\mathcal{A}_S:= \langle \mathcal{P}_S \rangle,
\end{equation}
the smallest $C^*-$algebra containing $\mathcal{P}_S$. The algebra $\mathcal{A}_S$ is called the \textit{"algebra of observables"} defining $S$; (possibly a misnomer, because, a priori, only the elements of $\mathcal{P}_S$ correspond to observable physical quantities - but let's not worry about this). For physical systems with finitely many degrees of freedom, $\mathcal{A}_S$ is usually a type-I $C^*-$algebra. 

We would like to have some natural notions  of symmetries of a system $S$, including time evolution. Here we encounter, for the first but not the last time, the complication that $S$ is usually in contact with some \textit{environment}, $E$, which may also include experimental equipment used to measure some observables of $S$. The environment is a physical system, too, and there usually are interactions between $S$ and $E$; in fact, only thanks to such interactions  is  it possible to retrieve information from $S$, i.e., measure a potential property $a_i$, $i \in I_S$, of $S$ in a certain interval of time. One typically chooses $E$ to be the \textit{smallest system} with the property that the \textit{composed system}, $S \vee E$, characterized by 
\begin{equation}
\label{2.13} 
\mathcal{P}_{S \vee E}= \{ a,b \mid a \in \mathcal{P}_S, b \in \mathcal{P}_E \},
\end{equation} 
can be viewed as a \textit{"closed physical system"}.

What is a "closed physical system"? Let $\overline{S}:=S \vee E$, and let $\mathcal{A}_{\overline{S}}$ denote the $C^*-$algebra generated by $\mathcal{P}_{S \vee E}$; i.e., $\mathcal{A}_{\overline{S}}= \langle \mathcal{P}_{S \vee E} \rangle$. 
We say that $\overline{S}$ is a \textit{closed (physical) system} if the \textit{time evolution} of physical quantities characteristic of $\overline{S}$ is given in terms of $^*$\textit{automorphisms} of $\mathcal{A}_{\overline{S}}$; i.e., given two times, $s$ and $t$, $\tau_{t,s}$ is a $^*$automorphism of $\mathcal{A}_{\overline{S}}$ that associates with every physical quantity in $\mathcal{A}_{\overline{S}}$ specified at time $s$ an operator in $\mathcal{A}_{\overline{S}}$ representing the \textit{same} physical quantity at time $t$. We must require that 
\begin{equation}
\label{2.14}
\tau_{t,s} \circ \tau_{s,u}=\tau_{t,u},
\end{equation}
for any triple of times $(t,s,u)$.

Given a physical system, $S$, we choose its environment $E$ such that, within a prescribed precision,
 $\overline{S}=S \vee E$ can be considered to be a closed physical system. "For all practical purposes" (FAPP, see \cite{Bell}), i.e., within usually astounding precision, $\overline{S}$ is much ... much smaller than the entire universe; it does usually \textit{not} include the experimentalist in the laboratory observing $S$ or the laptop of her theorist colleague next door, etc.. To say that $\overline{S}$ is a \textit{closed} physical system does, however, \textit{ not } exclude that $\overline{S}$ is entangled with another physical system, $S'$.

Given $S$ and $\overline{S}=S \vee E$, as above, we call $\mathcal{A}_{\overline{S}}$ the \textit{"dynamical} $C^*-$\textit{algebra"} of $S$. 

Let $\mathcal{G}_S$ denote a group of \textit{symmetries} of $S$. We will assume that every element $g \in \mathcal{G}_S$ can be represented by a $^*$automorphism, $\sigma_g$, of $\mathcal{A}_{\overline{S}}$, with the property that
\begin{equation}
\label{2.15}
\sigma_{g_1} \circ \sigma_{g_2}=\sigma_{g_1 \circ g_2},
\end{equation}
i.e., $\sigma: \mathcal{G}_S \longrightarrow \text{  }^* \text{Aut}(\mathcal{A}_{\overline{S}})$ is a representation of $\mathcal{G}_S$ in the group, $^* \text{Aut}(\mathcal{A}_{\overline{S}})$, of $^*$automorphisms of  $\mathcal{A}_{\overline{S}}$. We say that $\mathcal{G}_S$ is a  group of \textit{dynamical} symmetries of $S$ iff $\sigma_g$ and time evolution $\tau_{t,s}$ commute, for all $g \in \mathcal{G}_S$ and arbitrary pairs of times $(t,s)$.

By a \textit{"state of a physical system"} $S$ we mean a state on the $C^*-$algebra $\mathcal{A}_{\overline{S}}$, in the sense of  Eqs. (\ref{2.5}) and (\ref{2.6}) in Subsect. \ref{sec2.1}. (This will turn out to be a misnomer when we deal with quantum systems. But the expression appears to be here to stay.) The set of all states of $S$ is denoted by $\mathcal{S}_{\overline{S}}$. 

To summarize, \textit{ a (model of a) physical system}, $S$, is specified by the following data. 
\begin{definition}[Algebraic data specifying a model of a physical system]
\label{def2.1}
\end{definition}

\begin{enumerate}[(I)]
\item A list of \textit{physical quantities}, or \textit{observables}, $\mathcal{P}_S= \{ a_i=a_{i}^*\}_{i \in I_S}$,  generating a $C^*-$algebra, $\mathcal{A}_S$, of "observables", that is contained in the $C^*-$algebra $\mathcal{A}_{\overline{S}}$ (the "dynamical $C^*$-algebra" of $S$) of a \textit{closed} system, $\overline{S}=S \vee E$, containing $S$.

\item The convex set, $\mathcal{S}_{\overline{S}}$, of \textit{states of } $S$, interpreted as states on the  $C^*-$algebra $\mathcal{A}_{\overline{S}}$.

\item \textit{Time translations of } $\overline{S}$, represented as $^*$automorphisms $\{\tau_{t,s}\}_{t, s \in \mathbb{R}}$ on $\mathcal{A}_{\overline{S}}$ satisfying \eqref{2.14}, and a group, $\mathcal{G}_S$, of \textit{symmetries of }$S$ represented by $^*$automorphisms, $\{\sigma_g\}_{g \in \mathcal{G}_S}$, of $\mathcal{A}_{\overline{S}}$; (see \eqref{2.15}). 
\end{enumerate}
\vspace{1mm}

We should explain what is meant by \textit{"time translations"}: For each time $t \in \mathbb{R}$, we have copies $\mathcal{P}_{S}(t)$ and $\mathcal{A}_S(t)=\langle \mathcal{P}_S(t) \rangle$  $^*$isomorphic to $\mathcal{P}_S$ and $\mathcal{A}_S$, respectively, which are contained in $\mathcal{A}_{\overline{S}}$. If $a(s) \in \mathcal{P}_{S}(s)$ and $a(t) \in \mathcal{P}_S(t)$ are the operators in $\mathcal{A}_{\overline{S}}$ representing an arbitrary potential property, or observable, $a \in \mathcal{P}_S$,  of $S$ at times $s$ and $t$, respectively, then
\begin{equation}
\label{2.16}
a(t)=\tau_{t,s}(a(s)),
\end{equation}
with $\tau_{t,s}=\tau_{t,u} \circ \tau_{u,s}$, for arbitrary times $t,u$ and $s$ in $\mathbb{R}$.  

We say that the system $\overline{S}=S \vee E$ is \textit{autonomous} iff
\begin{equation}
\label{2.17}
\tau_{t,s}=\tau_{t-s}
\end{equation}
where $\{\tau_t\}_{t \in \mathbb{R}}$ is a one-parameter group of $^*$automorphisms of $\mathcal{A}_{\overline{S}}$.

We say that a system $S$ is a \textit{subsystem} of a system $S'$ iff
\begin{equation} 
\label{2.18}
\mathcal{P}_S \subset \mathcal{P}_{S'}
\end{equation}
and
\begin{equation} 
\label{2.19}
\mathcal{A}_{\overline{S}} \subseteq  \mathcal{A}_{\overline{S'}}.
\end{equation}

The composition, $S_1 \vee S_2$, of two systems, $S_1$ and $S_2$, can be defined by choosing
\begin{equation}
\label{2.20}
\mathcal{P}_{S_1 \vee S_2}:=\mathcal{P}_{S_1} \cup \mathcal{P}_{S_2}
\end{equation}
and $\mathcal{A}_{\overline{S_1 \vee S_2}}$ to contain the $C^*-$algebra generated by $\mathcal{A}_{\overline{S_1 }}$ and $\mathcal{A}_{\overline{S_2}}$. (A more precise discussion would lead us into the theory of tensor categories.)

\subsection{Potential properties, information loss and possible events}
\label{sec2.3}
Let $S$ be a physical system coupled to an environment $E$ and described, mathematically, by data
\begin{equation}
\label{data}
(\mathcal{P}_S, \mathcal{A}_{\overline{S}=S \vee E}, \{ \tau_{t,s}\}_{t,s \in \mathbb{R}}, \mathcal{G}_S, \mathcal{S}_{\overline{S}})
\end{equation}
with properties as specified in points (I) through (III) of Definition \ref{def2.1}, Subsect. \ref{sec2.2}. 

A "potential property" of $S$ is represented by an element $a \in \mathcal{P}_S$ or, more generally, by a self-adjoint operator $a=a^*$ in the algebra $\mathcal{A}_S$. An observation of a potential property, $a$, of $S$ at time $t$ will be described in terms of the operator $a(t)=\tau_{t,t_0}(a) \in \mathcal{A}_{\overline{S}}$, where $t_{0}$ is a fiducial time at which the state of $S$ is specified. Next, we have to clarify in which sense \textit{information} is \textit{lost}, as time increases. In \textit{local, relativistic} quantum theory, a distinction between $S$ and $\overline{S}$ becomes superfluous, and one may usually identify $S$ with $\overline{S}$. Moreover, the finiteness of the speed of light, i.e., of the speed of propagation of arbitrary signals, and locality lead to an intrinsic notion of \textit{information loss} \cite{J.F2,J.F.,Buchh-Roberts} -- at least in theories with massless particles that satisfy Huyghens' Principle \cite{Buchholz} and are allowed to escape to spatial $\infty$ (or fall into black holes). This is not so when one considers non-relativistic models of physical systems, with signals propagating arbitrarily fast ("Fernwirkung"). Nevertheless, one may argue that whenever properties of $S$ are observed successfully, thanks to interactions of $S$ with some environment/equipment $E$, then, as the price to pay, information is lost irretrievably: It disperses into the environment $E$, where it becomes inaccessible to experimental observation. Of course, this idea is plausible \textit{only} if the cut between "system $S$" and "environment $E$", given a closed system $\overline{S}$, is made at the right place. 
To determine this cut, one must specify the list $\mathcal{P}_{S}$ of physical quantities characterising 
$S$ that are measurable in experiments, using $E$. Mathematically, the cut is determined by specifying the pair $(\mathcal{A}_{S},\mathcal{A}_{\overline{S}})$ of algebras.

For the purpose of this essay, we adopt the point of view that the only properties of $\overline{S}$ that can potentially be observed, experimentally, are properties of $S$ represented by self-adjoint operators
\begin{equation}
\label{2.22}
a(t)=a^{*}(t), \text{ } \quad \text{with } a \in \mathcal{P}_S, t \in \mathbb{R}. 
\end{equation}
In order to arrive at a mathematically precise concept of information loss (as time goes by), it is convenient to introduce the following algebras.

\begin{definition}
\label{def2.2}
The algebra, $\mathcal{E}_{\geq t}$, of potential properties observable after time $t$ is the $C^{*}$-subalgebra of $\mathcal{A}_{\overline{S}}$ generated by arbitrary  finite linear combinations of arbitrary finite products
$$a_1(t_1)...a_n(t_n), n=1,2,3,...,$$
where $t_i \geq t$ and $a_i \in \mathcal{A}_S$, $i=1,...,n$, (with $a(s)$ the operator in  $\mathcal{A}_{\overline{S}}$ representing the operator $a \in \mathcal{A}_S$ at time $s$).
\end{definition}

\noindent It follows from this definition that 
\begin{equation}
\label{2.23}
\mathcal{E}_{\geq t} \subseteq \mathcal{E}_{\geq t'}
\end{equation}
whenever $t>t'$, with $\mathcal{E}_{\geq t} \subseteq \mathcal{A}_{\overline{S}}$, for all $t \in \mathbb{R}$. We speak of \textit{loss of information} iff
\begin{equation}
\label{2.24}
\mathcal{E}_{\geq t} \subsetneq \mathcal{E}_{\geq t'},
\end{equation}
for some times $t$ and $t'$, with $t >t'$.
We define an algebra $\mathcal{E}_{S}$ by
\begin{equation}
\mathcal{E}_{S} := \overline{\bigvee_{t\in\mathbb{R}} \mathcal{E}_{\geq t}}^{\parallel \cdot \parallel}
\end{equation}
It is one of the notorious problems in most approaches to a "quantum theory of experiments" that it is left unclear which self-adjoint operators in some very large algebra of operators correspond to potential properties of a quantum system that can actually be $measured$ or observed. Most authors consider \textit{far too many} operators as corresponding to potential properties of the system that are potentially measurable. As we will discuss in Section 5, it appears to be a general principle ("Duality between Observables and Indeterminates") that $\mathcal{E}_{S} \subsetneq \mathcal{A}_{\overline{S}}$ and that the relative commutant of $\mathcal{E}_{S}$ inside $\mathcal{A}_{\overline{S}}$ contains a subalgebra isomorphic to $\mathcal{E}_{S}$. (Obviously, for classical systems -- $\mathcal{A}_{\overline{S}}$ abelian, the commutant of $\mathcal{E}_{S}$ is all of $\mathcal{A}_{\overline{S}}$.)

Let $\omega \in \mathcal{S}_{\overline{S}}$ be a state of the system. Let $(\mathcal{H}_{\omega},\pi_{\omega},\Omega)$ denote the Hilbert space, the representation of  $\mathcal{A}_{\overline{S}}$ on $\mathcal{H}_{\omega}$, and the cyclic vector in $\mathcal{H}_{\omega}$, respectively, associated to the pair $(\mathcal{A}_{\overline{S}},\omega)$ by the GNS construction; see Sect. \ref{sec2.1}, \eqref{2.7}. By $\mathcal{A}_{\overline{S}}^{\omega}$ we denote the von Neumann algebra corresponding to the weak closure of $\pi_{\omega}(\mathcal{A}_{\overline{S}})$ in the algebra, $\mathcal{B}(\mathcal{H}_{\omega})$, of all bounded operators on $\mathcal{H}_{\omega}$.

\begin{definition}
\label{poss}
Given  a physical system $S$, as in Definition \ref{def2.1}, (I)-(III), above, and a state $\omega \in \mathcal{S}_{\overline{S}}$, a possible event in $S$ observable at time $t$ is a spectral projection,
\begin{equation}
\label{2.25}
P_{a(t)}(I),
\end{equation}
of the operator $\pi_{\omega}(a(t)) \in  \mathcal{A}_{\overline{S}}^{\omega}$ associated with a measurable subset $I \subseteq \text{spec } \pi_{\omega}(a(t)) \subseteq \mathbb{R}$, where $a=a^* \in \mathcal{P}_S$ and $t \in \mathbb{R}$. (Here $\text{spec }A$ denotes the spectrum of a self-adjoint operator $A$ on $\mathcal{H}_{\omega}$.)
\end{definition}

\begin{definition}
\label{possal}
The algebra, $\mathcal{E}^{\omega}_{\geq t}$, of all possible events observables at times $\geq t$, is the von  Neumann algebra corresponding to the weak closure of $\pi_{\omega}(\mathcal{E}_{\geq t})$ in $B(\mathcal{H}_{\omega})$. The von Neumann algebra $\mathcal{E}_{S}^{\omega}$ is defined similarly.
\end{definition}

Note that if $\omega'$ is a state that is normal with respect to the state $\omega$ then $\mathcal{A}_{\overline{S}}^{\omega'} = \mathcal{A}_{\overline{S}}^{\omega}$, etc. The algebra $\mathcal{E}^{\omega}_{\geq t}$ contains the spectral projections $P_{a(s)}(I)$ describing possible events at times $s \geq t$; (see \eqref{2.25}). It is therefore justified to call  $\mathcal{E}^{\omega}_{\geq t}$ the \textit{"algebra of possible events observable at times $\geq t$"}. Loss of information may manifest itself in the property that the relative commutant 
\begin{equation}
\label{2.26}
(\mathcal{E}^{\omega}_{\geq t})' \cap \mathcal{E}^{\omega}_{\geq t'}
\end{equation}
is \textit{non-trivial}, for some $t>t'$.

We note that the algebra $\mathcal{E}_{S}$
carries an action of the group, $\mathbb{R}$, of time translations by $^*$automorphisms, $\{\overline{\tau}_t \}_{t \in \mathbb{R}}$, defined as follows: For $a_1(t_1)...a_n(t_n) \in  \underset{t \in \mathbb{R}}{  \vee} \mathcal{E}_{\geq t} $, with $t_i 
\in \mathbb{R}, a_i \in \mathcal{A}_S, i=1,...,n$,
\begin{equation}
\label{2.28}
\overline{\tau}_{t}(a_1(t_1)...a_n(t_n)):=a_{1}(t_1+t)...a_{n}(t_n+t).
\end{equation}
The definition of $\overline{\tau}_t$ extends to all of $\mathcal{E}_{S}$ by linearity and continuity. One then has that
\begin{equation}
\label{2.29}
\overline{\tau}_t: \mathcal{E}_{\geq t'} \longrightarrow  \mathcal{E}_{\geq t' + t} \subseteq  \mathcal{E}_{\geq t'},
\end{equation}
for arbitrary $t \geq 0$.

Let $a \in \mathcal{P}_S$ be a potential property of $S$, and let $\omega$ be a state of $S$ (i.e., $\omega \in \mathcal{S}_{\overline{S}}$). Depending on the experimental equipment available to observe $a$, i.e., depending on the choice of the time evolution of $\overline{S}=S \vee E$, and depending on the choice of a state 
$\omega \in \mathcal{S}_{\overline{S}}$, an observation of $a$ may have different alternative outcomes; in particular, the resolution in an observation of $a$ at some time $t_{*}$ will depend on the choice of $(E, \{ \tau_{t,s}\}_{t,s \in \mathbb{R}}, \omega)$. These alternative outcomes correspond to spectral projections $P_{a(t_{*})}(I_{\alpha})$, $\alpha=1,...,k$, where $I_{\alpha} \cap I_{\beta} = \emptyset$, for $\alpha \neq \beta$, and $\cup_{\alpha=1}^{k} I_{\alpha}  \supseteq \text{spec } \pi_{\omega}(a(t_{*}))$. Then
\begin{equation}
\label{2.30}
P_{a(t_{*})}(I_{\alpha}) P_{a(t_{*})}(I_{\beta}) =\delta_{\alpha \beta} P_{a(t_{*})} (I_{\alpha}),
\end{equation}
and
\begin{equation}
\label{2.31}
\sum_{\alpha=1}^{k}P_{a(t_{*})} (I_{\alpha})=\mathds{1},
\end{equation}
for an arbitrary $t_{*}$.

Traditionally, one says that the \textit{purpose of a model of a physical system}, $S$, is to enable us to make predictions of the following kind: Suppose  we are interested in testing some potential properties (or, put differently, measure some physical quantities) $a_1,...,a_n$ characteristic of $S$ during intervals of time $\Delta_1 \prec \Delta_2 \prec ... \prec \Delta_n$, where
\begin{equation}
\Delta \prec \Delta' \text{ iff, } \forall t \in \Delta, \text{ } \forall t' \in \Delta': \text{ } t \leq t'.
\end{equation}

We assume that $S$ is in a state $\omega \in \mathcal{S}_{\overline{S}}$. Then a model of $S$ ought to tell us whether $a_1,..., a_n$ will actually be measurable (i.e., are \textit{"empirical"} properties) and predict the \textit{ probability } (frequency) that, in a test or measurement of $a_i$ at some time $t_i  \in \Delta_i$, the event corresponding to the spectral projection  $P_{a_i(t_i)}(I^{i}_{\alpha_i})$, $\alpha_i=1,...,k_i$, is observed, (i.e., property $a_i(t_i)$ has a value in the interval $I^{i}_{\alpha_i}$), for all $i=1,...,n,$ given the state $\omega \in \mathcal{S}_{\overline{S}}$; (the properties of the projections $P_{a_i(t_i)}(I^{i}_{\alpha_i})$ are as in Eqs. (\ref{2.30}), (\ref{2.31})). 

We simplify our notation by setting 
\begin{equation}
\label{2.33}
\Pi^{(i)}_{\alpha_i} \equiv \Pi^{(i)}_{\alpha_i} (t_i) :=P_{a_i(t_i)} (I^{i}_{\alpha_i}),
\end{equation}
with $t_i \in \Delta_i, \text{ }a_i \in \mathcal{P}_S, \text{ }i=1,...,n$,  $\Delta_1 \prec \Delta_2 \prec ... \prec \Delta_n$.  The time-ordered sequence
\begin{equation}
\label{2.34}
h^{n}_{1}(\underline{\alpha}):=\{  \Pi_{\alpha_1}^{(1)},..., \Pi_{\alpha_n}^{(n)} \}
\end{equation}
of possible events $\Pi^{(i)}_{\alpha_i}$ (as in \eqref{2.33}) is conventionally called a \textit{"history"}. Given such a history, we define operators
\begin{equation}
\label{2.35}
H^{n}_{k}(\underline{\alpha}):= \Pi^{(n)}_{\alpha_n}...\Pi^{(k+1)}_{\alpha_{k+1}}\Pi^{(k)}_{\alpha_k},
\end{equation}
with $\Pi^{(i)}_{\alpha_i}$ as in \eqref{2.33}.\\

{\bf{Postulate 2.5}}  (see \cite{Schw,Wig,Omnes})
\textit{Given a model of a physical system $S$, as specified in points (I)-(III) of Definition \ref{def2.1}, Sect. \ref{sec2.2}, the \textit{probability of a history} $h^{n}_{1}(\underline{\alpha})=\{ \Pi^{(1)}_{\alpha_1},...,\Pi^{(n)}_{\alpha_n}\}$ in a state $\omega \in \mathcal{S}_{\overline{S}}$ is predicted to be given by}
\begin{equation}
\label{2.36}
\text{Prob}_{\omega} \text{ }h_{1}^{n}(\underline{\alpha}) \equiv \text{Prob}_{\omega} \{  \Pi_{\alpha_1}^{(1)},..., \Pi_{\alpha_n}^{(n)} \}:=\omega \left((H^{n}_{1}(\underline{\alpha}))^{*} H^{n}_{1}(\underline{\alpha}) \right),
\end{equation} 
\textit{with $H^{n}_{1}(\underline{\alpha})$ as in \eqref{2.35}}. (It is assumed here that $a_1,...,a_n$ are measurable, for the given time-evolution and state of the system; see Sect. 5).

Much discussion in the remainder of this essay is devoted to finding out under what conditions formula (\ref{2.36}), is meaningful, and -- if it is -- what it tells us about $S$. To give away our secrets, Postulate 2.5 is perfectly meaningful for  \textit{classical} models of physical systems, as discussed in Sect.3, and it is most often meaningless for \textit{quantum-mechanical} models. While FMPP ("for many practical purposes"), formula (\ref{2.36}) is useful in quantum mechanics, conceptually it is misleading and often nonsensical! It does, however, pass some tests indicating  that it defines a probability:

\begin{enumerate}[(1)]
\item $\text{Prob}_{\omega}$ satisfies
\begin{equation}
\label{2.37}
0 \leq \text{Prob}_{\omega} \{ \Pi^{(1)}_{\alpha_1},..., \Pi^{(n)}_{\alpha_n}\} \leq 1,
\end{equation}
for every state $\omega \in \mathcal{S}_{\overline{S}}$ and an arbitrary history $\{ \Pi^{(1)}_{\alpha_1},...,\Pi^{(n)}_{\alpha_n}\}$.

\item 
\begin{equation}
\label{2.38}
\underset{\alpha_i=1,...,k_i (i= 1,...,n)}{\sum} \text{Prob}_{\omega} \{ \Pi^{(1)}_{\alpha_1},..., \Pi^{(n)}_{\alpha_n}\} =1,
\end{equation}
for arbitrary operators $a_1,...,a_n$ and time intervals $\Delta_1 \prec ... \prec \Delta_n$, (with $\Pi_{\alpha_i}^{(i)}$ as in \eqref{2.33}).\\
Properties (1) and (2) show that $\text{Prob}_{\omega}$ is a probability functional.
\vspace{1mm}

\item As observed in \cite{Ish,Omnes} and refs. given there, formula (\ref{2.36}) represents the \textit{"only possible"} definition of a probability functional on the lattice of possible events.
\end{enumerate}

As already mentioned, formula  (\ref{2.36}) is perfectly adequate for an analysis of the predictions of \textit{classical} models of physical systems. \textit{Quantum-mechanically}, however, given 
\begin{equation*}
(\mathcal{A}_{\overline{S}}, \{ \tau_{t,s}\}_{t,s \in \mathbb{R}}, \omega\in \mathcal{S}_{\overline{S}}),  
\end{equation*}
one encounters plenty of sequences of potential properties,
\begin{equation*}
\{a_1(t_1),...,a_n(t_n)\}, 
\end{equation*}
with $a_i \in \mathcal{P}_S$, $t_i  \in \Delta_i$, $i=1,...,n$, $\Delta_1 \prec ... \prec \Delta_n$, which turn out to be \textit{incompatible} with one another. The question then arises which one among such sequences of potential properties of $S$ actually corresponds to a sequence of \textit{empirical properties} of $S$ observed in the course of time; (assuming that there is only \textit{one} rather than "many worlds".) Formula (\ref{2.36}) does not tell us much about the answer to this question; but the idea of \textit{loss of information}, as expressed in Eqs. (\ref{2.24}) and (\ref{2.26}), along with the phenomenon of \textit{entanglement, does}! This is discussed in Subsects. 5.3 and 5.4.

\section{Classical ("realistic") models of physical systems}
\label{section3}
We start this section by recalling the usual distinction between \textit{classical, realistic models} (abbreviated as "R-models") and \textit{quantum-mechanical-models} (abbreviated as "Q-models") of physical systems: An \textit{R-model} of a system $S$ is fully characterized by the property that its "dynamical $C^{*}$-algebra" $\mathcal{A}_{\overline{S}}$ (see Subsect. \ref{2.2}) is \textit{abelian} (commutative). Hence $\mathcal{A}_S$ is abelian, too. 

A \textit{Q-model} of a system $S$  differs from an R-model only in that the algebra $\mathcal{A}_S$ (and hence $\mathcal{A}_{\overline{S}}$) is \textit{non commutative}.  Apart from this crucial difference, the algebraic data defining an R- or a Q-model are as specified in points (I)-(III) of Definition \ref{def2.1}, Subsect. \ref{sec2.2}.

\subsection{General features of classical models}
We recall a well-known theorem due to I.M. Gel'fand. Let $\mathcal{B}$ be an abelian $C^*$-algebra. The \textit{spectrum}, $M$, of $\mathcal{B}$ is the space of all non-zero $^*$homomorphisms from $\mathcal{B}$ into $\mathbb{C}$ (the "characters" of $\mathcal{B}$); $M$ is a locally compact topological (Hausdorff) space. If  $\mathcal{B}$ contains an identity, $\mathds{1}$, then $M$ is compact.

\begin{theorem}[Gel'fand] If $\mathcal{B}$ is an abelian $C^*-$algebra then it is $^*$isomorphic to the $C^*$-algebra, $C_{0}(M)$, of continuous functions on $M$ vanishing at $\infty$, i.e.,
\begin{equation}
\label{3.1}
\mathcal{B} \simeq C_{0}(M).
\end{equation}
Furthermore, every \textit{state}, $\omega$, on $\mathcal{B}$ is given by a unique (Borel) \textit{probability measure}, $d\mu_{\omega}$, on $M$ (and conversely).
\end{theorem}

Every pure state is given by a Dirac $\delta-$function, $\delta_{x}$, on $M$, for some $x \in M$; i.e., the space of pure states can be identified with $M$, (which is why $M$ is called \textit{"state space"}). Thus, the set of pure states of $\mathcal{B}$ cannot be endowed with a linear or affine structure. 

If $\mathcal{B}_0 \subset \mathcal{B}$ is a subalgebra of $\mathcal{B}$ then any pure state of $\mathcal{B}$ is also a pure state of $\mathcal{B}_0$. If $\mathcal{B}=\mathcal{A}_{\overline{S}}$ is the dynamical $C^*-$algebra of a realistic (classical) model of a physical system, $S$, we call $M=:M_S$ the state space of $S$. It is homeomorphic to the space of pure states of $\overline{S}$ and does not have a linear structure, i.e. there is \textit{no superposition principle} for pure states. If $S=S_1 \vee S_2$ is the composition of two subsystems, $S_1$ and $S_2$, these systems are, of course, classical, too, and we have that any pure state of $S$ is also a pure state of $S_1$ and of $S_2$; i.e., there is \textit{no} interesting notion of \textit{entanglement}. 

\subsection{Symmetries and time evolution in classical models}
According to point (III) of Definition \ref{def2.1} in Subsect. \ref{sec2.2}, symmetries and time evolution of a system $S$ are given by *automorphisms of its dynamical $C^*-$algebra $\mathcal{A}_{\overline{S}}$. If $\mathcal{B}$ is an abelian $C^*$-algebra and $M$ denotes its spectrum then any $^{*}$automorphism, $\alpha$, of $\mathcal{B}$ corresponds to a \textit{homeomorphism}, $\phi_{\alpha}$, of $M$: If $a$ is an arbitrary element of $\mathcal{B}$, thus given by a bounded continuous function (also denoted by $a$) on $M$, then
\begin{equation}
\label{3.2}
\alpha(a) (\xi)=: a(\phi_{\alpha}^{-1}(\xi)), \qquad \xi \in M.
\end{equation}
Conversely, any homeomorphism, $\phi$, from $M$ to $M$ determines a $^*$automorphism, $\alpha_{\phi}$, by 
\begin{equation}
\label{3.3}
\alpha_{\phi}(a) (\xi):=a(\phi^{-1}(\xi)), \qquad \xi \in M.
\end{equation}
If $\{ \alpha_{t,s}\}_{t,s \in \mathbb{R}}$ is a groupoid of $^*$automorphisms of $\mathcal{B}$, with $\alpha_{t,s} \circ \alpha_{s,u}=\alpha_{t,u}$, then there exists a groupoid of homeomorphisms, $\{\phi_{t,s}\}_{t,s \in \mathbb{R}}$, of $M$, with $\phi_{t,s} \circ \phi_{s,u}=\phi_{t,u}$, such that 
\begin{equation}
\label{3.4}
\alpha_{t,s}(a)(\xi)=a(\phi_{s,t}(\xi)), \qquad \xi \in M,
\end{equation}
where $\phi_{s,t}=\phi^{-1}_{t,s}$.

Let us suppose that there is a subalgebra $\mathring{\mathcal{B}} \subset \mathcal{B}$ that is \textit{norm-dense} in $\mathcal{B}$ such that $\alpha_{t,s}(a)$ is continuously differentiable in $t$ (and in $s$), for arbitrary $a \in \mathring{\mathcal{B}}$. We define
\begin{equation}
\label{3.5}
\delta_s(a)=\frac{d}{dt} \alpha_{t,s}(a)_{\vert t=s}, \qquad a \in \mathring{\mathcal{B}} .
\end{equation}
Then $\delta_s$ is a $^{*}\textit{derivation}$ defined on $\mathring{\mathcal{B}}$. An operator $\delta: \text{Dom}_{\delta} \rightarrow \mathcal{B}$ is a $^*$derivation of $\mathcal{B}$ iff $\text{Dom}_{\delta} \subseteq \mathcal{B}$ is norm-dense in $\mathcal{B}$, $\delta$ is linear, $\delta(a^*)=(\delta(a))^*$, and 
\begin{equation}
\label{3.6}
\delta(a \cdot b)=\delta(a) \cdot b + a\cdot \delta(b) \qquad (Leibniz \text{ }rule),
\end{equation}
for arbitrary $a,b \in \text{Dom}_{\delta}$. If $\mathcal{B}$ is \textit{abelian} then a $^*$derivation $\delta$ of $\mathcal{B}$ corresponds to a \textit{vector field} $X$ on $M$, (assuming that $M$ admits some vector fields):
\begin{equation}
\label{3.7}
\delta(a)(\xi)=(Xa)(\xi),
\end{equation}
where $a$ corresponds to an arbitrary continuously differentiable function on $M$. If $\delta_s$ satisfies \eqref{3.5} then, for $a \in  \mathring{\mathcal{B}} \subseteq \text{Dom}_{\delta_s}$, 
\begin{equation}
\label{3.8}
\frac{d}{dt} \alpha_{t,s}(a)_{\mid t=s} = \delta_s(a)=X_sa,
\end{equation}
where, for each $s \in \mathbb{R}$, $X_s$ is a vector field on $M$. \eqref{3.8} can be rewritten as 
\begin{equation}
\label{3.9}
\frac{d}{dt} \phi_{t,s}(\xi)=-X_{t}(\phi_{t,s}(\xi)), \qquad \xi \in M. 
\end{equation}
Hence, at least formally, the homeomorphisms $\phi_{t,s}$ can be constructed from a family of vector fields $\{X_s\}_{s \in \mathbb{R}}$ by integrating the ordinary differential equations (\ref{3.9}). These observations can be made precise if the spectrum $M$ of $\mathcal{B}$ admits a tangent bundle, $TM$, and the vector fields $X_s$ are globally Lipschitz and continuous in $s$, for all $s \in \mathbb{R}$. If $X_s \equiv X$ is independent of $s$ then  $\phi_{t,s}=\phi_{t-s}$ is a one-parameter group of homeomorphisms of $M$, (and conversely). 

All these remarks can be applied to a classical (model of a) physical system, $S$, with an abelian dynamical $C^*$-algebra $\mathcal{A}_{\overline{S}}$. One may then interpret the parameters $t,s \in \mathbb{R}$ of a groupoid $\{ \tau_{t,s}\}_{t,s \in \mathbb{R}}$ of $^*$ automorphisms of $\mathcal{A}_{\overline{S}}$ as \textit{times}; and we say that $S$ is \textit{autonomous} iff $\tau_{t,s}=\tau_{t-s}$ belongs to a one-parameter group of $^*$automorphisms of $\mathcal{A}_{\overline{S}}$, or if the vector field $X$ on $M_S=\text{spec} \mathcal{A}_{\overline{S}}$ generating $\tau_t$ is \textit{time-independent}.  It is straightforward to describe general symmetries of $S$ in terms of groups of homeomorphisms of $M_S$.

\subsection{Probabilities of histories, realism and determinism}
A physical quantity or property of a classical physical system $S$ is given by a continuous function, $a$, on $M_S$. We denote the family of all properties of $S$ specified at a fiducial time $t_0$ by $\mathcal{P}_S=\{a_i\}_{i \in I_S}$. A \textit{possible event} in $S$ at a time $t$ corresponds to the characteristic function, $\chi_{\Omega_{i}^{I}(t)}$, of an open subset, $\Omega_{i}^{I}(t)$, of $M_S$ given by 
\begin{equation}
\label{3.10}
\xi \in \Omega_{i}^{I}(t) \Leftrightarrow a_i(t)(\xi) \in I,
\end{equation}
where $a_i \in \mathcal{P}_S$,  $a_i(t) = \tau_{t,t_0}(a_i)$, and $I$ is an open subset of $\mathbb{R}$; (see Definition 2.3 in Subsect. \ref{sec2.3}). 

Let $\phi_{t,s}$ denote the homeomorphism of $M_S$  corresponding to $\tau_{t,s}$. Setting $\Omega_{i}^{I}:=\phi_{t_0,t}(\Omega_{i}^{I}(t))$, we have that  
\begin{align*}
\xi \in  \Omega_{i}^{I}(t) & \Leftrightarrow a_i(t)(\xi) \in I \Leftrightarrow \tau_{t,t_0}(a_i)(\xi) \in I\\
& \Leftrightarrow a_i(\phi_{t_0,t}(\xi)) \in I \Leftrightarrow \eta:= \phi_{t_0,t} (\xi) \in \Omega_{i}^{I}.
\end{align*}
 We choose $n$ properties, $a_1,...,a_n,$ of $S$ to be measured at times $t_1 \leq t_2 \leq ... \leq t_n$, with the measured value of $a_i$ contained in the interval $I_i$, $i=1,...,n$. We let $\Omega_{i}(t_i)$ be the open subset of $M_S$ given by 
 \begin{equation}
 \label{3.11}
 \xi \in  \Omega_{i}(t_i) \Leftrightarrow a_i(t_i)(\xi) \in I_i,
 \end{equation}
 $i=1,...,n$, and $\Omega_i =\phi_{t_0,t_i}(\Omega_{i}(t_i))$. 
 
 Let $\mu$ be a state of $S$, i.e., a probability measure on $M_S$. Every theoretical prediction concerning $S$ is the prediction of the \textit{probability of a history}, $\{ \xi_{t_i}:=\phi_{t_0,t_i}(\xi) \in \Omega_i \}_{i=1}^{n}$:
 \begin{equation}
 \label{3.12}
 \begin{split}
 \text{Prob}_{\mu} \{ \chi_{\Omega_1(t_1)}, ...,  \chi_{\Omega_n(t_n)} \}&:= \int_{M_S} d \mu (\xi) \prod_{i=1}^{n} \chi_{\Omega_i(t_i)} (\xi)\\
 &=  \int_{M_S} d \mu (\xi) \prod_{i=1}^{n} \chi_{\Omega_i} (\phi_{t_0,t_i}(\xi)).
 \end{split}
 \end{equation}
 If $\mu$ is a pure state, i.e., $\mu=\delta_{\xi_0}$, for some $\xi_0 \in M_S$ then
  \begin{equation}
 \label{3.13}
 \begin{split}
 \text{Prob}_{\delta_{\xi_0}} \{ \chi_{\Omega_1(t_1)}, ...,  \chi_{\Omega_n(t_n)} \} =  \prod_{i=1}^{n} \chi_{\Omega_i(t_i)} (\xi_0)=   \prod_{i=1}^{n} \chi_{\Omega_i} (\phi_{t_0,t_i}(\xi_0)),
 \end{split}
 \end{equation}
 i.e., the possible values of $ \text{Prob}_{\delta_{\xi_0}} $ are $0$ and $1$, for any $\xi_{0} \in M_S$ and \textit{all} histories. If $\xi_t:=\phi_{t_0,t}(\xi_0)$ is the trajectory of states with initial condition $\xi_0$ at time $t_0$ then
 \begin{equation}
 \label{3.14}
  \text{Prob}_{\delta_{\xi_0}} \{ \chi_{\Omega_1(t_1)}, ...,  \chi_{\Omega_n(t_n)} \}=1 \Longleftrightarrow \xi_{t_i} \in \Omega_i,
 \end{equation}
 for all $i=1,...,n$; otherwise, $ \text{Prob}_{\delta_{\xi_0}} $ vanishes. If $\xi_0 \notin \Omega_i$ then the event $\{ \phi_{t_0,t}(\xi) \in \Omega_i \}$ is first observed at time $t=\underline{t}_i$, where 
 \begin{equation}
 \label{3.15}
 \underline{t}_i:= \text{inf } \{ t \mid \xi_{0,t} =\phi_{t_0,t}(\xi_0) \in \Omega_i \},
 \end{equation}
 and it is last seen at time $\overline{t}_i$, where 
   \begin{equation}
 \label{3.16}
 \overline{t}_i:= \text{sup } \{ t \mid \xi_{0,t} =\phi_{t_0,t}(\xi_0) \in \Omega_i \}.
 \end{equation}
 These features of classical physical systems, in particular the "0-1 laws" in \eqref{3.14},  are characteristic of \textit{realism} and \textit{determinism}: Given that we know the state, $\xi_0$, of a system $S$ at some time $t_0$, we know its state, $\xi_t=\phi_{t_0,t}(\xi_0)$, and the value, $a_i(\xi_t)$, of an arbitrary property, $a_i \in \mathcal{P}_S$, of $S$, at an \textit{arbitrary (earlier or later) time} $t$.
 \begin{remark}
 (i)A straightforward extension of \eqref{3.12} is the basis for a definition of the dynamical (Kolmogorov-Sinai) entropy of the state $\mu$; see \cite{Kol,Sin}. 
 
 (ii)A special class of classical systems are \textit{Hamiltonian systems}, $S$, for which $M_S$ is a symplectic manifold, and the homeomorphisms $\phi_{t,s}$ are symplectomorphisms.
 \end{remark}

\section{Physical systems in quantum mechanics}
\label{sec4}
As indicated in the last section, the only feature distinguishing a quantum-mechanical model of a physical system $S$ (a $Q-$model) from a classical model (an $R-$model) is that, in a $Q-$model, $\mathcal{A}_S$ and hence $\mathcal{A}_{\overline{S}}$ are \textit{non-commutative} algebras. This has profound consequences! In this section, we recall some of the better known ones among them; in particular those that concern problems with the Schwinger-Wigner formula; see Postulate 2.5, \eqref{2.36}
.

\subsection{Complementary possible events do not necessarily exclude one another}
\label{sec4.1}
Let us recall the main task we are  confronted with: We have to clarify what the mathematical data (see Definition \ref{def2.1}, Sect \ref{sec2.2})
\begin{equation}
\label{4.1}
(\mathcal{P}_S, \mathcal{A}_{\overline{S}}, \{ \tau_{t,s}\}_{t,s \in \mathbb{R}},  \omega \in \mathcal{S}_{\overline{S}})
\end{equation}
tell us about the "behaviour" of the system $S$, as time goes by; in particular about the empirical properties displayed by $S$ and the events happening in $S$. This task will be shouldered for quantum-mechanical models in Sect. \ref{sec5}; it has been dealt with for classical models in the last section, (see also \cite{Fr-Schub}). To set the stage for the analysis of Sect. \ref{sec5}, it is useful to return to formulae (\ref{2.30}), (\ref{2.31}), (\ref{2.35}) and, in particular, formula (\ref{2.36}) for the probability of histories; see Subsect. \ref{sec2.3}. Thus, we consider $n$ possible events associated with physical quantities/potential properties, $a_i \in \mathcal{P}_S$, of $S$ measured at times $t_i \in \Delta_i \subset \mathbb{R}$, $i=1,...,n$, with $\Delta_1 \prec \Delta_2 \prec ... \prec \Delta_n$. Given a state $\omega$ on $\mathcal{A}_{\overline{S}}$, possible events are represented by spectral projections, $\Pi^{(i)}_{\alpha_i} \in \mathcal{A}_{\overline{S}}^{\omega}$, of the operators $a_i(t_i) \in \mathcal{A}_{\overline{S}}$. The projections $\Pi_{\alpha_i}^{(i)}$ are given by
\begin{equation}
\label{4.2}
\Pi^{(i)}_{\alpha_i} \equiv \Pi^{(i)}_{\alpha_i}(t_i):=P_{a_i(t_i)}(I^{i}_{\alpha_i}),
\end{equation}
$\alpha_i=1,...,k_i$, $i=1,...,n$, where $I^{i}_{\alpha_i}$ are disjoint measurable subsets of $\mathbb{R}$ with $\cup_{\alpha_i=1}^{k_i} I^{i}_{\alpha_i} \supseteq \text{ spec } \pi_{\omega}(a_i(t_i))$. It follows that 
\begin{equation}
\label{4.3}
\sum_{\alpha_i=1}^{k_i} \Pi_{\alpha_i}^{(i)}=\mathds{1},
\end{equation}
for all $i$. As in \eqref{2.35}, we set
\begin{equation}
\label{4.4}
H^{n}_{k}(\underline{\alpha}):= \Pi_{\alpha_n}^{(n)} \text{ } ... \text{ } \Pi_{\alpha_k}^{(k)}, \quad 1 \leq k \leq n. 
\end{equation}
A stretch, $h_{l}^{k}(\underline{\alpha})$, of a history $h_{1}^{n}(\underline{\alpha})$ is defined by 
\begin{equation}
\label{4.5}
h_{l}^{k}(\underline{\alpha}):=\{ \Pi^{(l)}_{\alpha_l}, ...,  \Pi^{(k)}_{\alpha_k}\}, \quad 1 \leq l \leq k \leq n,
\end{equation}
with $h^{n}:=h^{n}_1(\underline{\alpha})$. Furthermore, we set 
\begin{equation}
\label{4.6}
h^{n}_{\check{k}}:= \{ \Pi^{(1)}_{\alpha_1}, ...,  \Pi^{(k-1)}_{\alpha_{k-1}},  \Pi^{(k+1)}_{\alpha_{k+1}}, ...,  \Pi^{(n)}_{\alpha_n}\}.
\end{equation}
In the Schwinger-Wigner formula (\ref{2.36}), the \textit{probability of a history}, $h^n$, of $S$, given a state $\omega$, has been defined by 
\begin{equation}
\label{4.7}
\text{Prob}_{\omega} \{ \Pi_{\alpha_1}^{(1)}, ..., \Pi^{(n)}_{\alpha_n}\} := \omega\left( (H_{1}^{n}(\underline{\alpha}))^* H_{1}^{n}(\underline{\alpha}) \right)= \omega( \Pi^{(1)}_{\alpha_1}  \Pi^{(2)}_{\alpha_2} ...  \Pi^{(n)}_{\alpha_{n}} ...  \Pi^{(2)}_{\alpha_{2}}   \Pi^{(1)}_{\alpha_1}),
\end{equation}
with properties (1)-(3), (see Eqs. (\ref{2.37}) and (\ref{2.38})).

Here we wish to point out some \textit{fundamental problems} with formula (\ref{4.7}) in quantum mechanics. Suppose that the complementary possible events $\Pi^{(i)}_{1},..., \Pi^{(i)}_{k_i}$ were \textit{mutually exclusive}, given that $\Pi^{(1)}_{\alpha_1},..., \Pi^{(i-1)}_{\alpha_{i-1}} \Pi^{(i+1)}_{\alpha_{i+1}},...,\Pi^{(n)}_{\alpha_{n}}$ are observed, for some $i<n$, then we would imagine that the \textit{"sum rule"}
\begin{equation}
\label{4.8}
\begin{split}
\sum_{\alpha_i=1}^{k_i} \text{Prob}_{\omega} \text{ }h_{1}^{n}(\underline{\alpha})&= \sum_{\alpha_i=1}^{k_i} \text{Prob}_{\omega}  \{ \Pi_{\alpha_1}^{(1)},..., \Pi_{\alpha_i}^{(i)},...,\Pi_{\alpha_n}^{(n)}    \}\\
&=  \text{Prob}_{\omega}  \{ \Pi_{\alpha_1}^{(1)},..., \Pi_{\alpha_{i-1}}^{(i-1)}, \Pi_{\alpha_{i+1}}^{(i+1)},...,\Pi_{\alpha_n}^{(n)}    \}\\
&= \text{Prob}_{\omega} \text{ }h_{\check{i}}^{n}(\underline{\alpha})
\end{split}
\end{equation} 
holds;  see \eqref{4.6}. If $\Pi_{\alpha_i}^{(i)}$ commuted with the operator $H^{n}_{i+1}(\underline{\alpha})$, for all $\alpha_i$ -- \textit{as is the case in every classical model} -- then \eqref{4.8} would hold true. However, because of the non-commutative nature of $\mathcal{A}_{\overline{S}}$, 
\begin{equation}
\label{4.9}
[\Pi^{(i)}_{\alpha_i}, H^{n}_{i+1}(\underline{\alpha})] \neq 0,
\end{equation}
in general. This leads to \textit{non-vanishing interference terms}, 
\begin{equation}
\label{4.10}
\omega\left( ( H_{1}^{i-1}(\underline{\alpha}))^* \Pi^{(i)}_{\alpha_i}  ( H_{i+1}^{n}(\underline{\alpha}))^*  H_{i+1}^{n}(\underline{\alpha}) \Pi_{\beta_i}^{(i)} H_{1}^{i-1}(\underline{\alpha})\right),
\end{equation}
with $\alpha_i \neq \beta_i$. In the presence of non-vanishing interference terms the sum rule (\ref{4.8}) is usually \textit{violated}. This means that the complementary possible events $\Pi_{1}^{(i)},...,\Pi_{k_i}^{(i)}$, do, apparently, \textit{not} mutually exclude one another, \textit{ given future events}  $\Pi_{\alpha_{i+1}}^{(i+1)},...,\Pi_{\alpha_n}^{(n)}$ that cause interference. Put differently, a history $h^n$ does, in general, \textit{not} result in the determination of a potential property $a_i$, of $S$ in the $i^{th}$ observation (or measurement), given the data in (\ref{4.1}) (the time evolution $\{\tau_{t,s}\}_{t,s \in \mathbb{R}}$, and a state $\omega$). If the sum rule (\ref{4.8}) is violated, then the operator $a_i(t_i)$ does not represent an \textit{empirical property} of $S$, given \textit{later} observations of physical quantities $a_{i+1},...,a_{n}$. Apparently, the operators $a \in \mathcal{P}_S$ do, in general, \textit{not} represent properties of $S$ that exist \textit{a priori}, but only \textit{potential properties} of $S$ whose \textit{empirical status} depends on the choice of the time evolution $\{\tau_{t,s}\}_{\tau,s \in \mathbb{R}}$ of $\overline{S}=S\vee E$ and of the state $\omega$. This will be made precise in Sect. \ref{sec5}.

\subsection{The problem with conditional probabilities} 
\label{sec4.2}
In Subsect. \ref{sec2.3}, (\ref{2.37}) and (\ref{2.38}), we have seen that 
\begin{equation}
\label{4.11}
\mu_{\omega}(\underline{\alpha}):= \text{Prob}_{\omega} \{ \Pi_{\alpha_1}^{(1)}, ..., \Pi_{\alpha_n}^{(n)} \}
\end{equation}
is a \textit{probability measure} on $\mathbb{Z}_{k_1} \times ... \times \mathbb{Z}_{k_n}$. Let us fix $\alpha_1,...,\alpha_{i-1}, \alpha_{i+1},...,\alpha_n$, and ask what the \textit{conditional probability} 
\begin{equation}
\label{4.12}
\text{Prob}_{\mu_{\omega}} \{\Pi^{(i)}_{\alpha_i} \mid h^{n}_{\check{i}}(\underline{\alpha})\} 
\end{equation}
of the possible event $\Pi^{(i)}_{\alpha_i} $ is, given $\mu_{\omega}$ and $h^{n}_{\check{i}}$; (see \eqref{4.6}).  Since (\ref{4.11}) defines a probability measure, we may define
\begin{equation}
\label{4.13}
\text{Prob}_{\mu_{\omega}} \{\Pi^{(i)}_{\alpha_i} \mid h^{n}_{\check{i}}(\underline{\alpha})\} := \frac{\mu_{\omega}(\alpha_1,...,\alpha_i,...,\alpha_n)}{\sum_{\beta_i=1}^{k_i} \mu_{\omega}(\alpha_1,...,\beta_i,...,\alpha_n)}.
\end{equation}
Unfortunately, there is a problem with this definition! Recall that $\Pi_{\beta_i}^{(i)}$ is a shorthand for the spectral projection $P_{a_i(t_i)}(I_{\beta_i}^{i})$. We fix a subset $I^{i}_{\alpha_i}$, but introduce a \textit{new} decomposition of $\text{spec }a_i$ into subsets
\begin{equation*}
\tilde{I}_{1}^{i}:=I_{\alpha_i}^{i}, \qquad \mathbb{R} \setminus I^{i}_{\alpha_i}= \cup_{\beta=2}^{m_i} \tilde{I}_{\beta}^{i},  
\end{equation*}
with $\tilde{I}_{\beta}^{i} \cap \tilde{I}_{\gamma}^{i} =\emptyset$, for $\beta \neq \gamma$, and define 
\begin{equation*}
\tilde{\Pi}^{(i)}_{\beta_i}:=P_{a_i(t_i)}(\tilde{I}^{i}_{\beta_i}),
\end{equation*} 
$\beta_i=1,...,m_i$. We define
\begin{equation*}
\tilde{\mu}_{\omega}(\alpha_1,...,\beta_i,...,\alpha_n):= \text{Prob}_{\omega} \{ \Pi^{(1)}_{\alpha_1},..., \tilde{\Pi}_{\beta_i}^{(i)},..., \Pi^{(n)}_{\alpha_n}\}. 
\end{equation*}
Then
$$\tilde{\mu}_{\omega}(\alpha_1,...,1,...,\alpha_n)=\mu_{\omega}(\alpha_1,...,\alpha_i,...,\alpha_n);$$
but, most often, the putative "conditional probabilities" are different, 
\begin{equation}
\label{4.14}
\text{Prob}_{\tilde{\mu}_{\omega}} \{ \Pi_{\alpha_i}^{(i)} \mid h^{n}_{\check{i}}(\underline{\alpha}) \} \neq  \text{Prob}_{\mu_{\omega}} \{ \Pi_{\alpha_i}^{(i)} \mid h^{n}_{\check{i}}(\underline{\alpha}) \},
\end{equation}
unless all possible interference terms \textit{vanish}. Thus, in general, there is \textit{no} meaningful notion of \textit{"conditional probability"} in quantum mechanics. 

It may be of interest to note that if the operators $a_i$ have pure-point spectrum with \textit{only two} distinct eigenvalues then
\begin{equation*}
\{ \Pi^{(i)}_{\alpha_i}\}_{\alpha_i=1,2}=\{ \tilde{\Pi}^{(i)}_{\beta_i}\}_{\beta_i=1,2},
\end{equation*}
and we have equality in \eqref{4.14}. These findings may be viewed as a general version of the \textit{Kochen-Specker theorem}, \cite{K-Sp}.

Let us recall a "test" for one of the possible events $\{ \Pi^{(i)}_{\alpha_i}\}_{\alpha_i=1}^{k_i}$  to materialize in a measurement at time $t_i$ of the potential property of $S$ represented by the operator $a_i  \in \mathcal{P}_S$; (see \cite{Fr-Schub} and refs. given there). For this purpose, we introduce the  matrix 
\begin{equation}
\label{4.15}
P^{\omega}_{\underline{\alpha}, \underline{\alpha}'}:= \omega( \Pi^{(1)}_{\alpha_1} ...  \Pi^{(n)}_{\alpha_n}  \Pi^{(n)}_{\alpha'_n} ... \Pi^{(1)}_{\alpha'_1}),
\end{equation}
with $\alpha_n=\alpha'_n$; see (\ref{1.14}). Classically, $P^{\omega}=(P^{\omega}_{\underline{\alpha}, \underline{\alpha}'})$ is always a diagonal matrix, because all the operators $\Pi^{(i)}_{\alpha_i}$ commute with one another and by \eqref{2.30}. We say that a family of histories $\{ h_1^n(\underline{\alpha}\}$ is $consistent$ iff the commutators
\begin{equation*}
[\Pi^{(i)}_{\alpha_i}, H^{n}_{i+1}(\underline{\alpha})]
\end{equation*}
\textit{vanish}, for all $\alpha_i, \underline{\alpha}$ and $i=1,...,n$; (see \cite{Griff}). If   $\{ h_1^n(\underline{\alpha})\}$ is consistent then $P^{\omega}_{\underline{\alpha}, \underline{\alpha}'}$ is diagonal, and the sum rules (\ref{4.8}) are valid for all $\underline{\alpha}$ and all $i=1,...,n$. We say that a family $\{ h_{1}^{n}(\underline{\alpha})\}$ of histories is  $\delta$-\textit{consistent}($0 \leq \delta \leq 1$) iff
\begin{equation}
\label{4.16}
\|  [ \Pi^{(i)}_{\alpha_i}, H^{n}_{i+1}(\underline{\alpha}) ]\| \leq 1- \delta.
\end{equation}
A $1-$consistent family of histories is consistent.  We define a diagonal matrix $\Delta^{\omega}$ by 
\begin{equation*}
\Delta_{\underline{\alpha},\underline{\alpha}'} := \left \{  \begin{array}{cc} P^{\omega}_{\underline{\alpha},\underline{\alpha}} & \text{ if } \underline{\alpha}=\underline{\alpha}'\\ 0 & \text{ else}  \end{array} \right.
\end{equation*}
Clearly inequality (\ref{4.16}) implies that
\begin{equation}
\label{4.17}
\|P^{\omega} - \Delta^{\omega} \| \leq \text{const.} (1- \delta).
\end{equation}
 This shows that, for a $\delta-$consistent family of histories, with $\delta \approx 1$, the sum rules (\ref{4.8}) are very nearly satisfied, meaning that the events $\Pi^{(i)}_{1},...,\Pi^{(i)}_{k_i}$ mutually exclude one another FAPP ("for all practical purposes", \cite{Bell}). In \cite{Fr-Schub}, we have called 
 \begin{equation*}
 e^{\omega}:=1- \| P^{\omega} - \Delta^{\omega}\|
 \end{equation*}
 the $"evidence"$ for $\Pi^{(i)}_1,...,\Pi^{(i)}_{k_i}$ to mutually exclude one another, FAPP, $i=1,...,n$. Apparently, if $e^{\omega}$ is very close to $1$, then everything might appear to be fine. -- Well, the appearance is deceptive, as we will explain below! 
 
 Dynamical mechanisms that imply that $ \| P^{\omega} - \Delta^{\omega}\|$ becomes small, i.e., $e^{\omega}$ approaches $1$, in suitable limiting regimes are known under the names of \textit{"dephasing"} and \textit{"decoherence"}; see \cite{Hepp, GM-Hartle, HMaass, Cop}. Understanding decoherence is clearly an important task. Here we summarize a few observations on those mechanisms; but see Subsects. 5.3 and 5.4. (Some instructive examples will be discussed elsewhere.)
 
 \subsection{Dephasing/Decoherence}
 \label{sec4.3}  
 In our discussion of near (i.e., $\delta-$) consistency of families of histories, $h^n$, operators $Q^{n}_{k}(\underline{\alpha})$, defined by 
 \begin{equation}
 \label{4.18}
 Q^{n}_{k}(\underline{\alpha}):=(H^{n}_{k}(\underline{\alpha}))^* H^{n}_{k}(\underline{\alpha})= \Pi^{(k)}_{\alpha_k}(t_k) ...  \Pi^{(n)}_{\alpha_n}(t_n)...  \Pi^{(k)}_{\alpha_k}(t_k),
 \end{equation}
 $t_k<t_{k+1}<...<t_n$, $1 \leq k \leq n$, play an important role. Inequality (\ref{4.16}) implies that
 \begin{equation}
 \label{4.19}
 \| [ \Pi^{(i)}_{\alpha_i}, Q^{n}_{i+1}(\underline{\alpha})] \| \leq 2 (1- \delta) \ll 1
 \end{equation}
 if $\delta$ is very close to $1$. Condition (\ref{4.19}) is slightly weaker than (\ref{4.16}), so we will work with (\ref{4.19}). If (\ref{4.19}) holds, for all $i$ and all $\underline{\alpha}$, the sum rules (\ref{4.8}) are satisfied, up to tiny errors, and the matrix $P^{\omega}$ is very nearly diagonal; so there is \textit{"decoherence"}. A (very stringent) sufficient condition for 
 \begin{equation}
 \label{4.20}
 [\Pi^{(i)}_{\alpha_i},Q^{n}_{i+1}(\underline{\alpha})]=0
 \end{equation}
 to hold, for all $i$ and all $\underline{\alpha}$, i.e., for perfect decoherence to hold, is the following one: We observe that 
 \begin{equation}
 \label{4.21}
 Q^{n}_{k}(\underline{\alpha}) \in \mathcal{E}^{\omega}_{\geq t_k}, \text{ for all } \underline{\alpha},
 \end{equation}
 where the von Neumann algebras $\mathcal{E}^{\omega}_{\geq t}$ of possible events observable at times $\geq t$ have been introduced in Definition 2.4, Subsect. \ref{sec2.3}. If there is \textit{loss of information}, in the sense of condition (\ref{2.26}), more precisely if the relative commutants
 \begin{equation}
 \label{4.22}
( \mathcal{E}^{\omega}_{\geq t_{i+1}})' \cap \mathcal{E}^{\omega}_{\geq \tilde{t}_i}, \qquad t_{i-1} < \tilde{t}_i \leq t_i,
 \end{equation} 
 are \textit{non-trivial}, for suitable choices of sequences of times $t_1<t_2<...<t_n$, $\tilde{t}_1<\tilde{t}_2<...<\tilde{t}_n$, and if the operator 
 \begin{equation}
 \label{4.23}
 a_i(t_i) \in (\mathcal{E}^{\omega}_{\geq t_{i+1}})' \cap \mathcal{E}^{\omega}_{\geq \tilde{t}_i},
 \end{equation}
 and hence $\Pi^{(i)}_{\alpha_i}$ belongs to $(\mathcal{E}^{\omega}_{\geq t_{i+1}})' \cap \mathcal{E}^{\omega}_{\geq \tilde{t}_i}$, for all $\alpha_i=1,...,k_i$, with $t_{i-1}< \tilde{t}_i \leq t_i$, then 
 \begin{equation}
 \label{4.24}
 [ \Pi^{(i)}_{\alpha_i},Q_{i+1}^{n}(\underline{\alpha})]=0,
 \end{equation}
 for all $\alpha_i$ and all $\underline{\alpha}$. If (\ref{4.23}) and hence \eqref{4.24} hold, for all $i \leq n$, then there is perfect decoherence, and the histories $\{h^{n}_1(\underline{\alpha})\}$ form a consistent family. 
 
 The scenario for decoherence described here is encountered in \textit{relativistic quantum field theories} with a massless particle (e.g., the photon), as can be inferred from results in \cite{Buchholz,Buchh-Roberts}. In \textit{non-relativistic} quantum mechanics, the above scenario for decoherence remains plausible, provided one allows for small changes of the operators $a_i(t_i)$ into operators $\tilde{a}_i(t_i)$ that belong to $(\mathcal{E}^{\omega}_{\geq t_{i+1}})' \cap \mathcal{E}^{\omega}_{\geq \tilde{t}_i}$. In this connection the following result may be of interest.
 
 \begin{theorem}
\label{th4.1}
Let $\Pi^{(1)}_{\alpha_1},...,\Pi^{(n)}_{\alpha_n}$ be orthogonal projections, and let the operators $Q_{k}^{n}(\underline{\alpha})$ be defined as in \eqref{4.18}. Suppose that 
\begin{equation}
\label{4.25}
\| [\Pi^{(i)}_{\alpha_i}, Q^{n}_{i+1}(\underline{\alpha})] \| < \epsilon,
\end{equation}
for all $i=1,...,n-1$ and all $\underline{\alpha}=(\alpha_1,...,\alpha_n)$, with $\epsilon$ sufficiently small (depending on  the total number, $\sum_{i=1}^{n} k_i$, of $n-$tuples $\underline{\alpha}$, with $\alpha_i=1,...,k_i$).
Then there exist orthogonal projections $\tilde{\Pi}^{(i)}_{\alpha_i}$, $\alpha_i=1,...,k_i$, $i=1,...,n$, with
\begin{equation}
\label{4.26}
\tilde{\Pi}^{(i)}_{\alpha_i} \tilde{\Pi}^{(i)}_{\beta_i}=\delta_{\alpha_i \beta_i} \tilde{\Pi}^{(i)}_{\alpha_i}, \qquad \sum_{\alpha_i=1}^{k_i} \tilde{\Pi}_{\alpha_i}^{(i)}=\mathds{1},
\end{equation}
such that 
\begin{equation}
\label{4.27}
\| \tilde{\Pi}^{(i)}_{\alpha_i} - \Pi^{(i)}_{\alpha_i}\| \leq C \epsilon,
\end{equation}
and
\begin{equation}
\label{4.28}
[\tilde{\Pi}^{(i)}_{\alpha_i}, \tilde{Q}_{i+1}^{n}(\underline{\alpha})]=0,
\end{equation}
for all $\underline{\alpha}$ and all $i \leq n-1$. The constant $C$ in \eqref{4.27} depends on $\sum_{i=1}^{n} k_i$, and $\epsilon$ must be chosen so small that $C \epsilon <1$; (in which case $\tilde{\Pi}^{(i)}_{\alpha_i}$ and $\Pi^{(i)}_{\alpha_i}$ are unitarily equivalent). 
  \end{theorem}
  
  \begin{remark}
  The operators $, \tilde{Q}_{k}^{n}(\underline{\alpha})$ are defined as in \eqref{4.18}, with $\Pi^{(i)}_{\alpha_i}(t_i) \equiv \Pi^{(i)}_{\alpha_i}$ replaced by $\tilde{\Pi}^{(i)}_{\alpha_i}$, for all $i$. 
   \end{remark}

  The proof of Theorem \ref{th4.1} can be inferred from Sect. 4.5  of \cite{Fr-Schub}, (Lemmata 7 and 8).
  \vspace{2mm}
 
 \noindent \textit{Interpretation of Theorem \ref{th4.1}} Apparently, dephasing/decoherence in the form of inequalities (\ref{4.25}) implies that if one reinterprets the measurements made at times $t_1<t_2<...<t_n$ as observations of events $\tilde{\Pi}^{(1)}_{\alpha_1},...,\tilde{\Pi}^{(n)}_{\alpha_n}$ that differ slightly from the spectral projections  $\Pi^{(1)}_{\alpha_1},...,\Pi^{(n)}_{\alpha_n}$ of potential properties $a_1,...,a_n$ of $S$ then all interference terms (see (\ref{4.10}), (\ref{4.15})) vanish, the matrix $P^{\omega}$ is diagonal, and the sum rules (\ref{4.8}) hold. The family of histories $\{  \tilde{\Pi}^{(1)}_{\alpha_1},...,\tilde{\Pi}^{(n)}_{\alpha_n}\}$ is consistent, and the complementary possible events  $\tilde{\Pi}^{(i)}_{1},...,\tilde{\Pi}^{(i)}_{k_i}$ mutually exclude one another.
 \vspace{2mm}
 
 \noindent { \bf{Critique of the concept of "consistent families of histories."}}
 \begin{enumerate}[(i)]
\item Given a measurement of a potential property $a_i \in \mathcal{P}_S$ of $S$ at some time $t_i$, the success of this measurement, as expressed in the decoherence of (absence of interference between) the events $\Pi^{(i)}_1,...,\Pi_{k_i}^{(i)}$, apparently \textit{not} only depends on the \textit{past} but seems to depend on the \textit{future}, namely on subsequent measurements of potential properties $a_{i+1},...,a_n$ at times $>t_i$. This is how conditions such as (\ref{4.16}), (\ref{4.19}) and (\ref{4.25}) must be interpreted. The consistency of a family $\{ h_1^{i} (\underline{\alpha})\}$ of stretches of histories (see \eqref{4.5} for the definition) can apparently only be assured if one also knows the family $\{ h_{i+1}^{n} (\underline{\alpha})\}$ of stretches of histories in the \textit{future} of $\{ h_1^{i} (\underline{\alpha})\}$. This may be a deep aspect of quantum mechanics; but it is \textit{more likely} an indication that there is something \textit{wrong} with the concept of "consistent (families of) histories" and with a formulation of decoherence in the form of inequalities (\ref{4.25}).
\item Accepting, temporarily, the idea of "consistent (families of) histories" -- e.g., in the appealing form of conditions (\ref{4.23}) -- we encounter the following problem:  Fixing the data
\begin{equation}
\label{4.29}
(\mathcal{P}_S, \mathcal{A}_{\overline{S}}, \{ \tau_{t,s}\}_{t,s \in \mathbb{R}}, \omega \in \mathcal{S}_{\overline{S}}),
\end{equation}
see (\ref{4.1}), we may consider two (or more) families of potential properties of $S$,
\begin{equation}
\label{4.30}
\{a_1,...,a_n \} \qquad \text{ and } \qquad  \{b_1,...,b_m \},
 \end{equation}
 measured at times $t_1<...<t_n$ and $t'_1<...<t'_m$, respectively, with $a_i \in \mathcal{P}_S$ and $b_j \in \mathcal{P}_S$, for all $i$ and $j$.  Both families may give rise to consistent families of histories (e.g., if conditions (\ref{4.23}) hold for the $a_i$'s and the $b_j$'s). Yet, there may \textit{not} exist \textit{any} family
 $$\{c_1,...,c_N\}, \text{ } N \geq n+m,$$
  of potential properties of $S$ ($c_j \in \mathcal{P}_S$, for all $j$) measured at times $T_1<...<T_N$, with
  $$\{T_1,...,T_N\} \supseteq \{t_1,...,t_n\} \cup  \{t'_1,...,t'_m\}, $$
  encompassing the two families in (\ref{4.30}) and giving rise to a consistent family of histories. Since the data (\ref{4.29}) are fixed, the confusing question arises \textit{which one} of the two or more incompatible families of potential properties $ \{a_1,...,a_n\}$, $ \{b_1,...,b_m\}$, ...  will actually be observed in the course of time, i.e., become $real$, (or, put differently, correspond to \textit{empirical} properties). Some people suggest, following Everett \cite{Everett}, that there is a world for \textit{every} family of potential properties of $S$ giving rise to a consistent family of histories to be observed. This is the \textit{"many-worlds interpretation of quantum mechanics"}, which we find entirely unacceptable!
  \item Unfortunately, the problem described in (ii) persists even in the decoherence scenario described in (\ref{4.20}) -- (\ref{4.24}), above, because the von Neumann algebras
  \begin{equation}
  \label{4.31}
  \mathcal{M}_i:=(\mathcal{E}^{\omega}_{\geq t_{i+1}})' \cap  \mathcal{E}^{\omega}_{\geq \tilde{t}_{i}} \qquad (t_{i-1} <\tilde{t}_i \leq t_i)
  \end{equation}
  are usually \textit{non-commutative}. If there are an $a_i$ and a $b_j$ from the sets of operators in (\ref{4.30}) belonging to the \textit{same} $\mathcal{M}_l$, and if 
  \begin{equation}
  \label{4.32}
  [a_i(t_i),b_j(t'_j)] \neq 0,
  \end{equation}
  then the problem described in (ii) appears on the scene. It could be avoided if one assumed that $a_i(t_i)$ and $b_j(t'_j)$ must belong to the center, $\mathcal{Z}_{\mathcal{M}_l}$, of $\mathcal{M}_l$, because then the commutators on the left side in (\ref{4.32}) would all vanish. The right version of something like this idea will be formulated in Subsects. 5.3 and 5.4.
    
  \item It has tacitly been assumed, so far, that the times at which quantum-mechanical measurements of potential properties of a system $S$ are carried out (we are talking of the times $t_i$ at which potential properties $a_i$ of $S$ are observed) can be \textit{fixed precisely} (by an "observer"?). -- Obviously, this assumption is \textit{nonsense} in quantum mechanics, (as opposed to classical physics); see Subsect. \ref{sec5.4}. 
 \end{enumerate}
 
 In an appendix, the reader may find some remarks on positive operator-valued measures (POVM) \cite{Perez} and their uses; (but see also the end of Subsect. \ref{5.4} and \cite{Fr-Schub3}).
 
\begin{subappendices}
\subsection{Appendix  to Section 4. Remarks on positive operator-valued measures, (POVM)}
\label{appA}
It may and will happen sometimes that the commutators
$$[\Pi_{\alpha_i}^{(i)}, Q_{i+1}^{n}(\underline{\alpha})]$$
are \textit{not} small in norm, and the matrix $P^{\omega}$ defined in \eqref{4.15} has "large" off-diagonal elements. Then some of the operators $a_i$ representing potential properties of $S$ are not measurable and do apparently \textit{not} represent \textit{empirical properties} of $S$, given the data 
\begin{equation*}
(\mathcal{P}_S, \mathcal{A}_{\overline{S}}, \{ \tau_{t,s}\}_{t,s \in \mathbb{R}},  \omega \in \mathcal{S}_{\overline{S}}). 
\end{equation*}
While this is a perfectly interesting piece of information, it raises the question whether formula (\ref{4.7}) continues to contain interesting information, although the sum rule (\ref{4.8}) may be strongly violated. A conventional answer to this question involves the notion of \textit{"positive operator-valued measures"} (POVM): For $k^{-} < k^{+}$, we define
\begin{equation}
\label{A.1}
H_{k^-}^{k^+}(\underline{\alpha}):= \Pi_{\alpha_{k^+}}^{(k^+)} \Pi_{\alpha_{k^{+}-1}}^{(k^{+}-1)} ...  \Pi_{\alpha_{k^{-}+1}}^{(k^{-}+1)}  \Pi_{\alpha_{k^{-}}}^{(k^{-})}.
\end{equation}
We observe that
\begin{equation}
\label{A.2}
\underset{\underline{\alpha}}{\sum} \left(  H_{k^-}^{k^+}(\underline{\alpha}) \right)^* H_{k^-}^{k^+}(\underline{\alpha})= \mathds{1},
\end{equation}
(and
\begin{equation*}
\label{A.3}
\underset{\underline{\alpha}}{\sum}   H_{k^-}^{k^+}(\underline{\alpha}) \left(  H_{k^-}^{k^+}(\underline{\alpha}) \right)^*= \mathds{1}.)
\end{equation*}
Consider
\begin{equation}
\label{A.4}
\begin{split}
\text{Prob}_{\omega} \{ \Pi^{(1)}_{\alpha_1},..., \Pi^{(n)}_{\alpha_n}\} & \equiv  \text{Prob}_{\omega} \{ h_1^{k^- -1}(\underline{\alpha}),  h_{k^-}^{k^+ }(\underline{\alpha}),  h_{k^+ +1 }^{n}(\underline{\alpha}) \}\\
&:= \omega \left( (H_{1}^{k^- -1})^{*}  (H_{k^- }^{k^+})^{*}  (H_{k^+ +1}^{n})^{*} H_{k^+ +1}^{n} H_{k^- }^{k^+}  H_{1}^{k^- -1} \right)
\end{split}
\end{equation}
We may say that  $  h_{k^-}^{k^+ }(\underline{\alpha})$ represents a \textit{single experiment} on the system $S$ if the sum rule (\ref{4.8}) is \textit{violated substantially}, for all $i=k^-,k^-+1,...,k^+$,  but
\begin{equation}
\label{A.5}
\underset{\alpha_{k^-},...,\alpha_{k^+}}{\sum}  \text{Prob}_{\omega} \{ h_1^{k^- -1}(\underline{\alpha}),  h_{k^-}^{k^+ }(\underline{\alpha}),  h_{k^+ +1 }^{n}(\underline{\alpha}) \}  \approx  \text{Prob}_{\omega} \{ h_1^{k^- -1}(\underline{\alpha}),  h_{k^+ +1 }^{n}(\underline{\alpha}) \},
\end{equation}
up to an error that is so small that it is below the experimental resolution. In view of \eqref{A.2}, our discussion can be formalized as follows.

\begin{definition}
The "square root" of a positive operator-valued measure (POVM) is a (finite) family of operators 
\begin{equation}
\label{A.6}
\underline{X}= \{ X_{\alpha}\}_{\alpha=1}^{N}
\end{equation}
with the property that 
\begin{equation}
\label{A.7}
\sum_{\alpha=1}^{N} X_{\alpha}^{*} X_{\alpha}  = \mathds{1}.
\end{equation}
\end{definition}

The "positive operator-valued measure" is then given by the operators $\{X_{\alpha}^{*}X_{\alpha}\}_{\alpha = 1}^{N}$.

Given a time-ordered sequence of ("square roots" of) POVM's, $\underline{X}^{(1)},...,\underline{X}^{(n)}$, the probability of observing a "generalized history"
\begin{equation}
\label{A.8}
h^{n}_{1}(\underline{\alpha})= \{ X_{\alpha_1}^{(1)},...,X_{\alpha_n}^{(n)} \}
\end{equation}
is given by 
\begin{equation}
\label{A.9}
\text{Prob}_{\omega} \{ X_{\alpha_1}^{(1)},...,  X_{\alpha_n}^{(n)}\}:= \omega \left( (X_{\alpha_1}^{(1)})^* ... (X_{\alpha_n}^{(n)})^*   X_{\alpha_n}^{(n)}  ... X_{\alpha_1}^{(1)} \right).
\end{equation}
The probabilities of such generalized histories have the desirable properties (\ref{2.37}) and (\ref{2.38}). We say that $\{ \underline{X}^{(1)},..., \underline{X}^{(n)} \}$, with $\underline{X}^{(i)}$ (the square root of) a POVM, for all $i$, describes a time-ordered sequence of $n$ successful experiments, or observations, iff
\begin{equation}
\label{A.10}
\underset{\alpha_i}{\sum} \text{Prob}_{\omega} \{ X_{\alpha_1}^{(1)},..., X_{\alpha_i}^{(i)},..., X_{\alpha_n}^{(n)} \}  \approx   \text{Prob}_{\omega} \{ X_{\alpha_1}^{(1)},..., X_{\alpha_{i-1}}^{(i-1)},X_{\alpha_{i+1}}^{(i+1)},..., X_{\alpha_n}^{(n)} \} ,
\end{equation}
up to a tiny error below the experimental resolution, for all $i=1,...,n$ and all $\underline{\alpha}$. 
An example of events described by POVM's is described in Subsect. 5.4; (see also \cite{Fr-Schub3}).

All the concepts and notions introduced in Sect.\ref{sec4} can be carried over to this generalized setup, after replacing $a_i$ by $\underline{X}^{(i)}$ and $\Pi_{\alpha_i}^{(i)} \equiv \Pi_{\alpha_i}^{(i)}(t_i)$ by $X_{\alpha_i}^{(i)} \equiv X_{\alpha_i}^{(i)}(\Delta_i)$ (or their adjoints),  $i=1,...,n$, with $\Delta_1 \prec ... \prec \Delta_n$.  Wherever possible, we will, however, consider self-adjoint operators and their spectral projections, instead of POVM's, throughout this essay; (but see Remark 5.8, Subsect. 5.4).

\end{subappendices}
\section{Removing the veil: Empirical properties of physical systems in quantum mechanics}
\label{sec5}
In a classical model of a physical system, $S$, properties of $S$ exist \textit{a priori}. They are represented by real-valued continuous (or measurable) functions on the state space, $M_S$, of the system. In contrast, in a quantum-mechanical model of a physical system, the system can still be characterized by a list, $\mathcal{P}_{S}$, of \textit{potential properties} (represented by self-adjoint operators); but these properties do \textit{not} exist \textit{ a priori}. Whether they correspond to \textit{empirical properties} of $S$, or not, depends on the choice of the environment $E$; (e.g., on the experiments that are made). The question then arises what the \textit{empirical properties} are that will be observed  in the course of time, given the time evolution $\{\tau_{t,s}\}_{t,s \in \mathbb{R}}$ of $\overline{S}=S \vee E$ and its state $\omega \in \mathcal{S}_{\overline{S}}$; (see Definition \ref{def2.1}, Subsect. \ref{sec2.2}). In (\ref{data}), we have identified the fundamental data underlying a model of $S$,   
\begin{equation}
\label{5.1}
(\mathcal{P}_S, \mathcal{A}_{\overline{S}}, \{ \tau_{t,s}\}_{t,s \in \mathbb{R}} \subset \text{  } ^*\text{Aut}(\mathcal{A}_{\overline{S}}),  \omega \in \mathcal{S}_{\overline{S}}),
\end{equation}
see also (\ref{4.1}) and Subsect.\ref{sec2.2}. These data ought to \textit{determine} which \textit{empirical properties} S exhibits and what $family$ of histories of events (but, of course, not $which$ history) will be recorded in the course of time if $S$ is monitored/observed when coupled to a given environment/equipment $E$. We have seen in Sect.\ref{sec4} that the answer to the question of what exactly the data in (\ref{5.1}) determine is \textit{not obvious}. 

\subsection{Information loss and entanglement} 
\label{sec5.1}
Let $a$ be a potential property of $S$ ($a=a^* \in \mathcal{P}_S$). We assume, for simplicity, that $\text{spec } a$ consists of finitely many eigenvalues, $\alpha_1, \alpha_2,..., \alpha_k$. Let $\omega$ be the state of $\overline{S}=S \vee E$, and let us suppose that, thanks to an appropriate choice of $E$, the potential property $a$ is \textit{observed} (i.e., becomes an empirical property of  $S$) around some time $t$. According to almost everybody's understanding of quantum mechanics, the following claim appears to  be reasonably plausible: After the observation of $a$ at a time $\approx t$, $S$ evolves  \textit{as if } its state where given by 
\begin{equation}
\label{5.2}
\omega \longrightarrow \sum_{i=1}^{k} p_i \omega_i,
\end{equation}
where $p_i$ is the probability to observe the value $\alpha_i$ of $a$, and $\omega_i$ is a state with the property that if $a$ were observed in a system prepared in the state $\omega_i$ at time $\approx t$ then its value would be $\alpha_i$ with certainty. If no measurements are made before $a$ is observed then, according to Born \cite{Born}, 
\begin{equation*} 
p_i=\omega(\Pi_{i}(t)), 
\end{equation*}
where $\Pi_i(t)$ is the spectral projection of the operator $a(t)=\tau_{t,t_0}(a)$ corresponding to the eigenvalue $\alpha_i$, (with $t$ the time of measurement of $a$). Note that the state in (\ref{5.2}) is usually  a \textit{mixed state}, i.e., an incoherent superposition of the states $\omega_i$, \textit{even} if $\omega$ is a \textit{pure} state. It is perceived as one aspect of the \textit{"measurement problem"} to understand how a pure state can evolve into a mixture. (Another aspect is to understand why the state of $S$ is given by $\omega_i$, right after the measurement of $a$, if $a$ is measured to have the value $\alpha_i$, for some $i=1,...,k$. This will be discussed in Sects. \ref{sec5.4} and \ref{sec5.5}). 

In order to explain why the first aspect of the measurement problem does not represent a serious problem, we have to return to an analysis of two fundamental phenomena:  ($\mathcal{L}o\mathcal{I}$) \textit{Loss of information} into $E$; and ($\mathcal{E}$) \textit{Entanglement} between $S$ and $E$. 

In Definition \ref{def2.2} of Subsect.\ref{sec2.3}, we have introduced \textit{algebras}, $\mathcal{E}_{\geq t}$, \textit{of potential properties of }$S$ \textit{ observable/measurable after time} $t$. These algebras are $C^*-$subalgebras of the algebra $\mathcal{A}_{\overline{S}}$. We have denoted by $\mathcal{E}_S$ the smallest $C^*-$algebra containing $\mathcal{E}_{\geq t}$, for all $t \in \mathbb{R}$; see Eq. (2.26). Clearly $\mathcal{E}_S \subset \mathcal{A}_{\overline{S}}$. As indicated in Subsection 2.3, it is the consequence of a $\bf{general}$ $\bf{principle}$ -- "Duality between Observables and "Indeterminates -- that 
$\mathcal{E}_S$ is $properly$ contained in $\mathcal{A}_{\overline{S}}$ (and that the relative commutant of $\mathcal{E}_S$ inside $\mathcal{A}_{\overline{S}}$ contains a sub algebra isomorphic to $\mathcal{E}_S$). This principle will be discussed in the context of examples in a forthcoming communication. 

The algebra  $\mathcal{E}_S$ carries an action of the group $\mathbb{R}$ of time translations by $^*$automorphisms $\{ \overline{\tau}_t\}_{t \in \mathbb{R}}$,  where $\overline{\tau}_t$ determines $*$morphisms
\begin{equation*}
\overline{\tau}_t: \mathcal{E}_{\geq t'} \longrightarrow \mathcal{E}_{\geq t + t'} \subseteq \mathcal{E}_{\geq t'},
\end{equation*}
for all $t' \in \mathbb{R}$ and all $t \geq 0$; see (\ref{2.28}) and (\ref{2.29}).

Thus, in hindsight, the mathematical data enabling one to predict the behavior of a physical  system $S$ in the course of time, given its state, can be chosen to consist of the \textit{filtration} of algebras
\begin{equation}
\label{5.3}
\mathcal{A}_{\overline{S}} \supset \mathcal{E}_S \supseteq \mathcal{E}_{\geq t} \supseteq \mathcal{E}_{\geq t'} \supset \{ \mathbb{C} \mathds{1}\}, \qquad t' \geq t,
\end{equation}
along with a specification of $^*$\textit{morphisms} (\textit{time translations})
\begin{equation}
\label{5.4}
\overline{\tau}_t : \mathcal{E}_{\geq t'} \longrightarrow \mathcal{E}_{\geq t + t'} \subseteq \mathcal{E}_{\geq t'},
\end{equation}
for $t' \in \mathbb{R}$, $t \geq 0$, and of a \textit{state} $\omega$,
\begin{equation}
\label{5.5}
\omega: \text{state on } \mathcal{A}_{\overline{S}}.
\end{equation}

In Definition 2.4, Subsect.\ref{2.3}, we have introduced the von Neumann algebras $\mathcal{E}^{\omega}_{\geq t}$, $t \in \mathbb{R}$, and $\mathcal{E}^{\omega}_{S}$. (We recall that if $\omega'$ is an arbitrary state on $\mathcal{E}_S$ \textit{normal} with respect to $\omega$ then $\mathcal{E}^{\omega'}_{S}=\mathcal{E}^{\omega}_{S}$ and  $\mathcal{E}^{\omega'}_{\geq t}=\mathcal{E}^{\omega}_{\geq t}$, for all $t$). 

\textit{Loss of information} ($\mathcal{L}o\mathcal{I}$) is the phenomenon that if successful measurements of potential properties of $S$ have been made between some times $t$ and $t'>t$ then $\mathcal{E}_{\geq t'}$ is \textit{strictly contained} in $\mathcal{E}_{\geq t}$. Together with the phenomenon of \textit{entanglement} ($\mathcal{E}$), this may entail that the \textit{restriction of the state} $\omega$ \textit{to the algebra} $\mathcal{E}_{\geq t'}$ \textit{ is a mixture} (i.e., an incoherent superposition of approximate eigenstates of some physical quantity, as in (\ref{5.2})), $even$ if $\omega$ is a \textit{pure} state of $\mathcal{E}_S$.

While ($\mathcal{L}o\mathcal{I}$) is common to classical and quantum-mechanical models of physical systems, ($\mathcal{E}$) and (\ref{5.2}) (with $p_i>0$, for \textit{two} or \textit{more} choices of $i$) is specific to quantum-mechanical models. We have seen in Sect.\ref{sec2.3} that, quantum-mechanically, $(\mathcal{L}o\mathcal{I})$ may manifest itself in the property that some of the relative commutants,
\begin{equation}
\label{5.6}
(\mathcal{E}^{\omega}_{\geq t})' \cap \mathcal{E}^{\omega}_{\geq t'}
\end{equation}
are non-trivial, for some $t > t'$; (see (\ref{2.26})). If $\mathcal{E}^{\omega}_{\geq t}$ is a factor (i.e., a von Neumann algebra with trivial center, as defined in Sect.\ref{sec2.1}, (\ref{2.4})) then (\ref{5.6}) implies that
\begin{equation}
\label{5.7}
\mathcal{E}^{\omega}_{\geq t} \subsetneq \mathcal{E}^{\omega}_{\geq t'}.
\end{equation}
 
\subsection{Preliminaries towards a notion of "empirical properties" of quantum mechanical systems}
\label{sec5.2}
Let $a=a^* \in \mathcal{E}_S$ be an operator representing a potential property (or physical quantity) of $S$ (i.e., $a=\tau_{t',t_0}(c), \text{ } c \in \mathcal{P}_S$), and let $\omega$ denote the state of $S$. We assume that $a$ has a finite spectrum,
\begin{equation}
\label{5.8}
a=\sum_{i=1}^{k} \alpha_i \Pi_i, \qquad k < \infty,
\end{equation}
where $\alpha_1,...,\alpha_k$ are the eigenvalues of $a$ (now viewed as a self-adjoint operator in the von Neumann algebra $\mathcal{E}_{S}^{\omega}$), and $\Pi_i \in \mathcal{E}_{S}^{\omega}$ is the spectral projection of $a$ corresponding to $\alpha_i$, $i=1,...,k$. How should we define \textit{empirical properties} of $S$? To say that $a$ is an empirical property of $S$ at some time $t'$ earlier than $t$, i.e., that $a$ is measured (or observed) before time $t$, means that 
\begin{equation}
\label{5.9}
\omega(b)\approx \sum_{i=1}^{k} \omega(\Pi_i b \Pi_i),
\end{equation}
for all $b \in \mathcal{E}_{\geq t}$; i.e., $\omega_{\vert \mathcal{E}_{\geq t}}$ is close  to an incoherent superposition (mixture) of eigenstates, $p_{i}^{-1} \omega(\Pi_i (\cdot) \Pi_i)$ ($p_i \neq 0$), of $a$, where $p_i=\omega(\Pi_i)$, (and $p_i>0$, for at least one choice of $i$). A sufficient condition for \eqref{5.9} to hold is that 
\begin{equation}
\label{5.10}
a \in (\mathcal{E}_{\geq t})' \cap \mathcal{E}^{\omega}_S.
\end{equation} 
If there existed a sequence of times, $t_1<t_2<...<t_n$, and self-adjoint operators $a_1,...,a_n$, with finite point spectra, as above, and 
\begin{equation*}
a_{l} \in (\mathcal{E}_{\geq t_l+1})' \cap \mathcal{E}^{\omega}_{\geq t_l},
\end{equation*}
$l=1,...,n-1$, $a_n \in \mathcal{E}_{\geq t_n}$, then the family of histories
\begin{equation*}
h^{n}_1(\underline{j})=\{ \Pi_{j_1}^{(1)},...,\Pi_{j_n}^{(n)}\},
\end{equation*}
where $\Pi_{j_l}^{(l)}$ is the spectral projection of $a_l$ corresponding to the eigenvalue $\alpha_{j_l}^{(l)}$ of $a_l$, $l=1,...,n$, is \textit{consistent}; see (\ref{4.20}) -- (\ref{4.24}), Sect.\ref{sec4.3}. For this observation to be interesting, the relative commutants $(\mathcal{E}_{\geq t_l+1})'\cap \mathcal{E}^{\omega}_{\geq t_l}$ would have to be non-trivial and if we wish to escape from critique (iii) at the end of Sect.4.3 the algebras $(\mathcal{E}_{\geq t_l+1})'\cap \mathcal{E}^{\omega}_{\geq t_l}$  would have to be \textit{abelian}, for all $l$. This does \textit{not} look like a satisfactory or plausible assumption, and we have to continue our search for a good notion of "empirical properties"!

\begin{definition}
\label{def5.1}
\begin{enumerate}[(i)]
\item Given  von Neumann algebras $\mathcal{M} \subseteq \mathcal{N}$, a state $\omega$ on $\mathcal{N}$ and an operator $a \in \mathcal{N}$, we define $\{a,\omega]_{\mathcal{M}}$ to be the bounded linear functional on $\mathcal{M}$ defined by 
\begin{equation}
\label{5.11}
\{a,\omega]_{\mathcal{M}}(b):=\omega([a,b]), \qquad b \in \mathcal{M}.
\end{equation}
\item The centralizer (or stabilizer), $\mathcal{C}^{\omega}_{\mathcal{M}}$, of $\omega$ is the subalgebra of $\mathcal{M}$ defined by
\begin{equation}
\label{5.12}
\mathcal{C}^{\omega}_{\mathcal{M}}:=\{ a \in \mathcal{M} \mid \{a,\omega]_{\mathcal{M}}=0\}.
\end{equation}
\end{enumerate}
\end{definition}

It is easy to see that $\omega$ defines a \textit{trace} on $\mathcal{C}^{\omega}_{\mathcal{M}}$. This means that $\mathcal{C}^{\omega}_{\mathcal{M}}$ is a direct sum (or integral) of finite-dimensional matrix algebras, type-$II_1$ factors and abelian algebras. 
\begin{remark}
Centralizers of states or weights on von Neumann algebras play an interesting role in the classification of von  Neumann algebras, (in particular in the study of type-III factors); see \cite{Haagerup},\cite{Connes}. In an appendix to Sect.\ref{sec5}, we recall a few relevant results on centralizers. 
\end{remark}

Obviously, strict equality in \eqref{5.9} follows from the assumption that 
\begin{equation}
\label{5.13}
\{a,\omega]_{\mathcal{E}^{\omega}_{\geq t}} =0, \qquad a \in \mathcal{E}^{\omega}_{\geq t}.
\end{equation}
In other words, condition (\ref{5.13}) implies that, as a state on the algebra $\mathcal{E}^{\omega}_{\geq t}$ of possible events in $S$ observable after time $t$, $\omega$ is an incoherent superposition of eigenstates of $a$, \textit{even} if, as a state on $\mathcal{E}_S$, $\omega$ is \textit{pure}. However, to convince oneself that $\omega$ is a mixture (incoherent superposition) it is often enough to assume that the norm of the linear functional $\{a,\omega]_{\mathcal{E}^{\omega}_{\geq t}}$, with $a \in \mathcal{E}^{\omega}_{\geq t}$, is \textit{small}. Let us suppose that $a$ is self-adjoint and that its spectrum consists of finitely many eigenvalues $\alpha_1>\alpha_2>...>\alpha_k$. Then
\begin{equation*}
a=\sum_{i=1}^{k} \alpha_i \Pi_i,
\end{equation*}
where $\Pi_1,...,\Pi_k$ are the spectral projections of $a$ satisfying $\Pi_i=\Pi_{i}^{*}$, $\Pi_i \Pi_l=\delta_{il} \Pi_l$, for all $i,l=1,...,k$, and $\sum_{i=1}^{k} \Pi_i=\mathds{1}$. The following result is easily proven.
\begin{lemma}
\label{lem5.2}
The following assertions are equivalent:

\noindent (i) $\vert \{a,\omega]_{\mathcal{E}^{\omega}_{\geq t}} (b)\vert< \epsilon \|b \|, \text{ } \forall b \in \mathcal{E}^{\omega}_{\geq t}$

\noindent (ii) $\vert \omega(b)- \sum_{i=1}^{k} \omega(\Pi_i b \Pi_i) \vert \leq \text{const.} \epsilon \|b \|, \text{ } \forall b \in \mathcal{E}^{\omega}_{\geq t}$.
\end{lemma}

In view of Lemma \ref{lem5.2}, one might be tempted to identify elements of the centralizer
\begin{equation}
\label{5.14}
\mathcal{C}_{\geq t}^{\omega}:=\mathcal{C}_{\mathcal{E}^{\omega}_{\geq t}}^{\omega}
\end{equation}
with \textit{empirical properties of } $S$ \textit{observable at times} $\geq t$. Yet, this is not quite the right idea!
\begin{enumerate}[(1)]
\item A family of operators, $a_1,...,a_n$, with 
$$a_i \in \mathcal{C}^{\omega}_{\geq t_i},$$
$i=1,...,n$, $t_1<t_2<...<t_n$, does not necessarily give rise to a family of consistent histories. The reason is exceedingly simple: Let $\Pi^{(i-1)}_{l}$, $l=1,...,k_{i-1}$, be the spectral projections of $a_{i-1} \in \mathcal{C}_{\geq t_{i-1}}$. Let $\omega_l$ denote the state
$$\omega_l(b)=p_{l}^{-1} \omega(\Pi_{l}^{(i-1)} b \Pi_{l}^{(i-1)}),$$
where $p_l=\omega(\Pi^{(i-1)}_l)>0$. Let us assume that $p_l>0$ for at least \textit{two} distinct values of $l$. The problem is that, in general, the assumption that $a_i \in \mathcal{C}^{\omega}_{\geq t_i}$ does \textit{not} imply that $a_i \in  \mathcal{C}^{\omega_l}_{\geq t_i}$, for all $l=1,...,k_{i-1}$ for which $p_l >0$; this is the phenomenon of \textit{"spontaneous symmetry breaking"}. This means that the "sum rule" (\ref{4.8}), Sect.\ref{sec4.1}, may be violated at the $i^{th}$ slot, for some $1<i<n$. Hence the family $a_1,...,a_n$ may \textit{not} give rise to a family of consistent histories.
 
 \item In general, the centralizers $\mathcal{C}^{\omega}_{\geq t}$ are \textit{non-abelian} algebras. If the centralizers $\mathcal{C}^{\omega}_{\geq t}$ are non-commutative algebras then identifying empirical properties of $S$ observable at times $\geq t$ with elements of $\mathcal{C}^{\omega}_{\geq t}$ is subject to critique (ii), Subsect.\ref{4.3}. Our task is then to find out \textit{which elements of }$\mathcal{E}^{\omega}_{\geq t}$  \textit{may correspond to empirical properties of }$S$. (The center of $\mathcal{C}^{\omega}_{\geq t}$ is denoted by $\mathcal{Z}^{\omega}_{\geq t}$. If  $\mathcal{Z}^{\omega}_{\geq t}$ were known to contain operators representing potential properties of $S$ then these operators could be interpreted as \textit{empirical properties} of $S$ observed at some times $\geq t$, and critique (ii) of Subsect. \ref{4.3} would not apply, anymore.)
 
\end{enumerate}

\subsection{So, what are "empirical properties" of a quantum-mechanical system?}
\label{sec5.3}
Consider the data characterizing a physical system as specified in (\ref{5.3}) -- (\ref{5.5}). Let 
$\mathcal{E}_{\geq t}$ be the algebra of physical quantities pertaining to a system $S$ that can be observed at times $\geq t$, and let $\mathcal{E}_S$ be the $C^{*}-$algebra obtained as the norm closure of $\underset{t \in \mathbb{R}}{\vee} \mathcal{E}_t$. Let $\omega$ be a state on $\mathcal{E}_S$. 
By $\mathcal{C}^{\omega}_{\geq t}$ we have denoted the centralizer of the state $\omega$ (viewed as a state on the von Neumann algebra $\mathcal{E}^{\omega}_{\geq t}$ corresponding to the weak closure of  $\mathcal{E}_{\geq t}$ in the GNS representation associated with $(\mathcal{E}_S,\omega)$). We have seen, after definition (\ref{5.12}), that $\omega_{\vert \mathcal{C}^{\omega}_{\geq t}}$ is a \textit{trace} on $ \mathcal{C}^{\omega}_{\geq t}$. This implies that 
\begin{equation}
\label{5.15}
 \mathcal{C}^{\omega}_{\geq t} = \int_{\Lambda}^{\oplus}  \mathcal{C}^{\omega}_{\geq t,\lambda} \text{ },
\end{equation}
where every algebra $ \mathcal{C}^{\omega}_{\geq t,\lambda}$, $\lambda \in \Lambda \equiv \Lambda_{\omega}$, is either a finite-dimensional matrix algebra, $\approx \mathbb{M}_{n_{\lambda}}(\mathbb{C})$, of $n_{\lambda} \times n_{\lambda}$ matrices, with $1 \leq n_{\lambda} < \infty$, or a type-$II_1$ factor;  (see \cite{Opalg}, Theorem 8.21
 in Chapter 4, and Theorem 2.4 in Chapter 5). If  $\mathcal{C}^{\omega}_{\geq t,\lambda}$ is isomorphic to $\mathbb{M}_{n_{\lambda}}(\mathbb{C})$ then
\begin{equation}
\label{5.16}
\omega_{\mid  \mathcal{C}^{\omega}_{\geq t,\lambda}} \propto \text{tr}_{\mathbb{C}^{n_{\lambda}}} (\cdot).
\end{equation}
Let us assume, temporarily, that $\Lambda$ is discrete, and 
\begin{equation}
\label{5.17}
 \mathcal{C}^{\omega}_{\geq t} = \underset{\lambda \in \Lambda}{\oplus}  \mathcal{C}^{\omega}_{\geq t,\lambda},
\end{equation}
with 
\begin{equation}
\label{5.18}
 \mathcal{C}^{\omega}_{\geq t,\lambda} \simeq  \mathbb{M}_{n_{\lambda}}(\mathbb{C}), \qquad n_{\lambda}< \infty,
\end{equation}
for all $\lambda \in \Lambda$. Then $\mathcal{E}_{\geq t}^{\omega}$ is a von Neumann algebra of type I and 
\begin{equation}
\label{5.19}
\omega_{\mid \mathcal{E}^{\omega}_{\geq t}} =: \rho^{\omega}_{\geq t},
\end{equation}
where  $\rho^{\omega}_{\geq t}$ is a \textit{density matrix}, so that
\begin{equation}
\label{5.20}
 \rho^{\omega}_{\geq t}=\sum_{\lambda \in \Lambda} p_{\lambda}(t) \Pi_{\lambda}(t),
\end{equation}
and the operators $\Pi_{\lambda}(t)\equiv \Pi_{\lambda}^{\omega}(t)$ are the eigenprojections of $ \rho^{\omega}_{\geq t}$, with $\text{dim}(\Pi_{\lambda}(t))=n_{\lambda} < \infty$, the weights $p_{\lambda}(t) \equiv p_{\lambda}^{\omega}(t) \geq 0$ are the eigenvalues of $ \rho^{\omega}_{\geq t}$, arranged in decreasing order, and 
\begin{equation*}
\text{tr}(\rho^{\omega}_{t})= \sum_{\lambda \in \Lambda} p_{\lambda}(t) \text{dim}(\Pi_{\lambda}(t))=1.
\end{equation*}
Then $\mathcal{C}^{\omega}_{\geq t,\lambda} \simeq  \mathbb{M}_{n_{\lambda}}(\mathbb{C})$ is the algebra of all bounded operators from the eigenspace $\text{Ran } \Pi_{\lambda}(t)$ to itself, and
\begin{equation*}
\omega_{\mid  \mathcal{C}^{\omega}_{\geq t,\lambda}} = p_{\lambda}(t) \text{tr}(\Pi_{\lambda}(t)(\cdot)).
\end{equation*}
Any operator $a \in  \mathcal{E}^{\omega}_{\geq t}$ commuting with all the projections $\Pi_{\lambda}(t)$, $\lambda \in \Lambda$, belongs to $ \mathcal{C}^{\omega}_{\geq t}$, and any operator in the center 
$\mathcal{Z}^{\omega}_{\geq t}$ of $ \mathcal{C}^{\omega}_{\geq t}$ is a function of the projections $\Pi_{\lambda}(t)$, $\lambda \in \Lambda_{\omega}$. In particular $\Pi_{\lambda}(t) \in \mathcal{Z}^{\omega}_{\geq t}\subset \mathcal{C}^{\omega}_{\geq t}$, for all $\lambda$, (and hence the eigenprojections of $\rho^{\omega}_{\geq t}$ might qualify as empirical properties of $S$).

 Henceforth, we consider the special case specified in Eqs. (\ref{5.17}) -- (\ref{5.20}); (but see Remark (1) of Subsection 5.5, and Appendix \ref{AppB}).

\begin{definition}
\label{def5.3}
Let $a=a^*$ be an operator in $\mathcal{E}^{\omega}_{\geq t}$. We define 
\begin{equation}
\label{5.21}
\overline{a}^{\lambda}:= \frac{1}{n_{\lambda}} \text{tr}(\Pi_{\lambda}(t)a).
\end{equation}
If $\lambda$ is such that $p_{\lambda}(t)>0$ then
\begin{equation*}
\overline{a}^{\lambda}:= \frac{1}{ p_{\lambda} (t) n_{\lambda}}  \omega(\Pi_{\lambda}(t)a).
\end{equation*}
Note that $\overline{\mathds{1}}^{\lambda} =1$. We set
\begin{equation}
\label{5.22}
a^{\omega}:=\sum_{\lambda \in \Lambda} \overline{a}^{\lambda} \Pi_{\lambda}(t) \in  
\mathcal{Z}^{\omega}_{\geq t} \subset \mathcal{C}^{\omega}_{\geq t}
\end{equation}
and define the "variance of $a$ in $\omega$" by 
\begin{equation}
\label{5.23}
\Delta^{\omega}_{t} a :=  \sqrt{\sum_{\lambda \in \Lambda} p_{\lambda}(t) \text{Tr}(\Pi_{\lambda}(t) (a-\overline{a}^{\lambda} \cdot \mathds{1})^2)}=\sqrt{\omega((a-a^{\omega})^2)}.
\end{equation}
\end{definition}

We observe that if $\Delta^{\omega}_{t} a=0$ then $a \in  \mathcal{C}^{\omega}_{\geq t}$, and, on the range of $\rho^{\omega}_{t}$, $a\vert_{\text{Ran} \rho_{t}^{\omega}} =a^{\omega}\vert_{\text{Ran} \rho_{t}^{\omega}}$ is a function of $\rho^{\omega}_{t}$, i.e., $a\vert_{\text{Ran} \rho_{t}^{\omega}}\in \mathcal{Z}^{\omega}_{\geq t}$. For a general element, $a$, of $ \mathcal{E}^{\omega}_{\geq t}$, 
\begin{equation}
\begin{split}
\label{5.24}
\vert \{ a, \omega]_{\mathcal{E}_{t}^{\omega}} (b) \vert = \vert \omega([a,b]) \vert& =   \vert \omega([a-a^{\omega},b]) \vert\\
&\leq 2 \sqrt{\omega((a-a^{\omega})^2) \omega(b^*b)} \leq 2 \Delta^{\omega}_{t} a  \text{ }\|b\|,
\end{split}
\end{equation}
for arbitrary $b \in \mathcal{E}^{\omega}_{\geq t}$. Thus, if $\Delta^{\omega}_{t} a$ is small then $\|\{ a, \omega]_{\mathcal{E}^{\omega}_{\geq t}} \|$ is small, too, and Lemma \ref{lem5.2} then tells us that 
$\omega\vert_{\mathcal{E}^{\omega}_{\geq t}}$ is close to an incoherent superposition of eigenstates of $a$.\\

Let $d\mu_{\lambda}(\alpha)$ denote the spectral measure of the operator $a=a^* \in \mathcal{E}^{\omega}_{\geq t}$ in the state 
\begin{equation*}
n_{\lambda}^{-1} \text{tr}(\Pi_{\lambda}(t) (\cdot)). 
\end{equation*}
Then 
\begin{align*}
0& \leq \omega((a-a^{\omega})^2) = \sum_{\lambda \in \Lambda} p_{\lambda}(t) n_{\lambda}
\frac{1}{n_{\lambda}} \text{tr}(\Pi_{\lambda}(a-\overline{a}^{\lambda})^2) =\sum_{\lambda \in \Lambda} p_{\lambda}(t) n_{\lambda} \int d\mu_{\lambda}(\alpha) (\alpha-\overline{a}^{\lambda})^2.
\end{align*}
Thus, 
\begin{equation*}
p_{\lambda}(t) n_{\lambda} \int d\mu_{\lambda}(\alpha) (\alpha-\overline{a}^{\lambda})^2 \leq (\Delta^{\omega}_{t}a)^2,
\end{equation*}
for every $\lambda \in \Lambda$. We conclude  that if, for some $\lambda \in \Lambda$,
\begin{equation*}
\frac{1}{ p_{\lambda}(t) n_{\lambda}} (\Delta_{t}^{\omega} a)^2 <\epsilon^2,
\end{equation*}
for some $\epsilon>0$, then $a$ has spectrum at a distance less than $\epsilon$ from $\overline{a}^{\lambda}$. In particular, if $a$ has discrete spectrum then $a$ has at least one eigenvalue $\alpha_{\lambda}$, with
\begin{equation}
\label{5.25}
\vert \alpha_{\lambda} - \overline{a}^{\lambda} \vert < \epsilon.
\end{equation}
Next, let $a \in \mathcal{P}_S$ be the operator representing some \textit{potential property} of $S$. Then $a(t):=\tau_{t,t_0}(a) \in \mathcal{E}^{\omega}_{\geq t}$.
\begin{definition}
\label{def5.4}
We say that a potential property of $S$ represented by an operator $a \in \mathcal{P}_S$ is an \textit{empirical property of  S at time t within an uncertainty (of size)} $\delta \geq 0$ iff
\begin{equation}
\label{5.26}
\Delta^{\omega}_{t} a(t) \leq \delta.
\end{equation}
\end{definition}
\begin{remark}
If $\delta$ is below the resolution threshold of the equipment used to monitor $S$ then, FAPP, $a(t)$ indeed represents an empirical property of $S$ at time $t$, in the following sense:
\begin{enumerate}
\item $\| \{a(t), \omega]_{\mathcal{E}^{\omega}_{\geq t}} \|$ is so small that it cannot be distinguished from $0$;
\item $\omega(b) \approx \sum_i \omega(\Pi_i(t) b \Pi_i(t))$, for all $b \in \mathcal{E}^{\omega}_{\geq t}$,
where $\Pi_1(t),\Pi_2(t)$,... are the spectral projections of $a(t)$, (assuming $a=a^{*}$ has discrete spectrum; see Lemma \ref{lem5.2} for a precise statement); 
\item on the range of the density matrix $\rho^{\omega}_{\geq t}$, $a(t)$ is "close" to the operator $a(t)^{\omega} \in \mathcal{Z}_{\geq t}^{\omega}$;
\item $a$ has eigenvalues near the numbers $\overline{a(t)}^{\lambda}$, for all $\lambda \in \Lambda_{\omega}$ for which $(p_{\lambda}(t) n_{\lambda})^{-1} \delta^2$ is small. 
\end{enumerate}
\end{remark}

One may then argue that if $\Delta^{\omega}_{t} a(t)$ is very small, and if a measurement or observation of $a \in \mathcal{P}_S$ at a time $\approx t$ indicates that it has a value $\alpha \approx \overline{a(t)}^{\lambda}$ then one may use the state 
\begin{equation}
\label{5.27}
\omega_{\lambda}:= \frac{1}{n_{\lambda}} \text{tr}(\Pi_{\lambda}(t) (\cdot))
\end{equation}
to predict the behavior of the system $S$ at times later than $t$. This idea, reminiscent of "state collapse", will be further discussed below. 

Note that the maximal uncertainty $\delta$ admissible in statement (2) above depends on the spectrum of the operator $a$.

\subsection{When does an observation or measurement of a physical quantity take place?}
\label{sec5.4}
Let $a=a^* \in \mathcal{P}_S$ represent a potential property of a quantum-mechanical system $S$, which is assumed to be prepared in a state $\omega$ on the algebra $\mathcal{E}_S$. We propose to analyze \textit{whether} and  \textit{when} $a$ corresponds to an \textit{empirical property} of $S$, in the sense that, given the time evolution $\{\tau_{t,s}\}_{t,s \in \mathbb{R}}$ of $S$ and the state $\omega$, 
$a$ is measurable (i.e., the value of $a$ can be measured or observed) at some finite time.  Definition \ref{def5.4} and the discussion thereafter suggest to consider the variance $\Delta^{\omega}_{t} a(t)$ ($a(t)=\tau_{t,t_0}(a)$), of $a(t)$ as a \textit{function of time }$t$. This function is non-negative and bounded. Let $\delta$ be some non-negative number below the resolution threshold of the equipment used to monitor $S$. Let $t_{*}$ be defined as the smallest time such that 
\begin{equation}
\label{5.28}
\Delta^{\omega}_{t_*}a(t_{*}) \leq \delta.
\end{equation}
Then it is reasonable to say that $a$ is observed/measured -- put differently, $a$ becomes an empirical property of $S$ within an uncertainty of size $\delta$ -- at a time $\apprge t_*$. If the equipment $E$ used to monitor $S$ is only sensitive to observing the eigenvalue $\alpha_i$ of $a$, i.e., to the possible event $\Pi_i$ (spectral projection of $a$ corresponding to the eigenvalue $\alpha_i$) then one may plausibly say that the possible event $\Pi_i$ is observed at a time $\apprge t_*$ iff
\begin{equation*}
\Delta^{\omega}_{t_*} a(t_*) +1-\omega(\Pi_i(t_*))
\end{equation*}
is \textit{very small}. In this case, we say that the equipment $E$ prepares the state of $S$ to lie in the range of the projection $\Pi_i(t) \approx \sum_{\lambda\in\Lambda_{\omega}^{(i)}} \Pi_{\lambda}(t)$, with $t \apprge t_*$, where $\Lambda_{\omega}^{(i)}$ is defined by the property that $\vert \alpha_{i} - \overline{a(t)}^{\lambda}\vert < \delta$, for all $\lambda \in \Lambda_{\omega}^{(i)}$. Thus, the function
\begin{equation}
\label{5.29}
\mathcal{T}_{\omega,a}(t):=\Delta^{\omega}_t a(t)
\end{equation}
contains all important information on the time around which the potential property $a$ of $S$ becomes an empirical property; and the function
\begin{equation}
\label{5.30}
\mathcal{T}_{\omega,a}^{i}(t):= \Delta^{\omega}_t a(t) +1-\omega(\Pi_i(t))
\end{equation}
tells us when (around which time) a detector sensitive to the possible event $\Pi_i$ "clicks"; (see also \cite{Bru-Fred, RWe} for some ideas on this matter that will not be pursued here). 

Next, we analyze \textit{repeated observations/measurements}, as in Sect.\ref{sec4.1}. It suffices to consider only \textit{two} subsequent measurements. Let $a=a^* \in \mathcal{P}_S$ represent a potential property of $S$, and let $\delta \geq 0$ be a measure for the resolution of the equipment $E$ used to monitor $S$ in a measurement of $a$.

\begin{definition}
\label{def5.5}
For $a=a^* \in \mathcal{P}_S$, $\delta \geq 0$, and a time $t_{*}>-\infty$, we define a subset of states on $\mathcal{A}_{\overline{S}}$ (or on $\mathcal{E}_{S} \subset \mathcal{A}_{\overline{S}}$) by 
\begin{equation}
\label{5.31}
\mathcal{S}(a,\delta,t_{*}):=\{ \omega \in \mathcal{S}_{\overline{S}} \mid \underset{t \geq t_*}{\text{inf}} \Delta^{\omega}_{t} a(t) < \delta\},
\end{equation}
where $\delta$ is so small that properties (1) through (4) in  Remark 5.6, above, are valid.
\end{definition}

Apparently, $\mathcal{S}(a,\delta,t_*)$ is the set of states of $S$ with the property that, given the time evolution $\{\tau_{t,s}\}_{t,s \in \mathbb{R}}$, the operator $a$ corresponds to an $empirical$ $property$ of $S$, within an uncertainty of size $\delta$,  that is measurable at some time after $t_*$.

Next, we consider two potential properties of $S$ represented by two self-adjoint operators, $a_1$ and $a_2$, and we suppose that, first, $a_1$ and, afterwards, $a_2$, are measured. For simplicity we suppose that the spectra of $a_1$ and $a_2$ consist of finitely many eigenvalues $\alpha_j^{(i)}$, $j=1,...,k_i <\infty$, $i=1,2$. We  assume that the state, $\omega$, of $S$ before the measurement of $a_1$, belongs to $\mathcal{S}(a_1,\delta_1,t_{1*})$, for a sufficiently small number $\delta_1$ (below a threshold of resolution). Then $\Delta^{\omega}_{t_1} a_1(t_1) \leq \delta_1$, at some time t$_1 \geq t_{1*}$. A successful measurement of $a_1$ around some time $t_1 \geq t_{1*}$ results in the assignment of a value $\alpha_j^{(1)} \approx \overline{a_1(t_1)}^{\lambda}$, $\lambda \in \Lambda_{\omega}^{(j)}$, to the physical quantity represented by $a_1$, where
\begin{equation}
\label{5.32}
\Lambda_{\omega}^{(j)}:=\{ \lambda \in \Lambda_{\omega} \mid \vert \overline{a_1(t_1)}^{\lambda}- \alpha_j^{(1)} \vert < \delta_1\}.
\end{equation}
 (For consistency, we assume that $\underset{j \neq l}{\text{min}} \vert \alpha_{j}^{(1)} - \alpha_{l}^{(1)} \vert > 2 \delta_1.$)
The \textit{probability} of this measurement outcome is given by 
\begin{equation}
\begin{split}
P_{j}^{(1)} (t_1) &= \sum_{\lambda \in \Lambda_{\omega}^{(j)}} \omega(\Pi^{\omega}_{\lambda}(t_1)) =  \sum_{\lambda \in \Lambda_{\omega}^{(j)}}  p^{\omega}_{\lambda}(t_1)n^{\omega}_{\lambda} =\omega(\Pi^{(1)}_{j}(t_1)) + \mathcal{O}(\delta_1), 
\end{split}
\end{equation}
where $p_{\lambda}(t_1) \equiv p_{\lambda}^{\omega}(t_1)$, $n_{\lambda} \equiv n_{\lambda}^{\omega} =\text{dim} \Pi_{\lambda}^{\omega}(t_1)$, and $\Pi_{\lambda}(t_1)\equiv \Pi_{\lambda}^{\omega}(t_1)$ are as defined in Eqs. (\ref{5.19}) and (\ref{5.20}), (the superscript "$\omega$" is supposed to highlight the dependence on the state $\omega$), and $\Pi_{j}^{(1)}(t_1)$ is the eigenprojection of the operator $a_1(t_1)$ corresponding to the eigenvalue $\alpha_{j}^{(1)}$. If $P_{j}^{(1)}(t_1)$ is very small one can ignore the possibility that, for a system $S$ prepared in the state $\omega$, an observation/measurement of $a_1$ will yield a value $\approx \alpha_{j}^{(1)}$.

Let $\omega_j$ denote the state
\begin{equation}
\label{5.34}
\omega_j (b)= \frac{\sum_{\lambda \in \Lambda_{\omega}^{(j)}} \omega(\Pi_{\lambda}^{\omega}(t_1) b \Pi_{\lambda}^{\omega}(t_1))}{P_{j}^{(1)}(t_1)} =\frac{ \omega(\Pi_{j}^{(1)}(t_1) b \Pi_{j}^{(1)}(t_1) )}{\omega(\Pi_{j}^{(1)}(t_1))} + \mathcal{O}(\delta_1),
\end{equation}
for an arbitrary operator $b \in \mathcal{E}^{\omega}_{\geq t}$, with $t \geq t_1$; (recall that $ \mathcal{E}^{\omega}_{\geq t} \subseteq  \mathcal{E}^{\omega}_{\geq t_1}$, for $t \geq t_1$.)

Let us suppose that, for all $j \in \{ 1,...,k_1\}$ for which $P_{j}^{(1)}(t_1) >\delta_2 >0$,
\begin{equation}
\label{eq5.35}
\omega_j \in \mathcal{S}(a_2,\delta_2,t_{2*}^{(j)}),
\end{equation}
for some time $t_{2*}^{(j)} >t_1$. If $\delta_2$ is chosen small enough one may expect to be able to successfully measure the quantity represented by $a_2$ at a time $t_2 \geq t_{2*}^{(j)}$, assuming that, at a time $t_1<t_{2*}^{(j)}$, $a_1$ was found to have a value $\approx \alpha_{j}^{(1)}$.

The \textit{joint probability} to find a value $\approx \alpha_{j}^{(1)}$ in a measurement of $a_1$ around some time $t_1$ and, in a subsequent measurement around a time $t_2>t_1$, a value $\approx \alpha^{(2)}_{l}$ of the quantity represented by $a_2$, (with $l \in \{ 1,...,k_2\}$), is given by
\begin{equation}
\label{5.36}
\begin{split}
\text{Prob}_{\omega} \{ \Pi_{j}^{(1)} (t_1),   \Pi_{l}^{(2)} (t_2)\} &= P_{j}^{(1)} (t_1) \sum_{\lambda \in \Lambda_{\omega_j}^{(l)}} \omega_j(\Pi_{\lambda}^{\omega_j}(t_2))\\
&= \omega(\Pi_{j}^{(1)} (t_1) \Pi_{l}^{(2)}(t_2) \Pi_{j}^{(1)}(t_1)) + \mathcal{O}(\delta_1 \vee \delta_2),
\end{split}
\end{equation}
where $\Lambda_{\omega_j}^{(l)}=\{ \lambda \in \Lambda_{\omega_j} \mid \vert \overline{a_2(t_2)}^{\lambda} - \alpha_l^{(2)} \vert < \delta_2\}$, and $\delta_1 \vee \delta_2 = \max\{ \delta_1,\delta_2\}$. 

The definitions of centralizers, $\mathcal{C}^{\omega}_{\geq t_1}$, etc., and of the variance $\Delta_{t}^{\omega}a(t)$ readily imply that 
\begin{equation}
\label{5.37}
\sum_{j=1}^{k_1} \omega(\Pi_{j}^{(1)} (t_1) \Pi_{l}^{(2)}(t_2) b \Pi_{l}^{(2)}(t_2)  \Pi_{j}^{(1)}(t_1)) = \omega( \Pi_{l}^{(2)}(t_2) b \Pi_{l}^{(2)}(t_2)  ) + \mathcal{O}(\delta_1), 
\end{equation}
and if $\omega_j \in \mathcal{S}(a_2,\delta_2,t_{2*}^{(j)})$ then
\begin{equation}
\label{5.38}
\sum_{j=1}^{k_2} \omega(\Pi_{j}^{(1)} (t_1) \Pi_{l}^{(2)}(t_2) b \Pi_{l}^{(2)}(t_2)  \Pi_{j}^{(1)}(t_1)) = \omega( \Pi_{j}^{(1)}(t_1) b \Pi_{j}^{(1)}(t_1)  ) + \mathcal{O}(\delta_1 \vee \delta_2),
\end{equation}
for an arbitrary operator $b \in \mathcal{E}^{\omega}_{\geq t}$, with $t> \max_{j} t_{2*}^{(j)}$. It is clear how to extend our discussion to an arbitrary chronological (time-ordered) sequence of measurements of quantities $a_1,...,a_n$, ($a_i \in \mathcal{P}_S, \forall i$). Moreover, the mathematical relationship between Eqs. (\ref{5.37}) and (\ref{5.38}), on one side, and $\delta-$consistent families of histories  -- see (\ref{4.16}) and (\ref{4.17}), Sect.\ref{sec4.2} -- on the other side, is easy to unravel. We do not wish to discuss further details.

\begin{remark}[Remark on the role of POVM's]
It may and \textit{will} occasionally happen that, given that a quantity represented by an operator $a_1$ has been observed/measured, the quantity represented by the operator $a_2$ can be measured, subsequently,  \textit{only for certain}, but  \textit{not all}, outcomes of the measurement of $a_1$. More precisely, it may happen that, for some eigenvalues $\alpha_{j}^{1}$, $j \in G$, of $a_1$, $\omega_j \in \mathcal{S}(a_2,\delta_2,t^{(j)}_{2*})$, while, for $i\in B:=\{1,...,k_1\} \setminus G$,
\begin{equation}
\label{5.39}
\omega_i \notin \mathcal{S}(a_2,\delta_2,t_{2*}),
\end{equation}
for any $t_{2*} <\infty$; ($\delta_1$ and $\delta_2$ being chosen appropriately, depending on the resolution of the corresponding measurements, as discussed above). 

If $B \neq \emptyset$ then one must take the position that the observations of $a_1$ and $a_2$ represent \textit{one single} measurement, which must be described using "\textit{positive operator-valued measures}" (POVM's) -- see Appendix \ref{appA}, \eqref{A.6} and \eqref{A.7}:
\begin{equation}
X=\{ X_{jl}, X_i \mid j \in G, l=1,...,k_2, i \in B\}
\end{equation}
where, for $j \in G$,
\begin{equation}
\label{5.41}
X_{jl}= \sum_{\lambda_1 \in \Lambda^{(j)}_{\omega}} \sum_{\lambda_2 \in \Lambda^{(l)}_{\omega_j}}  \Pi_{\lambda_2}^{\omega_j}(t_{2}^{(j)}) \Pi^{\omega}_{\lambda_1}(t_1) \approx   \Pi_{l}^{(2)}(t_{2}^{(j)}) \Pi^{(1)}_{j}(t_1),  
\end{equation}
(up to a small perturbation of $\mathcal{O}(\delta_1 \vee \delta_2)$), while, for $i \in B$,
\begin{equation}
\label{5.42}
X_{i}=\sum_{\lambda_1 \in \Lambda_{\omega}^{(i)}} \Pi^{\omega}_{\lambda_1}(t_1) \approx \Pi_{i}^{(1)} (t_1),
\end{equation}
where $t_1$ and $t_{2}^{(j)}$ are the times of measurement of $a_1$ and $a_2$, respectively. Then
\begin{equation}
\label{5.43}
\begin{split}
\sum_{j \in G} \sum_{l=1}^{k_2} X^{*}_{jl} X_{jl} + \sum_{i \in B} X^{*}_i X_i = \mathds{1}.
\end{split}
\end{equation}
\end{remark}

The use of POVM's will be discussed in more detail and in connection with concrete examples elsewhere. Here we just remark that simple examples showing why one needs to introduce POVM's are encountered in the analysis of repeated Stern-Gerlach measurements of atomic spins (followed by detectors sensitive to the arrival of the atoms).

\subsection{Generalizations and summary}
\begin{enumerate}[(1)]
\item In order to keep our exposition reasonably simple, we have made the simplifying assumptions (\ref{5.17}) and (\ref{5.18}). It is, however, not very hard to develop our ideas in full generality. For this purpose, we must return ro formula (\ref{5.15}): The space $\Lambda=\Lambda_{\omega}$ appearing in (\ref{5.15}) is the \textit{spectrum} of the \textit{center}, $\mathcal{Z}^{\omega}_{\geq t}$, of the centralizer, $\mathcal{C}^{\omega}_{\geq t}$, of the state $\omega$, viewed as a state on the algebra $\mathcal{E}^{\omega}_{\geq t}$. The theory of "conditional expectations" \cite{MTak2} enables us (under fairly general hypotheses) to construct a conditional expectation $\dot{\epsilon}_{\geq t} : \mathcal{E}^{\omega}_{\geq t} \rightarrow \mathcal{Z}^{\omega}_{\geq t}$, which permits us to associate with every operator $a \in \mathcal{E}^{\omega}_{\geq t}$ an operator $a^{\omega} \in \mathcal{Z}^{\omega}_{\geq t}$. The map $a \mapsto a^{\omega}$ is linear, and $(a^{\omega})^{\omega}=a^{\omega}$. (In the special case where Eqs. (\ref{5.17}) and (\ref{5.18}) hold it is given by formula (\ref{5.22}).) Having constructed $a^{\omega}$, we set 
\begin{equation*}
\Delta^{\omega}_{t}a:= \sqrt{\omega((a-a^{\omega})^2)}.
\end{equation*}
From this point on, we may follow the arguments from (\ref{5.24}) onwards, and in Subsection \ref{sec5.4}. 

\item In our approach to the "quantum theory of experiments/quantum measurement theory", the \textit{"ontology"} underlying a quantum-mechanical model of a physical system $S$ is represented by 
\begin{enumerate}[(a)]
\item a set, $\mathcal{P}_S$, of physical quantities characterizing $S$;
\item a filtration of $C^*-$algebras
\begin{equation*}
\mathcal{E}_S \supseteq \mathcal{E}_{\geq t} \supseteq \mathcal{E}_{\geq t'} \supset \{ \mathbb{C} \cdot \mathds{1}\}, \qquad t \leq t',
\end{equation*}
and $*$morphisms
\begin{equation*}
\overline{\tau}_t: \mathcal{E}_{\geq t'} \longrightarrow \mathcal{E}_{\geq t'+t} \subseteq \mathcal{E}_{\geq t'},
\end{equation*}
for $t \geq 0$;

\item a state $\omega$ on $\mathcal{E}_S$;
\item the centralizers $\mathcal{C}^{\omega}_{\geq t}$ of $\omega_{\vert \mathcal{E}_{\geq t}}$ and their centers $\mathcal{Z}^{\omega}_{\geq t}$.\\
\end{enumerate}
\begin{noindent}
If $S$ is prepared in a state $\omega$ before one attempts to measure a physical quantity represented 
by an operator $a \in \mathcal{P}_S$ then the measurement is successful around some time $t$ if $a(t)=\tau_{t,t_0}(a)$ is "close" to an operator in $\mathcal{Z}_{\geq t }^{\omega}$, in the sense that the variance, $\Delta_{t}^{\omega} a (t)$, of $a(t)$ in $\omega$ is small.
\end{noindent}

\item Let us return to the special situation described in \eqref{5.17} through \eqref{5.20}. Suppose that all the algebras  $\mathcal{E}_{\geq t}^{\omega}$, $t \in \mathbb{R}$, are isomorphic to a fixed factor
$\mathcal{E} \simeq B(\mathcal{H})$ of type $I_{\infty}$. Then our approach is "dual" to one where the density matrices $\{ \rho^{\omega} _{\geq t}\}_{t \in \mathbb{R}}$ are interpreted as states on $\mathcal{E}$ and are considered to be the fundamental objects, and time evolution is described in terms of completely positive maps on the space of density matrices. With the idealization/approximation that time evolution is given by a groupoid of completely  positive maps, this is the point of view popular among quantum information scientists; (see, e.g., \cite{Lindblad}).\\
The trajectories of density matrices $\{\rho^{\omega}_{\geq t}\}_{t\in\mathbb{R}}$ are then what replaces the trajectories $\{\xi_{t} = \phi_{t,t_0}(\xi_0)\}_{t\in\mathbb{R}}$ of a classical system (as discussed in Sect. 3). However, because of the phenomena of information loss and entanglement, the density matrices 
$\rho^{\omega}_{\geq t}$ tend to describe \textit{mixed} states, $even$ if the state $\omega$ is a pure state of the algebra $\mathcal{E}_S$, and hence only yield probabilistic predictions, while the states $\xi_{t}$ of a classical system are $pure$, for all $t$, provided the initial state is pure, and hence yield deterministic predictions.
\item It is clearly important to extend our theory to local relativistic quantum theory (LRQT). In LRQT, the algebras $\mathcal{E}_{\geq t}$, $t \in \mathbb{R}$, are replaced by algebras, $\mathcal{E}_{P}$, of "observables" localized inside the  \textit{forward light cone} of a point $P$ (the momentary position of an observer) on a time-like curve in space-time, (the obsever's world line). If the theory describes a massless photon and if $\omega$ is a state normal to the vacuum then the von Neumann algebras $\mathcal{E}^{\omega}_{P}$ are all isomorphic to the hyperfinite factor of type $\text{III}_1$, as discussed
 in \cite{Buchh-Roberts}. Hence the algebras $\mathcal{E}^{\omega}_{P}$ do \textit{not} have \textit{any pure states}, and the principle of Loss of Information ($\mathcal{L}o\mathcal{I}$) is a fundamental feature of the theory. We will return to this topic elsewhere.

\item It is clearly important to understand how quantum-mechanical systems can be prepared in \textit{specific states} ("preparation of states"). This topic will be discussed in \cite{Fr-Schub2}; but see also (\ref{5.27}) and the remark right above (\ref{5.29}). Moreover, it is quite crucial to back up the general analysis presented in this essay with simple models of "information loss" and "decoherence/dephasing". This will be done in a forthcoming publication.
\end{enumerate}

The last topic we briefly address in this essay is a theory of weak (non-demolition) experiments, following \cite{Ba-Be}. This theory explains why in many experiments, the system ends up being in an eigenstate of the operator representing the quantity that is measured, i.e., why \textit{"facts"} emerge in non-demolition measurements.

\subsection{Non-demolition measurements}
\label{sec5.5}
After having presented a long and rather abstract discussion of "direct (or von Neumann) measurements", in Subsections \ref{sec5.3}-5.5, we wish to sketch the theory of "indirect (non-demolition) measurements". The main results described here have recently appeared in \cite{Ba-Be};  see also \cite{Adler}. The practical importance of these results comes from recent experiments; see, e.g.,  \cite{Guerlin}. 

We consider a physical system $S$ (e.g., the quantized electromagnetic field in a cavity). We wish to measure a physical quantity represented by an operator $a=a^* \in \mathcal{A}_S$ (e.g., the photon number inside the cavity) with the help of "non-demolition measurements". For this purpose, we bring $S$ into contact with a sequence, $E_1$, $E_2$, $E_3$,..., of identical "probes" (e.g., excited atoms passing through the cavity); the interaction of $E_k$ with $S$ is supposed to take place in the time interval $[k-1,k]$ and is supposed to be turned off during all other times. Actually, after some direct measurement of a property $b_k=b_{k}^{*} \in \mathcal{A}_{E_k}$ at a time later than $k$ -- as described abstractly in Subsection \ref{sec5.4} -- probe $E_k$ "gets lost for ever", in the sense that no further information about $E_k$ can be retrieved, anymore.

Let $\rho$ denote the initial state of $S$ and $\psi^{(k)}:= \psi$ the initial state of probe $E_k$, (the same for all $k$). For simplicity, we assume that the spectrum of the operator $a$ representing the physical property of $S$ to be measured is finite pure-point spectrum. We denote the spectral projection corresponding to an eigenvalue $\alpha$ of $a$ by $\Pi_{\alpha}=\Pi_{\alpha}^{*}$. Then
$$\Pi_{\alpha} \Pi_{\beta}= \delta_{\alpha \beta } \Pi_{\alpha}, \qquad \sum_{\alpha} \Pi_{\alpha}=\mathds{1}.$$
Next, we specify the time-evolution of the composed system $S \vee E_1 \vee E_2 \vee ... :$ Up to time $k=1,2,3,...,$, the time evolution of $E_j$ is assumed to be \textit{trivial}, for all $j>k$. For the subsystem $S \vee E_1 \vee .... \vee E_k$ it is specified as follows : Let $A_{\alpha,\alpha'}$ be an arbitrary operator in $\mathcal{A}_S$ mapping $\text{Ran } \Pi_{\alpha'}$ to  $\text{Ran } \Pi_{\alpha}$, with $\Pi_{\beta} A_{\alpha,\alpha'} \Pi_{\beta'}= \delta_{\alpha \beta }  \delta_{\alpha' \beta' } A_{\alpha,\alpha'}$. Let $B_j$ be an operator in $\mathcal{A}_{E_j}$, $j \leq k$. Then the time-evolution of $A_{\alpha,\alpha'} \otimes B_1 \otimes ... \otimes B_k$ from time $0$ to time $k$ in the Heisenberg picture is given by 
\begin{equation*}
\tau_{k,0} (A_{\alpha,\alpha'} \otimes B_1 \otimes ... \otimes B_k):= A_{\alpha,\alpha'} \otimes U_{\alpha} B_1  U_{\alpha'}^{*} \otimes ... \otimes U_{\alpha} B_{k} U^{*}_{\alpha'},
\end{equation*}
where $U_{\alpha}$ is a unitary operator in $\mathcal{A}_{E_k} \simeq \mathcal{A}_E$, for all $\alpha \in \text{spec }a$. Defining 
$$U(i,i-1):=\sum_{\alpha} \Pi_{\alpha} \otimes \mathds{1} \otimes ... \otimes U_{\alpha} \otimes \mathds{1} \otimes ...,$$
with $U_{\alpha}$ inserted in the $(i+1)^{st}$ factor of the tensor product, we have that 
\begin{equation}
\label{5.44}
\begin{split}
\tau_{k,0} ( A_{\alpha,\alpha'}  \otimes B_1 \otimes ... \otimes B_{j}) &= \prod_{i=k}^{1} U(i,i-1) (A_{\alpha,\alpha'} \otimes B_1 \otimes ... \otimes B_j) \prod_{i=1}^{k} U(i,i-1)^{*} = \\
&\tau_{k,0} ( A_{\alpha,\alpha'}  \otimes B_1 \otimes ... \otimes B_{k})  \otimes B_{k+1} \otimes ... \otimes B_j,
\end{split}
\end{equation}
for arbitrary $j \geq k$. This is a typical (albeit highly idealized) example of time-evolution in a non-demolition measurement. Let $\Psi:= \rho \otimes \psi \otimes \psi \otimes ...$ denote the initial state of the composed system, $S \vee E_1 \vee E_2 \vee ...$. If we set 
$$B_1=B_2=...=B_{k_0}=\mathds{1},$$
for some $k_0 < \infty$ then
\begin{equation}
\label{5.45}
\begin{split}
\Psi &\left(\tau_{k,0}(A_{\alpha,\alpha'} \otimes \mathds{1} \otimes ... \otimes \mathds{1} \otimes B_{k_0+1} \otimes ... \otimes B_{k_0+l}) \right) =  \\
&\qquad \rho(A_{\alpha,\alpha'}) \psi(U_{\alpha} U^{*}_{\alpha'})^{k_0} \prod_{i=k_0+1}^{k} 
\psi(U_{\alpha} B_i U^{*}_{\alpha'}) \prod_{i=k+1}^{k_0+l} \psi(B_i),
\end{split}
\end{equation}
for $k_0 \leq k \leq k_0+l$. Because $U_{\alpha}$ is unitary, for all $\alpha \in \text{spec }a$,
$$\vert \Psi(U_{\alpha} U^{*}_{\alpha'}) \vert \leq 1, \qquad \text{ for all } \alpha,\alpha',$$
by the Cauchy-Schwarz inequality. We assume that 
\begin{equation}
\label{5.46}
\vert \Psi(U_{\alpha} U^{*}_{\alpha'}) \vert \leq \mu <1, \qquad \text{ for } \alpha \neq \alpha'.
\end{equation}
Then
\begin{equation}
\label{5.47}
\vert \Psi \left( \tau_{k,0}(A_{\alpha,\alpha'} \otimes \mathds{1} \otimes ... \otimes \mathds{1} \otimes B_{k_0+1} \otimes ... \otimes B_{k_0+l}) \right) \vert \leq \mu^{k_0},
\end{equation}
which, by \eqref{5.46}, tends to $0$ exponentially fast, as $k_0 \rightarrow \infty$, for arbitrary $A_{\alpha,\alpha'}$, $B_{k_0+1}$,...,$B_{k_0+l}$, with $\|A_{\alpha,\alpha'}\|$,  $\|B_{k_0+1}\|$,..., $\| B_{k_0+l}\|$ bounded by $1$. This is \textit{"decoherence"} over the spectrum of the operator $a$ representing the quantity to be measured: 
\begin{equation}
\label{5.47'}
\Psi_{\mid \mathcal{E}_{\geq k_0}} \longrightarrow \sum_{\alpha} \Psi(\Pi_{\alpha} ( \cdot) \Pi_{\alpha})_{\mid \mathcal{E}_{\geq k_0}},
\end{equation}
as $k_0 \rightarrow \infty$, where $ \mathcal{E}_{\geq k_0}$ is the algebra introduced in Definition \ref{def2.2}.  Henceforth, we choose an initial state, $\rho$, for $S$ satisfying
\begin{equation*}
\rho=\sum_{\alpha} \rho(\Pi_{\alpha} (\cdot) \Pi_{\alpha}) = \sum_{\alpha} p_{\alpha} \rho_{\alpha},
\end{equation*} 
where 
\begin{equation}
\label{5.48}
p_{\alpha}=\rho(\Pi_{\alpha}), \text{ }\rho_{\alpha}=p_{\alpha}^{-1} \rho(\Pi_{\alpha} (\cdot) \Pi_{\alpha}).
\end{equation}
We assume that (after many identical probes have interacted with $S$, so that decoherence over the spectrum of the observable $a$ has set in) 
a \textit{direct measurement} of a physical quantity represented by an operator $b=b^* \in \mathcal{A}_E$ is carried out on \textit{every probe} $E_{k} \simeq E$, \textit{after} it has interacted with $S$. We assume that the spectrum of $b$ is pure-point, with eigenvalues denoted by $\xi$ and corresponding spectral projections written as $\pi_{\xi}$. Then $\pi_{\xi}=\pi_{\xi}^{*}$ and 
\begin{equation}
\label{5.49}
\pi_{\xi}\pi_{\xi'}= \delta_{\xi \xi'} \pi_{\xi}, \qquad \sum_{\xi} \pi_{\xi}=\mathds{1}.
\end{equation}

The probability, $\mu( \underline{\xi}_k \vert \alpha)$, of a history 
\begin{equation}
\label{5.50}
\underline{\xi}_{k}:=\{ \pi_{\xi_1},...,\pi_{\xi_k}\}
\end{equation}
of possible outcomes of those direct measurements in the state $\Psi_{\alpha}$ defined by 
\begin{equation*}
\Psi_{\alpha}:=\rho_{\alpha} \otimes \psi \otimes \psi \otimes  ...,
\end{equation*}
with $\rho_{\alpha}$ as in \eqref{5.48}, is given by 
\begin{equation}
\label{5.51}
\mu( \underline{\xi}_k \vert \alpha) =\prod_{i=1}^{k} p( \xi_i \vert \alpha),
\end{equation}
where 
\begin{equation}
\label{5.52}
p(\xi \vert \alpha):= \psi(U_{\alpha} \pi_{\xi} U^{*}_{\alpha}).
\end{equation}
Note that $\sum_{\xi} p(\xi \vert \alpha)=1$, by \eqref{5.49} and the unitarity of $U_{\alpha}$. In the following, we identify $\pi_{\xi} $ with $\xi$ and use the notation $\underline{\xi}_k=(\underline{\xi}_{k-1},\xi_k)$. In the initial state $\Psi$, the probability of the history $\underline{\xi}_k$ is then given by 
\begin{equation}
\label{5.53}
\mu(\underline{\xi}_k)=\sum_{\alpha} p_{\alpha} \mu(\underline{\xi}_k \vert \alpha ).
\end{equation}
Next, we calculate the probability, $p^{(k)}(\alpha \vert \underline{\xi}_k)$ of the possible event $\Pi_{\alpha}$, given that a history $\underline{\xi}_k$ is observed on the first $k$ probes, and given the initial state $\Psi$. By \eqref{5.48} and \eqref{5.51}-(\ref{5.53}),
\begin{equation}
\label{5.54}
p^{(k)}(\alpha \vert \underline{\xi}_k)=p_{\alpha} \frac{\mu(\underline{\xi}_k \vert \alpha )}{\mu(  \underline{\xi}_k)},
\end{equation}
(with $p_{\alpha}=\rho(\Pi_{\alpha})$; see \eqref{5.48}). These probabilities have the following properties:
\begin{enumerate}[(i)]
\item 
\begin{equation*}
0 \leq p^{(k)} (\alpha \vert \underline{\xi}_k) \leq 1 , \qquad \text{and} \text{  }\sum_{\alpha} p^{(k)}(\alpha \vert \underline{\xi}_k)=1.
\end{equation*}
\item  
\begin{equation}
\label{5.55}
\begin{split}
p^{(k)}(\alpha \vert \underline{\xi}_k) &= p_{\alpha} \frac{\mu(\alpha \vert \underline{\xi}_{k-1})}{\mu(\underline{\xi}_k)} p (\xi_k\vert \alpha)\\
&= p^{(k-1)} (\alpha \vert \underline{\xi}_{k-1}) \frac{\mu(\underline{\xi}_{k-1})}{\mu(\underline{\xi}_k)} p( \xi_k \vert \alpha)\\
&=p^{(k-1)} (\alpha \vert \underline{\xi}_{k-1}) \frac{p(\xi_k\vert \alpha)}{\sum_{\beta} p^{(k-1)}(\beta \vert \underline{\xi}_{k-1}) p (\xi_k\vert \beta)},
\end{split}
\end{equation}
because, by \eqref{5.53}, \eqref{5.51} and \eqref{5.54},
\begin{equation}
\label{5.56}
\begin{split}
\frac{\mu(\underline{\xi}_k)}{\mu(\underline{\xi}_{k-1})} &= \sum_{\beta} p_{\beta} \frac{\mu(\underline{\xi}_{k-1} \vert \beta)}{\mu(  \underline{\xi}_{k-1})} p ( \xi_k \vert \beta)\\
&= \sum_{\beta} p^{(k-1)}(\underline{\xi}_{k-1} \vert \beta) p (\xi_{k}\vert \beta),
\end{split}
\end{equation}
\item The expectation, $\mathbb{E}_k$, of $p^{(k)}(\alpha \vert \underline{\xi}_k)$, given $\alpha$ and $\underline{\xi}_{k-1}$, satisfies
\begin{equation}
\label{5.57}
\begin{split}
\mathbb{E}_{k} p^{(k)} (\alpha \vert \underline{\xi}_k)&:= \sum_{\xi_k} p^{(k)} (\alpha \vert \underline{\xi}_{k-1},\xi_k) \frac{\mu(\underline{\xi}_{k-1}, \xi_k)}{\sum_{\xi_k} \mu(\underline{\xi}_{k-1},\xi_k)}\\
&= \sum_{\xi_k} p^{(k)} (\alpha \vert \underline{\xi}_{k-1}, \xi_k) \frac{\mu(\underline{\xi}_{k})}{\mu(\underline{\xi}_{k-1})}\\
&=\sum_{\xi_k} p_{\alpha}  \frac{\mu( \underline{\xi}_{k-1} \vert \alpha) p(\xi_k \vert \alpha)}{\mu(\underline{\xi}_k)} \frac{\mu(\underline{\xi}_{k})}{\mu(\underline{\xi}_{k-1})}\\
&= \sum_{\xi_k} p^{(k-1)} (\alpha \vert \underline{\xi}_{k-1}) p(\xi_k \vert \alpha) = p^{(k-1)}(\alpha \vert \underline{\xi}_{k-1}),
\end{split}
\end{equation}
(see below \eqref{5.52}).
\end{enumerate}
Properties (i) and (iii) identify the random variables $\{ p^{(k)} (\alpha \vert \underline{\xi}_k) \mid \alpha \in \text{spec } a\}$ as \textit{bounded martingales}. The Martingale Convergence Theorem (see e.g., \cite{MCT}) then implies that 
\begin{equation*}
p^{(k)}(\alpha \vert \underline{\xi}) \underset{k \rightarrow \infty}{ \longrightarrow} p^{(\infty)}(\alpha \vert \underline{\xi}),
\end{equation*}
where $\underline{\xi}=\underline{\xi}_{\infty}$, and $p^{(k)}(\alpha \vert \underline{\xi})$ does not depend on $\xi_{k+1},\xi_{k+2},...$. Property (ii) then implies that, for every $\xi_{\infty} \in \text{spec }b$,
\begin{equation}
\label{5.58}
p^{(\infty)}(\alpha \vert \underline{\xi})=p^{(\infty)}(\alpha \vert \underline{\xi}) \frac{p(\xi_{\infty} \vert \alpha)}{\sum_{\beta} p^{(\infty)}(\beta \vert \underline{\xi}) p( \xi_{\infty}\vert \beta)}.
\end{equation} 
If, for every $\alpha \in \text{spec }a$ with $p^{(\infty)}(\alpha \vert \underline{\xi}) \neq 0$ and for every $\xi_{\infty}$, there is some $\beta \in \text{spec }a$ such that $p(\xi_{\infty} \vert \alpha) \neq p(\xi_{\infty} \vert \beta)$ then \eqref{5.58} and (i) obviously imply that 
\begin{equation}
\label{5.59}
p^{(\infty)}(\alpha \vert \underline{\xi}) = \delta_{\alpha \alpha_0},
\end{equation}
for some $\alpha_0$ (depending on $\underline{\xi}$). 

Thus, for almost every history $\underline{\xi}=\underline{\xi}_{\infty}$ of outcomes of "von Neumann measurements" of the probes $E_1,E_2,....,$ the state $\Psi \circ \tau_{k,0}$, conditioned on 
$\underline{\xi}_{\infty}$, converges on $\mathcal{A}_S$ to an \textit{eigenstate} of the operator $a \in \mathcal{A}_S$ representing the physical quantity to be measured, as $k \rightarrow \infty$. The probability (with respect to the histories $\underline{\xi}_{\infty}$) of convergence to an eigenstate corresponding to the eigenvalue $\alpha$ of $a$ is given by $p_{\alpha}$; (see \eqref{5.48}). Stated differently, the range of values of the functions $p^{(\infty)}(\alpha \vert \cdot)$ on the space of histories consists of $\{0,1\}$, and, for almost every history $\underline{\xi}_{\infty}$, $\sum_{\alpha} p^{(\infty)} (\alpha \vert \underline{\xi}_{\infty}) =1$. These are the results that have been announced in Subsection 1.2; see \eqref{1.18}.

It is not hard to see that the approach of the state of $S$ to an eigenstate of $a$ is exponential in the time $k$. This is a "large-deviation estimate" established in \cite{Ba-Be}. It involves use of a "dynamical relative entropy". The techniques sketched in this subsection have interesting applications to Mott's problem of \textit{"particle tracks"} in quantum theory. 

For a mathematical theory of \textit{"preparation of states"} in quantum mechanics, see \cite{F-G-S, Fr-Schub3}. Simple models of "information loss" and "decoherence" will be proposed and studied in a separate publication. 
\begin{subappendices}
\subsection{Appendix  to Section 5}
\label{AppB}
The purpose of this appendix is to describe some mathematical structure useful to imbed the material in Subsections \ref{sec5.3} and \ref{5.4} into a more general context. In particular, we do not wish to assume that the algebras $\mathcal{E}^{\omega}_{\geq t}$ are type-I von Neumann algebras; (i.e., we do not start from Eqs. (\ref{5.17}) -- (\ref{5.19})). To begin with, we summarize some further basic facts concerning von Neumann algebras; (see also Subsection \ref{sec2.1}). Let $\mathcal{M}$ be a von Neumann algebra, and let $\omega$ be a normal state on $\mathcal{M}$. Then $(\pi_{\omega}, \mathcal{H}_{\omega},\Omega)$ stands for the representation, $\pi_{\omega}$, of $\mathcal{M}$ on the Hilbert space $\mathcal{H}_{\omega}$, with $\Omega$ the cyclic unit vector in $\mathcal{H}_{\omega}$ (unique up to a phase) such that 
\begin{equation}
\label{B.1}
\omega(a)=\langle \Omega, \pi_{\omega}(a) \Omega \rangle_{\mathcal{H}_{\omega}}.
\end{equation}
This is the GNS construction applied to $(\mathcal{M},\omega)$; see \eqref{2.7}, Subsection \ref{sec2.1}. We say that $\omega$ is \textit{separating} for $\mathcal{M}$ iff, for any $a \in \mathcal{M}$,
\begin{equation}
\label{B.2}
\omega(ba)=0, \text{ } \forall b \in \mathcal{M} \Longrightarrow a=0;
\end{equation}
or, equivalently, $\pi_{\omega}(a) \Omega=0$ (in $\mathcal{H}_{\omega}$) implies that $a=0$; (it is assumed that $\pi_{\omega}$ is \textit{faithful}, and we will henceforth write $a$ for $\pi_{\omega}(a)$). 

Given a separating state, $\omega$, on a von Neumann algebra $\mathcal{M}$, Tomita-Takesaki theory \cite{MTak2,Brat-Rob} guarantees that there is a one-parameter unitary group $\{\Delta_{\omega}^{i \sigma} \}_{\sigma \in \mathbb{R}}$, where $\Delta_{\omega} >0$ is a self-adjoint operator on $\mathcal{H}_{\omega}$ (the Tomita-Takesaki modular operator) and an anti-unitary involution, $J_{\omega}$, on $\mathcal{H}_{\omega}$, with the properties
\begin{equation}
\label{B.3}
\Delta_{\omega}^{i \sigma} a \Delta_{\omega}^{-i \sigma} \in \mathcal{M}, \quad J_{\omega} a J_{\omega} \in \mathcal{M}',
\end{equation}
for all $a \in \mathcal{M}$ and for all $\sigma \in \mathbb{R}$, ($\mathcal{M}'$ is the commutant of $\mathcal{M}$),
\begin{equation}
\label{B.4}
\Delta_{\omega}^{i \sigma} \Omega=\Omega, \qquad J_{\omega }\Omega =\Omega,
\end{equation}
for all $\sigma$, and 
\begin{equation}
\label{B.5}
\langle \Omega , ab \Omega \rangle_{\mathcal{H}_{\omega}} = \langle \Omega, b \Delta_{\omega}  a \Omega  \rangle_{\mathcal{H}_{\omega}} ,
\end{equation}
for arbitrary $a,b \in \mathcal{M}$; (KMS condition).
If $\varphi$ is a linear functional on $\mathcal{M}$ we define 
\begin{equation}
\label{B.6}
\|\varphi \|:= \underset{b \in \mathcal{M}}{\text{sup}} \frac{\vert \varphi(b) \vert}{\|b \|}
\end{equation}
Eqs. (\ref{B.1}) and (\ref{B.5}) then show that if $\omega$ is separating for $\mathcal{M}$, 
\begin{equation}
\label{B.7}
\| \{ a, \omega ]_{\mathcal{M}} \| < \epsilon \Longleftrightarrow \| (\Delta_{\omega} a -a) \Omega \|_{\mathcal{H}_{\omega}} < \epsilon,
\end{equation}
for any $a \in \mathcal{M}$; (recall that $\{a,\omega]_{\mathcal{M}}(b)= \omega ([a,b])$, $b \in \mathcal{M}$ -- see \eqref{5.11}, Subsection \ref{sec5.2}). In \eqref{5.12}, we have defined the centralizer of $\omega$ to be the subalgebra of $\mathcal{M}$ given by 
\begin{equation}
\label{B.8}
\mathcal{C}^{\omega}_{\mathcal{M}}:=\{ a \in \mathcal{M} \mid \{a,\omega]_{\mathcal{M}}=0 \}.
\end{equation}
We recall that $\omega$ defines a \textit{trace} on $\mathcal{C}^{\omega}_{\mathcal{M}}$. By (\ref{B.7}),
\begin{equation}
\label{B.9}
\mathcal{C}^{\omega}_{\mathcal{M}}=\{ a \in \mathcal{M} \mid \Delta_{\omega} a \Omega =a \Omega \},
\end{equation}
assuming that $\omega$ is separating for $\mathcal{M}$. The following claim is easy to verify (using Liouville's theorem for analytic functions of one complex variable, and \eqref{B.9}): If $\omega$ is separating for $\mathcal{M}$
\begin{equation}
\label{B.10}
\{ a, \omega ]_{\mathcal{M}} =0 \Longleftrightarrow \Delta_{\omega}^{i \sigma}  a \Delta_{\omega}^{-i \sigma} =a, \quad \forall \sigma \in \mathbb{R},
\end{equation}
for any $a \in \mathcal{M}$; (see, e.g., \cite{Araki}). The group, $\{\alpha_{\sigma}\}_{\sigma \in \mathbb{R}}$, of $^*$automorphisms of $\mathcal{M}$ defined by $\alpha_{\sigma }(a)=\Delta_{\omega}^{i \sigma}  a \Delta_{\omega}^{-i \sigma}$ is called the Tomita-Takesaki modular automorphism group. The equivalence in (\ref{B.10}) together with \eqref{B.8} show that if $\omega$ is separating for $\mathcal{M}$ then the centralizer, $\mathcal{C}_{\mathcal{M}}^{\omega}$, is nothing but the subalgebra of $\mathcal{M}$ of fixed points under the Tomita-Takesaki modular automorphism group.  The following result is due to Takesaki, \cite{MTak2}: Let $\mathcal{N}$ be a von Neumann subalgebra of $\mathcal{M}$, and let $\omega$ be a faithful, normal, separating state on $\mathcal{M}$. Then the following statements are equivalent:
\begin{enumerate}[(i)]
\item $\mathcal{N}$ is invariant under the modular automorphism group $\{\alpha_{\sigma} \}_{\sigma \in \mathbb{R}}$ associated with $(\mathcal{M}, \omega)$.
\item There exists a ($\sigma$-weakly) continuous projection, $\epsilon$, of norm $1$ (a "conditional expectation") of $\mathcal{M}$ onto $\mathcal{N}$ such that 
\begin{equation}
\label{B.11}
\omega(a)= \omega_{\mid \mathcal{N}} (\epsilon(a)),
\end{equation}
for all $a \in \mathcal{M}$.
\end{enumerate} 

\begin{remark}
For $a,b $ in $\mathcal{N}$ and $x \in \mathcal{M}$, we have that 
\begin{equation}
\label{B.12}
\left. \begin{array}{c}
\epsilon(x^*x) \geq \epsilon(x)^* \epsilon(x) \geq 0,\\
\epsilon(a x b) = a \epsilon(x) b.
\end{array} \right \}
\end{equation}
As a corollary of Takesaki's result on conditional expectations, we have that if $\omega$ is separating for $\mathcal{M}$ then 
\begin{enumerate}[(a)]
\item there is a conditional expectation, $\epsilon=\epsilon^{\omega}$, from $\mathcal{M}$ onto the centralizer $\mathcal{C}^{\omega}_{\mathcal{M}}$ of $\omega$ satisfying (\ref{B.11}); and 
\item there is a conditional expectation, $\dot{\epsilon}^{\omega}$, from $\mathcal{M}$ onto the center, $\mathcal{Z}^{\omega}_{\mathcal{M}}$, of $\mathcal{C}^{\omega}_{\mathcal{M}}$ satisfying (\ref{B.11}).
\end{enumerate}
\end{remark}
\begin{definition}
The variance of an operator $a \in \mathcal{M}$ in the state $\omega$ is defined by 
\begin{equation}
\label{B.13}
\Delta^{\omega}_{\mathcal{M}} a:= \sqrt{ \omega((a-a^{\omega}))},
\end{equation}
where $a^{\omega}:= \dot{\epsilon}^{\omega}(a)$.
\end{definition}

These general results can be applied to the considerations in subsections \ref{sec5.2}-\ref{sec5.4}, with the following identifications:
\begin{equation}
\mathcal{M} \rightarrow \mathcal{E}^{\omega}_{\geq t}, \quad  \mathcal{C}^{\omega}_{\omega} \rightarrow \mathcal{C}^{\omega}_{\geq t}, \quad \mathcal{Z}_{\mathcal{M}}^{\omega} \rightarrow \mathcal{Z}^{\omega}_{\geq t}.
\end{equation}
We then use the notations $\epsilon^{\omega} \rightarrow \epsilon^{\omega}_{\geq t}$, $\dot{\epsilon}^{\omega} \rightarrow \dot{\epsilon}^{\omega}_{\geq t}$ and $\Delta_{\mathcal{M}}^{\omega} a \rightarrow \Delta^{\omega}_{t} a$; (see \eqref{5.23}, Subsect. \ref{sec5.3}). 
Concerning the special case introduced in Eqs. (\ref{5.17})-(\ref{5.19}), we remark that $\omega$ is separating for $\mathcal{E}^{\omega}_{\geq t}$ iff all eigenvalues of the density matrix $\rho^{\omega}_{\geq t}$ introduced in \eqref{5.19} are strictly positive (which is generically the case). As an exercise, the reader may enjoy deriving the explicit formulae for $\epsilon^{\omega}_{\geq t}$ and $\dot{\epsilon}^{\omega}_{\geq t}$; (see \eqref{5.22}).
The material sketched here is important in relativistic quantum theory (LRQT).

\end{subappendices}
\nocite{*}
\bibliographystyle{plain}
\bibliography{main}

\end{document}